\documentclass[floats,aps,amsfonts,notitlepage,superscriptaddress,twocolumn,eqsecnum,nofootinbib]{revtex4-1}

\usepackage[retainorgcmds]{IEEEtrantools}

\usepackage{ifpdf}
\usepackage[utf8]{inputenc}
\usepackage{graphicx}
\usepackage[colorlinks]{hyperref}
\usepackage{amsmath, amsthm, amssymb}
\usepackage{multirow}
\usepackage{url}
\usepackage{subfigure}
\usepackage{color}
\usepackage[usenames,dvipsnames,svgnames,table]{xcolor}
\usepackage{tabularx}

\allowdisplaybreaks[1]

\definecolor{CiteColor}{rgb}{0,0.5,0}
\hypersetup{citecolor=CiteColor}
\definecolor{RefColor}{rgb}{0.55,0,0}
\hypersetup{linkcolor=RefColor}
\definecolor{darkgreen}{rgb}{0.2,0.7,0.2}

\newcommand{\ab}{{\bar{a}}}
\newcommand{\bb}{{\bar{b}}}
\newcommand{\cb}{{\bar{c}}}
\newcommand{\db}{{\bar{d}}}
\newcommand{\eb}{{\bar{e}}}
\newcommand{\fb}{{\bar{f}}}

\newcommand{\xb}{{\bar{x}}}

\newcommand{\xp}{{x'}}

\newcommand{\rb}{r_0}
\newcommand{\rbdot}{\dot{r}_0}

\newcommand{\ur}{\dot{r}_0}

\newcommand{\dur}{\ddot{r}_0}
\newcommand{\dup}{\ddot{\phi}_0}
\newcommand{\dut}{\ddot{t}_0}

\newcommand{\ddur}{r_0^{(3)}}
\newcommand{\ddup}{\phi_0^{(3)}}
\newcommand{\ddut}{t_0^{(3)}}

\newcommand{\dddur}{r_0^{(4)}}
\newcommand{\dddup}{\phi_0^{(4)}}
\newcommand{\dddut}{t_0^{(4)}}

\newcommand{\phb}{{\bar{\phi}}}
\newcommand{\thb}{{\bar{\theta}}}
\newcommand{\wob}{{\bar{w}_1}}
\newcommand{\wtb}{{\bar{w}_2}}

\newcommand{\zrho}{{\rho}}
\newcommand{\zrhoz}{{\rho_0}}

\newcommand{\adv}{{(adv)}}
\newcommand{\ret}{{(ret)}}
\newcommand{\sing}{{(S)}}

\newcommand{\geo}{\text{geo}}
\newcommand{\diff}[2]  {\frac{d #1}{d #2}}

\renewcommand{\c}{\hskip0.1cm,}
\newcommand{\p}{\hskip0.1cm.}

\newcommand{\PhiS}{\Phi^{\rm \sing}}

\newcommand{\lnpow}[1]{[\scalebox{0.85}{-}#1]}

\DeclareMathOperator{\sgn}{sgn}

\newcommand{\rn}{Reissner-Nordstr\"om }

\begin{document}
\title{Accelerated motion and the self-force in Schwarzschild spacetime}

\author{Anna Heffernan}
\affiliation{Department of Physics, University of Florida, 2001 Museum Road, Gainesville, FL 32611-8440, USA}
\affiliation{School of Mathematics and Statistics, University College Dublin, Belfield, Dublin 4, Ireland}
\affiliation{Advanced Concepts Team, European Space Research and Technology Centre (ESTEC), European Space Agency (ESA), Keplerlaan 1, 2201 AZ Noordwijk, Netherlands}

\author{Adrian C. Ottewill}
\affiliation{School of Mathematics and Statistics, University College Dublin, Belfield, Dublin 4, Ireland}

\author{Niels Warburton}
\affiliation{School of Mathematics and Statistics, University College Dublin, Belfield, Dublin 4, Ireland}

\author{Barry Wardell}
\affiliation{School of Mathematics and Statistics, University College Dublin, Belfield, Dublin 4, Ireland}

\author{Peter Diener}
\affiliation{Center of Computation \& Technology, Louisiana State University, Baton Rouge, LA 70803, USA}
\affiliation{Department of Physics \& Astronomy, Louisiana State University, Baton Rouge, LA 70803, USA}
\begin{abstract}
	
We provide expansions of the Detweiler-Whiting singular field for a particle with a scalar field moving along arbitrary, planar accelerated trajectories in Schwarzschild spacetime. We transcribe these results into mode-sum \emph{regularization parameters}, computing previously unknown terms that increase the convergence rate of the mode-sum. We test our results by computing the self-force along a variety of accelerated trajectories. For non-uniformly accelerated circular orbits we present results from a new 1+1D discontinuous Galerkin time-domain code which employs an effective-source. We also present results for uniformly accelerated circular orbits and accelerated bound eccentric orbits computed within a frequency-domain treatment. Our regularization results will be useful for computing self-consistent self-force inspirals where the particle's worldline is accelerated with respect to the background spacetime.
\end{abstract}

\maketitle

\section{Introduction} \label{sec:intro}
The era of both gravitational-wave \cite{Abbott:2016blz,Abbott:2016nmj,Abbott:2017vtc,Abbott:2017oio,TheLIGOScientific:2017qsa,Abbott:2017gyy} and multi-messenger \cite{GBM:2017lvd} astronomy  has begun. Combined with the successful operation and results of the LISA Pathfinder mission \cite{Armano:2016bkm, Armano:2018} and the prospect of multiband gravitational-wave astronomy \cite{AmaroSeoane:2009ui, Kocsis:2011jy, Sesana:2016ljz, Colpi:2016fup, Sesana:2017vsj} in delivering increased parameter estimation \cite{Vitale:2016rfr, Bonvin:2016qxr}, constraints on evolution channels \cite{Sesana:2009wg, Nishizawa:2016jji, Nishizawa:2016eza, Breivik:2016ddj, Inayoshi:2017hgw} and better tests of general relativity \cite{Barausse:2016eii, Vitale:2016rfr}, to name a few; this has solidified interest in a space-based gravitational-wave detector. To that end, the European Space Agency has formally selected the Laser Interferometer Space Antenna, LISA \cite{LISA}, for its third `Large-class' mission. With a proposed launch date in the early 2030s, there is much to do both experimentally and theoretically, to ensure a successful mission. A key target source for LISA is extreme mass ratio inspirals (EMRIs). These binary systems are expected to form in galactic centres, where a compact object (neutron star or stellar-mass black hole) is `bumped' into the grasp of the massive black hole living at the galactic core, thus forming a highly relativistic orbit lasting some hundreds of thousands of orbits which effectively maps out the Kerr spacetime geometry of the massive black hole \cite{Glampedakis:2005cf,Vigeland:2009pr,Moore:2017lxy}. Such sources not only provide us with pristine tests of general relativity \cite{Barausse:2014tra} but also give us the deepest view into the galactic centres, and with that comes important astrophysical information \cite{Babak:2017tow}.

Modeling binary black holes necessitates solving the two-body problem in the general relativistic context. For dynamic binaries there are no known closed form solutions to the Einstein field equations and so a variety of approximate solution methods have been developed. The self-force approach tackles this problem by expanding the field equations in the (small) mass ratio.  At zeroth order, the smaller object follows a geodesic of the background black hole spacetime. At first order (in the mass ratio) the body's worldline deviates from geodesic motion as it inspirals into the black hole.  This deviation arises from the smaller object interacting with its own field, which is seen as a, so-called, self-force pushing it off its geodesic. By design, the self-force approach is only applicable to systems with a small mass ratio, making it ideally suited to modelling EMRIs.  

With no weak-field or low velocity assumptions, the self-force approximation is applicable for the duration of the inspiral.  For EMRIs, post-Newtonian methods successfully capture the early inspiral, and there has been a plethora of work confirming the agreement of these two methods \cite{Detweiler:2008ft, Blanchet:2009sd, Blanchet:2010zd, Blanchet:2011aha, LeTiec:2011dp, Tiec:2013twa, Dolan:2013roa, Isoyama:2014mja, Akcay:2015pza, Akcay:2016dku}.  Numerical relativity results have also been successfully compared to self-force calculations \cite{Tiec:2013twa, Zimmerman:2018}, however more extreme mass ratios still lie beyond the scope of current state-of-the-art numerical relativity (see \cite{Tiec:2014lba} for a review of application and overlap of the approaches to the relativistic two-body problem). The effective-one-body approach, although applicable to the whole parameter space, requires calibration with results from the other techniques (see e.g., \cite{Damour:2009sm, Barack:2010ny, Akcay:2012ea} for calibration in the EMRI regime).  Thus, the self-force approach finds itself imperative to EMRI modelling and, by extension, the future LISA mission. 

The self-force approach models the smaller compact object as a point particle. As a consequence, a significant issue in self-force calculations arises from the singularity in the field at the particle's location. Much theoretical work has gone into identifying an appropriate singular field \cite{Mino:Sasaki:Tanaka:1996,Quinn:Wald:1997,Detweiler-Whiting-2003,Gralla:Wald:2008,Poisson:2003}. A particularly convenient formulation by Detweiler and Whiting \cite{Detweiler-Whiting-2003} gives a singular field which contains the metric singularity but has zero influence on the motion of the particle. This is then subtracted from the retarded field to leave a fully regular field which, by design, is wholly responsible for the motion of the particle. There are several practical methods available to carry out this removal of the singularity. In this work we concentrate on two: the mode-sum \cite{Barack:1999wf} and effective-source \cite{Vega:Detweiler:2008, Barack:Golbourn:2007} methods.

One key place within the self-force program that this work applies to is self-consistent inspiral evolution. To date, all gravitational self-force calculations have been made by fixing the particle's motion to a geodesic of the background spacetime \cite{Barack:Sago:2007,Barack:Sago:2010,Akcay:2013wfa,Osburn:2014hoa,Merlin:2014qda,vandeMeent:2016pee,vandeMeent:2017bcc}. Inspirals are then computed by solving the equations of motion and making the approximation that the self-force at each instance is that of a particle moving along a tangent to the inspiraling worldline \cite{Warburton:2011fk,Osburn:2015duj,Warburton:2017sxk}. Quantifying the phase error induced by making this `geodesic self-force approximation' requires comparison with a self-consistent inspiral where the self-force at each instance is the true self-force calculated as an integral over the past, inspiraling worldline. Thus far in these assessments, the case of a particle carrying scalar charge \cite{Diener:2011cc} was considered, however, the self-consistent inspiral did not include acceleration terms and so the phase error requires further work.
With the results of this work, these terms can now be included and the evaluations improved.

The results of this work will also find utility in other areas of the self-force program; one such example is when both compact objects are spinning. Modeled as a point particle endowed with a dipole moment, the inspiralling particle's worldline becomes accelerated with respect to the background spacetime due to the Mathisson-Papapetrou-Dixon \cite{Mathisson2010,Papa51,Dixo70} spin-curvature force (see \cite{Kyrian:2007} for a recent review). To-date only dissipative calculations, which do not require a local calculation of the self-force (and hence avoid regularization issues), have been made \cite{Harms:2015ixa,Harms:2016ctx,Han:2010tp}. Going beyond this and making local self-force calculations with a spinning particle \cite{Harte:2011ku} requires a regularization scheme that incorporates accelerated motion. A number of studies have suggested that it will be important to include the effect of the secondary's spin when modeling EMRI dynamics \cite{Burko:2004,Huerta:2011kt,Steinhoff:2012,Burko:2015}, enhancing the need for acceleration terms.

Taking a brief step back from gravitational-wave astronomy, we also note that non-geodesic
self-force calculations have been important for probing the cosmic censorship conjecture
\cite{Penrose:1969pc}. In the context of black hole perturbation theory, this question has received
much attention. Initially Wald showed that an extremal Kerr-Newman black hole could not be
overcharged \cite{Wald:1974}. Later, Hubeny showed a nearly-extremal \rn black hole could be
overcharged, if back reaction were ignored \cite{Hubeny:1998}. Following her work, several groups
carried out calculations including various pieces of the self-force in Kerr \cite{Jacobson:2009,
Barausse:2010, Barausse:2011, Colleoni:2015, Colleoni:2015b} and \rn \cite{Isoyama:2011,
Zimmerman:2012} spacetimes. For the latter, the motion of the charged particle is accelerated with
respect to the background spacetime. Consequently, to include conservative self-force corrections
these studies required a regularization procedure for accelerated motion. These works demonstrated
with explicit calculations how the overspinning or overcharging scenarios are averted once the full
effect of the self-force is taken into account (a result later rigorously proven in a more general
context \cite{Sorce:2017dst}).

The work presented in this article naturally follows on from \cite{HOW:2012, HOW:2014, Heffernan:2014}, hereafter referred to as Paper 1, Paper 2 and HT, respectively.  In Papers 1 and 2, the regularization methods used here were developed extensively following from the mode-sum scheme first introduced by Barack and Ori \cite{Barack:1999wf, Barack:Ori:2000, Barack:2001, Barack:2002mha, Barack:2002bt} and later refined by means of the Detweiler-Whiting singular field \cite{Detweiler-Whiting-2003, Detweiler:2002gi, Haas:2006ne}.  A vital and new component of our work was recognizing the simplification of the procedure by carrying out all calculations in coordinates most suited for regularizing, Riemann normal coordinates (RNC).  In HT, these methods were further modified to give the first four regularization parameters for accelerated motion, (previously, only the specific case of a static particle had been considered - see \cite{Casals:2012qq} for summary).  Linz and colleagues developed a somewhat different approach in \cite{Linz:2014}, hereafter referred to as LFW, using a refined method of that first used by Quinn \cite{Quinn:2000}, also carrying out the entire calculation in RNC.  Their work served to unify the different techniques in obtaining the singular field and resulting regularization parameters and, in doing so, produced the singular field and the first two regularization parameters for non-geodesic motion.  LFW improved on the work in HT by not only considering higher spins, i.e., the electromagnetic and gravitational cases (HT: scalar only), but also was applicable to Kerr (HT: Schwarzschild only). 

This work serves to further formalize and simplify the procedure first produced in HT. Where LFW gave a more broad calculation of the singular field and regularization parameters, we  concentrate on scalar-field perturbations of Schwarzschild spacetime, going to higher order in the singular field and explicitly calculating the resulting self-force.  Higher-order calculations of the singular field and resulting regularization parameters speed up calculations of the self-force significantly, an imperative when considering the tremendous size of the required waveform template banks for EMRI gravitational-wave astronomy.  

This paper is laid out as follows: In Sec.~\ref{sec:sing_and_mode_sum} we derive the singular field and associated regularization parameters for accelerated motion. In Sec.~\ref{sec:AcceleratedMotion} we describe a number of parameterizations for accelerated worldlines. In Sec.~\ref{sec:numerical_results} we give numerical results computed using a modified frequency-domain code and a new discontinuous Galerkin time-domain code. Finally, we give some concluding remarks in Sec.~\ref{sec:concluding_remarks} and the higher-order mode-sum regularization parameters in Appendix \ref{apdx:mode-sum_params}.

Throughout this work we use geometrized units such that the gravitational constant and the speed of light are both equal to unity. We shall denote the mass of the Schwarzschild black hole by $M$ and use standard Schwarzschild coordinates $(t,r,\theta,\phi)$.  In many of our calculations, we have several spacetime points to be considered. Our convention
is that
\begin{itemize}
\item the point $x$ refers to the point where the field is evaluated,
\item the point $\xb$ refers to an arbitrary point on the worldline,
\item the point $x'$ refers to an arbitrary spacetime point,
\item the point $x_{\rm \adv}$ refers to the advanced point of $x$ on the world line,
\item the point $x_{\rm \ret}$ refers to the retarded point of $x$ on the world line.
\end{itemize}
Where tensors or scalars are to be evaluated at these points, we decorate their indices (or themselves in the case of scalars)  appropriately using the relevant notation, e.g., $T^a$, $T^\ab$ and $T^{a'}$ refer to tensors at $x$, $\xb$ and $x'$, respectively.  In computing expansions, we use $\epsilon$ as an expansion parameter to denote the fundamental scale of separation, so that $\Delta x = x-\xb \approx \mathcal{O}(\epsilon)$.  When multiple occurances of the 4-velocities and/or $\Delta x$'s occur, we condense the vectors by $u^{ab} \equiv u^a u^b$ and $\Delta x^{ab} \equiv \Delta x^a \Delta x^b$.  Throughout, as is standard, we denote partial differentiation by a comma and covariant differentiation by a semi-colon.

\section{The Singular Field and Regularization via mode-sum}
\label{sec:sing_and_mode_sum}

The retarded scalar field, $\Phi_{\rm{ret}}(x)$, of an arbitrary point particle
satisfies the inhomogeneous scalar wave equation with a distributional source,
\begin{equation}
\label{eq:Wave}
	\Box \Phi_{\rm{\ret}} = q \int \sqrt{-g} \delta_4(x,z(\tau)) d\tau,
\end{equation}
where $\square \equiv g^{ab}\nabla_{a}\nabla_{b}$, 
$g^{ab}$ is the (contravariant) metric tensor, $\nabla_{a}$ is the standard covariant
derivative
, $q$ is the scalar charge
of the particle and $\delta_4 \left( x,z(\tau') \right)$ is an invariant Dirac functional in a four-dimensional curved spacetime (as defined in Eq.~(13.2) of Ref.~\cite{Poisson:2003}).  This is related to the standard Dirac distribution functional, $\delta_4 \left( x-x' \right)$, by
\begin{equation}
\delta_4 \left( x,x' \right) = \left(g g'\right)^{-1/4} \delta_4 \left( x-x' \right),
\end{equation}
where $g$ and $g'$ represent the metric determinant defined at $x$ and $x'$ respectively.  The retarded solutions to this equation give rise to a field which one can expect to `push' the smaller body off the background geodesic, a.k.a., exert a self-force,
\begin{equation}\label{eqn:FaS}
	F_a = \partial_a \Phi_{\rm{\ret}}.
\end{equation}

With such a distributional source, the self-force will clearly be divergent at the location of the
particle necessitating a regularisation scheme. The Detweiler-Whiting singular field is a solution to the same inhomogeneous scalar wave equation, Eq.~\eqref{eq:Wave}, and hence captures the singular behaviour of $\Phi_{\rm{\ret}}$.  The beauty of such a field becomes clear when subtracted from $\Phi_{\rm{\ret}}$, as it will leave a finite \emph{physical} field wholly responsible for the self-force.

\subsection{The singular field}
To calculate the singular field, the technique remains the same as described in Paper 1, 2 and HT.  We use the standard formula for the scalar Detweiler-Whiting singular field (see Sec.~14.1, 14.2, 14.5 and 17.5 of \cite{Poisson:2003} and Sec.~2.3 of HT for reviewed derivation),
\begin{equation} \label{eq:PhiS}
\PhiS(x) = \frac{q}{2} \Bigg[ \frac{U(x,x')}{\sigma_{c'} u^{c'}} \Bigg]_{x'=x_{\rm \ret}}^{x'=x_{\rm \adv}}
   + \frac{q}{2} \int_{\tau_{\rm \ret}}^{\tau_{\rm \adv}} V(x,z(\tau)) d\tau,
\end{equation}
where $\sigma$ refers to the Synge world function, 
while $U$ and $V$ are known as the direct and tail parts of the singular field respectively.  In the scalar case, $U$ is calculated from the Van Vleck determinant while $V$ is obtained via a Hadamard series, as previously described in Paper 1, 2 and HT.  It should be noted that $V(x,z(\tau))$ of \cite{Poisson:2003} differs from the $V(x,z(\tau))$ here (as well as Papers 1, 2 and HT) by a minus sign. This orginates from the ansatz used; where we have chosen to be consistent with \cite{Detweiler-Whiting-2003} and the Hadamard representation definitions \cite{Decanini:Folacci:2005a}.  From here onwards, for simplicity, we set $q=1$.

In obtaining an expression for the singular field via Eq.~\eqref{eq:PhiS}, we follow the procedure of HT.  This differs from Paper 1 and 2 only in the expressions introduced for the 4-velocity at the point $\xb$ when obtaining expressions for $x^{a'}$ in Eq.~(3.17) of Paper 1, that is
\begin{equation} \label{eqn:xp}
x^{{a'}}(\tau') = x^{\ab} + u^{\ab} \Delta \tau + \tfrac{1}{2!} \dot{u}^{\ab} \Delta \tau^2 + \tfrac{1}{3!} \ddot{u}^{\ab} \Delta \tau^3 +  \cdots,
\end{equation}
where $\bar{x}=x(\bar{\tau})$ is an arbitrary point on the worldline, $u^\ab$ is the contravariant 4-velocity evaluated at $\xb$, $\Delta \tau = \tau' - \bar{\tau}$ and an overdot denotes differentiation with respect $\tau$.  For generic motion in the equatorial plane, one simply keeps our expressions for the 4-velocity and its higher derivatives open, therefore
\begin{equation}
u^{\bar{t}} = \dot{t_0}, \quad u^{\bar{r}} = \rbdot, \quad u^{\bar{\theta}} = 0, \quad u^{\bar{\phi}} = \dot{\phi}_0,
\end{equation}
where we have taken motion to be in the $\theta_0 = \pi/2$ plane without loss of generality (a subscript $0$ denotes a quantity's value on the worldline).  
Practically implementing Eq.~\eqref{eqn:xp}, we find the resulting expressions greatly reduce in complexity if we rewrite the higher derivatives of the 4-velocity in terms of the 4-acceleration, $a^a$, 4-jerk, $\dot{a}^a$ and 4-jounce, $\ddot{a}^a$, explicitly that is,
\begin{eqnarray} 
\label{eqn:4v1}
\dot{u}^a&=&a^a-\Gamma^a{}_{bc} u^{bc},
\\
\ddot{u}^a&=&\dot{a}^a-\big[
\Gamma^a{}_{bc} a^b 
\nonumber \\
\label{eqn:4v2}
&& \quad
+ \left(\Gamma^a{}_{bc,d} - 2 \Gamma^a{}_{be} \Gamma^e{}_{cd} \right) u^{bd}
\big] u^c,
\\
\dddot{u}^a&=&\ddot{a}^a - \big[
3 \Gamma^a{}_{bc} a^c + \big(
5 \Gamma^a{}_{bc,d} + \Gamma^a{}_{cd,b} 
\nonumber \\
&& \qquad
- 5 \Gamma^a{}_{be} \Gamma^e{}_{cd} 
- 7 \Gamma^a{}_{ce} \Gamma^e{}_{bd}
\big) u^{cd}
\big] a^b
\nonumber \\
&&
- \big\{
4 \Gamma^a{}_{bc} \dot{a}^b 
+ \big[
\Gamma^a{}_{bc,de}
- \Gamma^f{}_{bc} \big(
\Gamma^a{}_{de,f} + 4 \Gamma^a{}_{df,e} 
\nonumber \\
&& \quad \quad \quad
- 2 \Gamma^a{}_{fg} \Gamma^g{}_{de}
\big)
- \Gamma^a{}_{bf} \big(
2 \Gamma^f{}_{cd,e}
\nonumber \\
\label{eqn:4v3}
&& \qquad \quad
-4 \Gamma^f{}_{cg} \Gamma^g{}_{de}
\big)
\big] u^{bde}
\big\} u^c. 
\end{eqnarray}
Despite this, we obtain extremely large expressions; as a guide, we include the first two terms below, where the order in $\epsilon$ is denoted in square brackets,
\begin{eqnarray} 
\PhiS_{\lnpow{1}} &=&\frac{1}{\rho},\label{eqn:phiSm1}\\
\PhiS_{[0]} &=&
	-\frac{1}{2 \rho} a_\ab \Delta x^a \nonumber \\
&&+\frac{1}{2 \rho^3} \Big(a_\ab u_{\bb \cb} - \Gamma_{\ab \bb \cb}-\Gamma^\db_{\ab \bb} u_{\cb \db}\Big) \Delta x^{a b c}, \label{eqn:phiS0}
\end{eqnarray}
where $\Delta x=x-\xb$, $a_\ab$ is the covariant 4-acceleration evaluated at $\xb$ while $\rho$ is given by,
\begin{equation} \label{eqn:rho}
\rho^2=(g_{\ab \bb}+u_{\ab \bb}) \Delta x^{ab}.
\end{equation}
It should be noted that Eqs.~\eqref{eqn:phiSm1} and \eqref{eqn:phiS0} are valid for any spacetime, not just Schwarzschild.

In order to obtain expressions usable by the mode-sum technique, previous calculations
\cite{Barack:1999wf, Detweiler:2002gi, Haas:2006ne} found it useful
to work in a rotated coordinate frame.  As previously mentioned in Sec.~\ref{sec:intro}, we found it most efficient to carry out this rotation
prior to doing any calculations, that is, all calculations are done in Riemann normal coordinates.  To this end, we introduce these Riemann normal coordinates on the 2-sphere at $\xb$ in the
form
\begin{equation}
w_{1} = 2 \sin\left(\frac{\alpha}{2}\right) \cos\beta, \quad\quad w_2 = 2 \sin\left(\frac{\alpha}{2}\right) \sin\beta,
\end{equation}
where $\alpha$ and $\beta$ are rotated angular coordinates given by
\begin{eqnarray}
\sin \theta \cos (\phi-\phi'_0) &=& \cos \alpha, \\
\sin \theta \sin (\phi-\phi'_0) &=& \sin \alpha \cos \beta, \\
\cos \theta &=& \sin \alpha \sin \beta.
\end{eqnarray}

\subsection{Mode-sum decomposition}
There are several methods of regularising the self-force, of which, the mode-sum scheme of Barack and Ori \cite{Barack:1999wf, Barack:2001, Barack:2002mha, Barack:2002bt} is one of the most successful.  This involves the decomposition of the singular field into spherical harmonics where each of the resulting multipole modes is finite at the particle, allowing us to conveniently subtract the singular field mode-by-mode.  

As with the singular field, the mode-sum decomposition follows much the same method as that described in papers 1, 2 and HT.  The reader is reminded of Eq.~\eqref{eqn:FaS}, i.e., that the self force for the scalar singular field is obtained by taking a partial derivative of $\PhiS$. A spherical decomposition of the singular and retarded fields allows one to write,
\begin{equation} \label{eqn:Fa}
F_a(\xb)=\sum^\infty_{\ell}\left(F_a^{\ell\rm{\ret}}(\xb) - F_a^{\ell \rm{\sing}} (\xb)\right),
\end{equation}
where 
\begin{eqnarray} \label{eqn:FRS}
F_a^{\ell\rm{\ret}/ \rm{\sing}}(\xb) &=&\lim_{\Delta r \rightarrow 0} \frac{2\ell+1}{4 \pi}  \int F_a^{\rm{\ret}/\rm{\sing}} (\rb+\Delta r,t_0,\alpha,\beta) \nonumber \\
&& \quad \times P_\ell(\cos{\alpha}) d \Omega.
\end{eqnarray}
Here we have evaluated $F_a$ at the particle, i.e., at $\xb=(\rb,t_0,\alpha_0, \beta_0)$ and rotated our system so that the particle lies at the pole (hence the only non-zero $Y^*_{lm}$'s to survive are at $m=0$).  One can see from Eq.~\eqref{eqn:Fa}, to calculate a smooth self-force, it is necessary to calculate the singular version of Eq.~\eqref{eqn:FRS} and in doing so one calculates the mode-sum \emph{regularisation parameters} (with each order in $\Delta x$ corresponding to one regularisation parameter).  As the singular field is separated into orders of $~\Delta x^n$, it is found (and shown later) that the regularisation parameter for each order scales as $\ell^{-n}$; this has the nice implication that the more regularisation parameters one calculates, the faster the convergence of the mode-sum, and by extension, the faster the overall self-force calculation.

From Eqs.~\eqref{eqn:phiS0} and \eqref{eqn:phiSm1}, one can see that taking a partial derivative will leave one with the singular self-force expression in the form,
\begin{equation} \label{eqn:fasum}
F^{\rm{\sing}}_a \left( r, t, \alpha, \beta \right) = \sum_{n=-1}^{\infty} \sum_{p=-n-2}^{\lfloor n/2 \rfloor} B_{a[n]}^{(n -2 p)} \zrho^{2 p - 1} \epsilon^{n-1},
\end{equation}
where 
\begin{eqnarray}
B_{a[n]}^{(k)} &=& b^{[n]}_{\ab \bar{c}_1 \dots \bar{c}_k}(\bar{x}) \Delta x^{c_1} \dots \Delta x^{c_k}. \nonumber \\ 
\end{eqnarray}
To see this explicitly, we give the first two orders of $F_a$, that is,
\begin{eqnarray}
F_{a \lnpow{1}}^{\rm{\sing}}&=&\frac{-\Delta x^b}{\rho^3} \left(g_{\ab \bb}+u_\ab u_\bb \right), \label{eqn:FaSO}\\
F_{a[0]}^{\rm{\sing}}&=&\frac{-a_\ab}{2 \rho}  \nonumber \\
&&+\frac{\Delta x^{bc}}{2\rho^3} \Big[g_{\ab \bb} a_\cb - \left( \Gamma^\db_{\bb \cb} u_\ab + 2 \Gamma^\db_{a \bb} u_\cb \right) u_\db \nonumber \\
&& \quad  -\Gamma_{\ab \bb \cb} - 2 \Gamma_{\bb \ab \cb}+\left(3 a_\cb u_\ab +a_\ab u_\cb \right)u_\bb \Big] \nonumber \\
&&+\frac{3 \Delta x^{bcde} u_\cb}{2\rho^5} \left(g_{\ab \bb} +u_{\ab \bb}\right) \left(\Gamma^\fb_{\db \eb} u_\fb-a_\db u_\eb \right), \nonumber
\end{eqnarray}
where one can read off the $b^{[n]}_{\ab \bar{c}_1 \dots \bar{c}_k}(\bar{x})$ coefficients, e.g., 
\begin{equation}
b^{[0]}_{\ab \bb \cb \db \eb}(\bar{x})=\frac{3 u_\cb}{2} \left(g_{\ab \bb} +u_{\ab \bb}\right) \left(\Gamma^\fb_{\db \eb} u_\fb-a_\db u_\eb \right).
\end{equation}
It should be noted, although written covariantly, $F_{a[0]}^{\rm{\sing}}$ is only valid for Schwarzschild spacetime as simplifications have been made that would not apply to a more complicated spacetime, e.g., Kerr.

In Eqs.~\eqref{eqn:FaSO}, $F_{a}^{\rm{\sing}}$ is still being evaluated at the field point, $x$; however, in Eq.~\eqref{eqn:FRS}, we need $F_{a}^{\rm{\sing}} (r,t_0,\alpha,\beta)$, i.e., on breaking covariance we can set $\Delta t=0$. It is therefore beneficial to have an expression for $\zrho\left(r, t_0, \alpha, \beta \right)$, particularly in the form

\begin{equation} \label{eqn:rho1}
 \zrho \left(r, t_0, \alpha, \beta \right)^2 = \nu^2 \Delta r^2 + \zeta^2 \left(\Delta w_1 - c \Delta r \right)^2 + \xi^2 \Delta w_2^2.
\end{equation} 
Using Eq~\eqref{eqn:rho}, the diagonality of  Schwarzschild spacetime and $u_\wtb=u_\thb=0$, these can be quickly calculated as,
\begin{eqnarray}
\zeta^2 &=& g_{\wob \wob} +u_\wob^2, \quad \xi^2 = g_{\wtb \wtb}, \label{eqn:zeta}
\\
\nu^2&=&\frac{g_{\bar{r} \bar{r}} g_{\wob \wob}}{\zeta^2} \left(1+\frac{u_\wob^2}{g_{\wob \wob}}+\frac{u_{\bar{r}}^2}{g_{\bar{r} \bar{r}}} \right),\nonumber \\
&=& \frac{g_{\bar{r} \bar{r}}{}^2 g_{\wob \wob} } {\zeta^2} u_{\bar{t}}^2, \label{eqn:nu}\\
c&=&-\frac{u_\wob u_{\bar{r}}}{\zeta^2}.\label{eqn:c}
\end{eqnarray}

Some care is required in order to obtain easily integrable expressions in the case of eccentric orbits; we use the approach of previous methods \cite{Detweiler:2002gi,Barack:1999wf,Mino:2001mq,Haas:2006ne} (also employed in Paper 1, 2 and HT).  By using the coordinate freedom of $\phi_0$, we redefine $\Delta w_1$ to ensure $\Delta r \Delta w_1$ cross terms in $\rho (r,t_0,\alpha, \beta)$ vanish; that is, we make the replacement $\Delta w_1 \rightarrow \Delta w_1 + c \Delta r$.

The first regularisation parameter (denoted $L A_a$ in the older notation of \cite{Detweiler:2002gi,Barack:1999wf,Mino:2001mq,Haas:2006ne} and $F^\ell_{a \lnpow{1}}$ in the later notation of Paper 1,2 and HT) can be calculated from Eq.~(2.35) of Paper 2 (similar derivation in HT), that is, 
\begin{IEEEeqnarray}{rCl} \label{eqn: Fa1}
F^\ell_{a \lnpow{1}}\left(r_0,t_0\right) &=& \left(\ell + \tfrac{1}{2} \right) \frac{\tilde{b}_{\ab \bar{r}}^{\lnpow{1}} \sgn \left(\Delta r\right)} {\zeta \nu \xi},
\end{IEEEeqnarray}
where $\xi^2=g_{\wtb \wtb}=\rb^2$ from above and $\tilde{b}_{\ab \bar{r}}$ are the coefficients of $\Delta r$ as described in Eq.~\eqref{eqn:fasum}, where (importantly) the tilde denotes that we have made the transformation  $\Delta w_1 \rightarrow \Delta w_1 + c \Delta r$. From Eq.~\eqref{eqn:FaSO}, one can read off $b_{\ab \bar{r}} = -g_{\ab \bar{r}}-u_\ab u_{\bar{r}}$, however taking the above transformation, this now becomes
\begin{equation}
\tilde{b}_{\ab \bar{r}}^{\lnpow{1}} = -g_{\ab \bar{r}}-u_\ab u_{\bar{r}} - c \left( g_{\ab \bar{w}_1}+u_\ab u_{\bar{w}_1}  \right),
\end{equation}
which explicitly gives,
\begin{equation} \label{eqn:bs}
\tilde{b}_{\bar{r} \bar{r}}^{\lnpow{1}}=-\nu^2, \quad \tilde{b}_{\bar{t} \bar{r}}^{\lnpow{1}}=-\frac{u^{\bar{r}}}{u^{\bar{t}}} \tilde{b}_{\bar{r} \bar{r}}^{\lnpow{1}}, \quad\tilde{b}_{\wob \bar{r}}^{\lnpow{1}}=0, \quad \tilde{b}_{\bar{w}_2 \bar{r}}^{\lnpow{1}}=0.
\end{equation}
Combining Eqs.~\eqref{eqn:zeta}, \eqref{eqn:nu}, \eqref{eqn: Fa1} and \eqref{eqn:bs} we arrive at
\begin{eqnarray}
F_{r \lnpow{1}}^\ell &=& - \left(\ell+\frac{1}{2} \right)\frac{\sgn\left(\Delta r\right)}{V} \frac{u^{\bar{t}}}{\left(g_{\phb \phb} g_{\thb \thb}\right)^{1/2}}, \nonumber \\
&=& \left(\ell+\frac{1}{2} \right)\frac{\sgn\left(\Delta r\right)}{V \rb^2} \frac{E}{g_{\bar{t} \bar{t} }},\\
F_{t \lnpow{1}}^\ell &=& \left(\ell+\frac{1}{2} \right) \frac{\sgn\left(\Delta r\right)}{V} \frac{u^{\bar{r}}}{\left(g_{\phb \phb} g_{\thb \thb}\right)^{1/2}}, \nonumber \\
&=& \left(\ell+\frac{1}{2} \right) \frac{\sgn\left(\Delta r\right) u^{\bar{r}}}{V \rb^2}, \\
F_{\theta \lnpow{1}}^\ell &=& 0, \qquad F_{\phi \lnpow{1}}^\ell = 0,
\end{eqnarray}
where $V=1+\frac{u_\phb^2}{g_{\phb \phb}}$ and we have transformed back to Schwarzschild coordinates (at the particle, $g_{\wob \wob}=g_{\phb \phb}$, $g_{\wtb \wtb}=g_{\thb \thb}$, and $u_\wob = u_\phb$).  Written in this form, the agreement with the $L A_a$'s ($\equiv F^\ell_{a \lnpow{1}}$'s here) in Eq.~(83a) of \cite{Barack:2002mha} is immediate.  Although their calculations are for a geodesic trajectory, it is unsurprising that the first regularisation parameters are the same at leading order, our non-geodesic motion is `hidden' in the undetermined nature of the 4-velocity.  Thus, the expected departure from geodesic behaviour emerges with terms involving the 4-acceleration, 4-jerk, etc.  As these only appear in the next to leading and higher orders in our expression derived for $\xp$, i.e., Eq.~\eqref{eqn:xp}, a deviation from the geodesic parameters is expected to appear at next to leading order, $F^l_{a[0]}$, and higher.

In \cite{Barack:1999wf}, it was shown that the integral and limit in Eq.~\eqref{eqn:FRS} are interchangeable for all orders except the leading order term, where the limiting $\Delta r/\rho^{3}$ is not integrable.  This means our higher-order expressions can be more easily integrated than the leading-order expression.  When one interchanges the limit and integral, we can set $\Delta r=0$ in  $F_{a}^{\ell (S)}$, necessitating expressions for $\rho_0=\rho(r_0,t_0,\alpha,\beta)$; from Eq.~\eqref{eqn:rho1},
\begin{eqnarray} \label{eqn:rho0}
	\rho_0 \left(\alpha, \beta \right)^2  &=& \zeta^2 \Delta w_1 ^2+\xi^2\Delta w_2^2, \nonumber \\
										  &=&2 \left(1 - \cos{\alpha}\right) \zeta^2 \chi(\beta),
\end{eqnarray}
where
\begin{equation} \label{eqn:k}
	\chi(\beta) \equiv 1 - k \sin^2 \beta, \qquad k = \frac{\zeta^2-\xi^2}{\zeta^2}.
\end{equation}
Following Sec.~\ref{sec:mode-sum_params} in Paper 2, we rearrange these and use $2 \left(1 - \cos{\alpha}\right) = \rho_0^2 / (\zeta^2 \chi)$ and $\sin^2 \beta = (1-\chi)/k$ to rewrite our $\Delta w$'s as,
\begin{align}
	\Delta w_1 ^2 &= \frac{\zrhoz^2}{\left(\zeta^2-\xi^2\right)\chi}\left[k-(1-\chi)\right], \\
	\Delta w_2^2 &= \frac{\zrhoz^2}{\left(\zeta ^2 - \xi^2 \right)\chi} (1 - \chi),
\end{align}
where we have used $\Delta w_n = w_n$ as $\bar{w}_n=0$ at the pole.  We can now pull out a factor of $\rho_0/L$ from our $\Delta w$'s (using $\zeta^2-\xi^2=L^2$ for Schwarzschild spacetime). Carrying out the above substitution allows us to rewrite Eq.~\eqref{eqn:fasum} in orders for $n \geq 0$ as
\begin{equation} \label{eqn:FaSimp}
F^{\rm{\sing}}_{a[n]} \left( r, t, \alpha, \beta \right) = \zrhoz^{n -1 } \epsilon^{n-1}\sum_{p=-n-2}^{\lfloor n/2 \rfloor} c_{a[n]}^{(n -2 p)} L^{2p-n},
\end{equation}
where $c_{a[n]}^{(k)} (\rb,\chi) \equiv (\rho^k / L^k) B_{a[n]}^{(k)}$ and we have eliminated all odd powers of $\Delta w_1$ and  $\Delta w_2$ (being odd functions ensures their disappearance on integration).  For illustration purposes, the $n=0$ coefficients are
\begin{eqnarray}
c_{a[0]}^{(0)} (\rb,\chi) &=&-\frac{a_\ab}{2},\\
c_{a[0]}^{(2)} (\rb,\chi) &=& \frac{1}{2 \zeta^2 \chi} \big\{  \left( \rb^2-\zeta^2 \chi \right) \big[\frac{2L^2 \delta_a^r}{\rb} - L^2 a_\ab \nonumber \\
&& - \delta_a^{w_1} \rb \left(a_\wob \rb + 2 L \ur \right) -3 L u_\ab a_\wob \big] \nonumber \\
&& - L^2 \rb \left( \delta_a^r  -  u_\ab \ur \right) \big\}, \nonumber \\
c_{a[0]}^{(4)} (\rb,\chi) &=& \frac{3 L}{2 \zeta^4 \chi^2} \left( \delta_a^{w_1} \rb^2 + L u_\ab\right) \left( \rb^2-\zeta^2 \chi \right) \big[ \rb \ur L^2 \nonumber \\
&& - L a_\wob \left( \rb^2-\zeta^2 \chi \right) \big].\nonumber
\end{eqnarray}

From Eq.~\eqref{eqn:FaSimp}, it is clearly beneficial to have an `integration friendly' $\rho_0$; following the methods of Paper 1, 2 and HT, we rewrite $\rho_0$ as
\begin{IEEEeqnarray}{rCl} \label{eqn:rhon}
	\zrhoz \left(r_0, t_0, \alpha, \beta \right)^n &=& \zeta^n \left[2\chi \left( 1 - 
	\cos \alpha \right)\right]^{{n}/{2}} \nonumber \\
&=&
	\zeta^n \left(2\chi \right)^{n/2} \sum_{\ell=0}^{\infty} \mathcal{A}_{\ell}^{{n}/{2}} (0) P_{\ell} 
	\left( \cos \alpha \right), \nonumber
\end{IEEEeqnarray}
where $\mathcal{A}_{\ell}^{-\frac{1}{2}} (0) = \sqrt{2}$, from the generating function
of the Legendre polynomials and, as given in Appendix~D of
\cite{Detweiler:2002gi}, for $(n+1)/2\in\mathbb{N}$,
\begin{align}
	\mathcal{A}^{n/2}_{\ell} \left(0 \right) =& 
		 \frac{\mathcal{P}_{n/2} \left( 2 \ell + 1\right) }{\left(2 \ell - n\right)\left(2 \ell - n +2\right) \cdots \left(2 \ell + n\right) \left(2 \ell + n +2 \right)},
\end{align}
where
\begin{align}
	\mathcal{P}_{n/2} = &\left(-1\right)^{(n+1)/2} 2^{1 + n/2} \left( n!! \right)^2. \nonumber \\
\end{align}
We now have $F^{\rm{\sing}}_{a[n]} $ written in an easily integratable format, by combining Eqs.~\eqref{eqn:FaSimp} and \eqref{eqn:rhon}, we have
\begin{eqnarray}
	F^{\rm{\sing}}_{a[n]} \left( r, t, \alpha, \beta \right)  &=& \zeta^{n-1} \left(2\chi \right)^{(n-1)/2} \sum_{\ell=0}^{\infty} \mathcal{A}_{\ell}^{{(n-1)}/{2}} (0)  \nonumber \\
	&& \times P_{\ell} \left( \cos \alpha \right) \epsilon^{n-1} c_{a[n]}(\rb,\chi),
\end{eqnarray}
where we have shortened our expressions by using the identity,
\begin{equation}
c_{a[n]}(\rb,\chi) \equiv \sum_{p=-n-2}^{\lfloor n/2 \rfloor} c_{a[n]}^{(n -2 p)} L^{2 p-n}
\end{equation}
When substituted into Eq.~\eqref{eqn:FRS} for the higher terms, the integration over $\alpha$ becomes trivial due to the orthogonal nature of the Legendre polynomials, while the $\beta$ integration reduces to known elliptic integrals, as discussed in detail in Papers 1, 2 and HT.
 
 \subsection{Mode-sum regularization parameters for accelerated motion}\label{sec:mode-sum_params}

 To aid comparison with the geodesic case we use the notations $E=-u_t = (1-2M/r) \dot{t} $
 and $L=u_\phi = \dot{\phi}/r^2$ where here and throughout this section an overdot denotes differentiation with respect to proper-time along the particle's worldline. Furthermore, by the definition of proper time
 \begin{align*}
	 \dot{r}^2 = E^2 - \left(1-\frac{2M}{r}\right)\left(1 + \frac{L^2}{r^2}\right)
 \end{align*}

We also define $\mathcal{K} = \int_0^{\pi/2}(1-k\sin^2\theta)^{-1/2}\,d\theta$ and $\mathcal{E} = \int_0^{\pi/2}(1-k\sin^2\theta)^{1/2}\,d\theta$ as the complete elliptic integrals of the first and second kind, respectively, where $k$, defined by Eq.~\eqref{eqn:k}, simplifies to $k=L^2/(r_0^2 + L^2)$ . In writing the regularization parameters we use the 4-acceleration, $a^a$ and its derivatives, the 4-jerk, $\dot{a}^a=D_U a^a \equiv \partial_t a^a + \Gamma^a_{b c} a^b u^c$, and the 4-jounce, $\ddot{a}^a = D^2_U a^a$, as calculated in Eqs.~(\ref{eqn:4v1}~-~\ref{eqn:4v3}).  All $a^a$'s below are evaluated at $\xb$. The explicit form of these in terms of $E,L, \rb$, the 4-velocity and its derivatives can be found in Appendix \ref{apdx:mode-sum_params}.  

The regularization parameters for the self-force and the self-field are given below. The additional parameters, $F_{a}{}_{[2]}$, that act to further regularize and improve the convergence rate of the mode sum are given in Appendix \ref{apdx:mode-sum_params}.  We follow the notation of Papers 1 and 2, where
\begin{align}
 F^l_{a}{}_{\lnpow{1}} = & (2l+1)  F_{a}{}_{\lnpow{1}}, \\
 F^l_{a}{}_{[0]} = &  F_{a}{}_{[0]} ,\\ 
 F^l_{a}{}_{[2]} = &  \frac{F_{a}{}_{[2]}}{(2l-1) (2l+3)}.
\end{align}
We then arrive with,

 \begin{gather}
 F_{t}{}_{\lnpow{1}} = \frac{ \ur \sgn(\Delta r)}{2(
    L^2+\rb^2)} , \\
 F_{r}{}_{\lnpow{1}} = -\frac{\sgn(\Delta r) E \rb
    }{2 (\rb-2 M )
    ( L^2+\rb^2)},\\
 F_{\theta}{}_{ \lnpow{1}} = F_{\phi}{}_{ \lnpow{1}} =0.
 \end{gather}

 \begin{equation}
 F_{t}{}_{[0]} = \frac{F_{t\,\mathcal{E}}{}_{[0]} \mathcal{E} + E \rb^2 F_{t\,\mathcal{K}}{}_{ [0]} \mathcal{K}} {\pi L \rb (L^2+\rb^2 )^{3/2}},
 \end{equation}

where

\begin{align}
F_{t\,\mathcal{E}}{}_{[0]}  =& a^t L \left(r-2 M \right)\left(L^2+\rb^2\right) \nonumber \\
		&-E \rb^2 \left[a^\phi \rb\left(L^2-\rb^2\right)+2 L \ur \right],\nonumber\\
F_{t\,\mathcal{K}}{}_{[0]}  =&L \ur-a^\phi \rb^3, \nonumber
\end{align}

 \begin{equation}
 F_{r\,[0]} = \frac{ F_{r\,\mathcal{E}\,[0]} \mathcal{E} + F_{r\,\mathcal{K}\,[0]} \mathcal{K} }{\pi  L \rb (\rb-2 M)(L^2+\rb^2){}^{3/2} },
 \end{equation}
 where
 \begin{align}
 F_{r\,\mathcal{E}}{}_{[0]} =&\,\, 2 L E^2 \rb^3 -L (L^2+\rb^2)\left(\rb-2 M +a^r \rb^2 \right)  \nonumber \\
  							&+a^\phi \rb^4 \ur(L^2-\rb^2), \nonumber\\
 F_{r\,\mathcal{K}}{}_{[0]} =& a^\phi \rb^6 \ur -L(L^2+\rb^2)(\rb-2 M) -L E^2 \rb^3, \nonumber
 \end{align}

 \begin{equation}
 F_{\theta}{}_{[0]} = 0,
 \end{equation}

\begin{equation}
 F_{\phi}{}_{[0]} =\frac{\rb \left(L \ur- \rb^3 a^\phi\right)}{\pi L^2 \left(L^2+\rb^2 \right)^{1/2}} \left(\mathcal{K} - \mathcal{E} \right).
\end{equation}

It is also beneficial to regularize the field directly - to this end, the regularization parameters for the self-field are given by
\begin{equation}
	\Phi_{[0]} = \frac{2\mathcal{K}}{\pi\sqrt{L^2 + \rb^2}},
\end{equation}
and
\begin{equation}
	\Phi_{[2]} = \frac{\Phi_{\mathcal{E}[2]} \mathcal{E}+\rb^2 \Phi_{\mathcal{K}[2]} \mathcal{K}}{3 \pi L^2 \rb^3 \left(L^2+\rb^2\right)^{3/2}},
\end{equation}
where
\begin{align}
\Phi_{\mathcal{E}[2]}=&L (L^2+\rb^2) \big[a^2 L \rb^3 \left( 8L^2+7 \rb^2 \right)-3 L\rb^3 \left(2 a^r  \rb +  1\right) \nonumber \\
				& - 4 \dot{a}^\phi \rb^5  \left( 2 L^2+\rb^2 \right) + 6 M L \left(2 L^2+3 \rb^2 \right) \big] \nonumber \\
				& -3 a^\phi \rb^6 (L^2-\rb^2) \left( a^\phi \rb^3-2 L \ur \right) +6 E^2 L^2 \rb^5, \nonumber\\
\Phi_{\mathcal{K}[2]}=&2 L (L^2+\rb^2) (2 \dot{a}^\phi \rb^5-2 a^2 L \rb^3-3 M L) \nonumber \\
				&-3 \rb^3 \left[a^\phi \rb^3\left(a^\phi \rb^3-2 L \ur \right)+L^2 E^2\right].\nonumber
\end{align}

\section{Specific examples of accelerated motion}
\label{sec:AcceleratedMotion}

Generic accelerated motion in a given spacetime can be described by the forced geodesic equation,
\begin{align}
	\label{eq:forced_geodesic} u^b \nabla_b (\mu u^a) = F^a = \mu a^a + \frac{d\mu}{d\tau} u^a, 
\end{align} 
where $F^a$ is the force and $\mu$ is the particle's rest mass. The force can be split into
a piece orthogonal to the 4-velocity (tangential to the 4-acceleration, $a^a$) and a
piece tangential to the 4-velocity, $u^a$ (orthogonal to the 4-acceleration). In the
case of self-accelerating electromagnetic and massive particles, the 4-velocity and self-force
must be orthogonal ($F^a u_a=0$) and there is no tangential component. For the case of a
particle carrying a scalar charge, orthogonality is not required and any tangential component of
the acceleration gives rise to a dynamically varying rest mass \cite{Quinn:2000}. Likewise, for a
generic external force it is possible to have both orthogonal and tangential components of the
force. In the examples studied in the remainder of this work we will avoid this complication by
only considering accelerations which are orthogonal to the 4-velocity. In practice, we achieve
this by prescribing a \emph{worldline} rather than an acceleration, and we define the mass to be a
constant. It should be emphasized, however, that this is not a fundamental restriction, and all
results up to this point are in fact valid for generic accelerating worldlines. Indeed, in a
forthcoming work \cite{InspiralComparison} we will present an application of the time domain code
(discussed in Sec.~\ref{sec:TD} below) to the problem of self-consistent orbital evolution
including self-force effects, a problem which does not fall into one of the simplified categories
considered here.

One of the aims of this work is to compare the self-force experienced by a particle with an accelerated
worldline with that of a particle moving along an associated geodesic. It is thus natural to parametrize the
particle's motion in a similar fashion to the way geodesic motion is often described. In
particular, we will explore worldlines that trace out the same spatial path as a 
geodesic, but allow the rate at which the path is traced to differ (this may be interpreted as
an external acceleration being imposed on the worldline). This choice and parametrization is
convenient as it allows for the easy modification of already existing self-force codes to explore
non-geodesic orbits.

In the following sections we consider two specific classes of orbit, which are of particular
interest: (i) circular orbits; and (ii) bound eccentric orbits. In both cases we shall for
simplicity assume the motion lies in the equatorial plane ($\theta_0=\pi/2,u^\theta=0$).

\subsection{Circular orbits}\label{sec:circ_accel_description}

For circular orbits we have $u^r=0$, hence the spatial trajectory of the orbit is uniquely specified by the orbital radius, $r_0$; and the rate at which the azimuthal angle accumulates uniquely defines the particle's worldline. To describe the azimuthal accumulation we define the function
\begin{align}\label{eq:Omega_phi_circ_accel}
	\Omega_\phi \equiv \Omega_\phi(t) = \frac{d \phi}{dt}.
\end{align}
Note that we are not defining an average rate of azimuthal accumulation, as is often done for
geodesic motion, hence, for non-uniformly accelerated orbits, this quantity should not be
interpreted as an orbital frequency. The two non-zero components of the 4-velocity are simply
related via
\begin{align}\label{eq:circ_u^phi}
	u^\phi(t) = \Omega_\phi(t) u^t(t).
\end{align}
Combining this with the normalization of the 4-velocity, $u^a u_a = -1$, we get
\begin{align}\label{eq:circ_u^t}
	u^t(t) = \left[f_0 - r_0^2 \Omega_\phi(t)^2\right]^{-1/2},
\end{align}
where $f_0 = 1-2M/r_0$. The particle's (specific) energy and angular-momentum are given in the usual way,
\begin{align}
	E(t) 	&\equiv -u_t(t) = f_0 u^t(t),																		\\
	L(t) &\equiv u_\phi = r_0^2 \Omega_\phi(t) u^t(t).		\label{eq:circ_L}
\end{align}
For geodesic motion the rate of azimuthal accumulation is constant and we have
\begin{align}\label{eq:Omega_varphi_geo}
	\Omega_\phi^\geo = \left(\frac{M}{r_0^3}\right)^{1/2},
\end{align}
where hereafter a superscript `$\geo$' denotes a quantity's value for geodesic motion. The other quantities associated with circular geodesic motion are obtained by substituting Eq.~\eqref{eq:Omega_varphi_geo} into Eqs.~\eqref{eq:circ_u^phi}-\eqref{eq:circ_L}.

Equation \eqref{eq:circ_u^t} restricts the values of $\Omega_\phi$ to:
\begin{align}
	0 \le | \Omega_\phi(t) | < \sqrt{f_0/r_0^2}.
\end{align}
The lower limit represents a static particle and the upper limit corresponds to the particle approaching the speed of light (which, as it has non-zero rest mass, it can never attain). 

\subsection{Bound eccentric orbits}\label{sec:bound_eccentric_orbits_parametrization}

For bound orbits there exist minimum and maximum orbital radii which we shall denote by $r_{\min}$
and $r_{\max}$, respectively. From these we can define the dimensionless semi-latus rectum, $p$, and orbital
eccentricity, $e$ via
\begin{align}
	p 	&= \frac{2r_{\max} r_{\min}}{M(r_{\max} + r_{\min})},				\\
	e	&= \frac{r_{\max} - r_{\min}}{r_{\max} + r_{\min}}.
\end{align}
To describe the particle's worldline we introduce the relativistic anomaly parameter
\cite{Darwin-1961}, $\chi$, in terms of which we write the radial and azimuthal angle of the
particle in the standard way \cite{Cutler-Kennefick-Poisson}
\begin{align}
	r_0(\chi) 							&= \frac{p M}{1+e \cos \chi}, \\
	\diff{\phi_0}{\chi} 	&= \sqrt{\frac{p}{p-6-2e\cos \chi}},
\end{align}
where we have chosen $\chi=0$ to be at periastron ($r=r_{\min}$). This parametrization of the motion allows us to describe all
accelerated orbits that follow the same spatial path as a corresponding bound eccentric geodesic.
The particular acceleration of the orbit will be specified by choosing the relation between $t_0$
and $\chi$. We will discuss various different choices for this relation in
Sec.~\ref{sec:results_ecc}. Once $dt_0/d\chi$ is specified, the complete description of the
particle's worldline is then given as follows.

From the normalization of the 4-velocity we have
\begin{align}
	\diff{\tau}{\chi} = \left[ f_0 \left(\diff{t_0}{\chi}\right)^2 - \frac{1}{f_0}\left(\diff{r_0}{\chi}\right)^2 - r_0^2 \left(\diff{\phi_0}{\chi}\right)^2 \right]^{1/2}\c
\end{align}
where $f_0 \equiv f_0(\chi) = 1-2 M/r_0(\chi)$ and
\begin{align}
	\diff{r_0}{\chi} = \frac{p M e\sin \chi}{(1+e\cos \chi)^2}\p
\end{align}

Unlike for geodesic orbits, the energy, $E$, and angular momentum, $L$, are functions of $\chi$. They are computed via
\begin{align}
\label{eq:NonGeodesicEL}
	E(\chi) 	&= f_0 \diff{t_0}{\chi}\bigg/\diff{\tau}{\chi}\c		\qquad L(\chi) 	= r_0^2 \diff{\phi_0}{\chi}\bigg/\diff{\tau}{\chi}\p
\end{align}
The 4-velocity is similarly computed via $u^a(\chi) = (dx_0^a/d\chi)/(d\tau/d\chi)$.
The 4-acceleration is given by $a^a = u^b\nabla_b u^a$, where $\nabla$ is the
covariant derivative. In Schwarzschild coordinates the three non-zero components of the
4-acceleration are given by
\begin{align}
	a^t 			&= (u^t)'/(d\tau/d\chi) + \frac{2}{f_0 r_0^2} u^t u^r\c		\\
	a^r 			&= (u^r)'/(d\tau/d\chi) + \frac{f_0}{r_0^2} (u^t)^2 - \frac{(u^r)^2}{f_0 r_0^2}  - f_0 r_0 (u^\phi)^2\c	\\
	a^\phi &= (u^\phi)'/(d\tau/d\chi) + \frac{2}{r_0}u^\phi u^r\c
\end{align}
where a $'$ denotes differentiation with respect to $\chi$.

If we assume that the acceleration is periodic with respect to $\chi$, i.e., $t_0(\chi) = t_0(\chi + 2\pi)$ then we can define orbital frequencies as follows. The azimuthal accumulation over one orbit is given by
\begin{align}
	\Delta\phi = \int^{2\pi}_0 \diff{\phi_0}{\chi}\,d\chi = 4\left(\frac{p}{p-6+2e}\right)^{1/2}\mathcal{K}\left(\frac{4e}{p-6+2e}\right)\c
\end{align}
where $\mathcal{K}(k) = \int^{\pi/2}_0(1-k\sin\theta)^{-1/2}\,d\theta$ is again the complete elliptic integral of the first kind. The coordinate time between successive periastron passages is given by
\begin{align}
	T_r = \int^{2\pi}_0 \diff{t_0}{\chi}\,d\chi\p
\end{align}
The orbital frequencies associated with the motion are then defined in the usual way:
\begin{align}
	\Omega_r  = \frac{2\pi}{T_r},\qquad \Omega_\phi = \frac{\Delta\phi}{T_r}\p
\end{align}

So far in this section all the equations given are valid for generic motion along the spatial
trajectory corresponding to a bound, eccentric orbit with the same values of $p$ and $e$. In
order to specify geodesic motion the relation between $t_0$ and $\chi$ is given by
\cite{Cutler-Kennefick-Poisson}
\begin{align}
	\left(\diff{t_0}{\chi} \right)_\geo =& \frac{M p^2}{(p-2-2e\cos\chi)(1+e\cos\chi)^2}		\nonumber\\
										 & \times \sqrt{\frac{(p-2-2e)(p-2+2e)}{p-6-2e\cos\chi}}.	\label{eq:dt_dchi_geo}
\end{align}	

\section{Numerical calculation of the scalar self-force along accelerated trajectories}
\label{sec:numerical_results}

With a regularization scheme appropriate for accelerated worldlines now at hand, we may explore
interesting orbital configurations which were previously inaccessible using a geodesic-based
regularization procedure. To that end, we have used two numerical codes to compute the self-force
in a variety of scenarios. The two codes are complementary: we use a highly-accurate frequency
domain code in scenarios where the Fourier spectrum is sufficiently narrow, and we use a time-domain code both as a check on our
frequency-domain results, and in scenarios that the frequency-domain approach is unable to tackle.
Both codes are discussed in detail in other works \cite{Warburton-Barack:eccentric,Diener_etal:dG_code} and, as such, we will only summarize their
key features below.

\subsection{Frequency-domain calculation}

Our frequency-domain approach follows very closely that of \cite{Warburton-Barack:eccentric}, in which the
field equation is decomposed into spherical harmonic and Fourier modes, resulting in a set of
uncoupled ordinary differential equations, one for each spherical-harmonic $(\ell,m)$ mode and for
each frequency, $\omega$. This frequency-domain approach is best suited to cases in which the
spectrum of frequencies in the solution is reasonably small. In terms of orbital configurations,
this restriction translates to exploring bound periodic orbits, which in turn implies that the
acceleration must also be periodic (though see \cite{Hopper:2017iyq} 
and \cite{Colleoni:Soichiro:Sago:Barack:2018}
for progress modeling unbound orbits in the frequency-
and time-domain
respectively). Furthermore, the convergence of the Fourier sum
over frequencies is very poor if the worldline has any rapidly-varying features. We, therefore, restrict our attention in the frequency-domain case to a subset of scenarios involving accelerating
worldlines in which the orbit remains periodic and does not vary rapidly.

We also make a further simplification: by only considering cases in which the orbital motion traces
out the same path as a corresponding geodesic orbit, but allowing the rate at which the orbit is
traced out to differ from the geodesic value by a \emph{constant} factor. Given these restrictions, the two scenarios
described in Sec.~\ref{sec:AcceleratedMotion} above admit special cases which are particularly
numerically tractable: uniformly accelerated circular orbits and eccentric orbits which modify the
geodesic relation
\eqref{eq:dt_dchi_geo} by a constant multiple. The only difference relative to the geodesic case discussed in \cite{Warburton-Barack:eccentric} is now:
\begin{enumerate}
  \item The 4-velocity appearing in the source term is now no longer that of a geodesic, but
        instead is the 4-velocity of the accelerating worldline.
  \item The frequencies that are excited now differ from those of the geodesic case.
\end{enumerate}
The consequences of both of these follow trivially from the fact that \eqref{eq:dt_dchi_geo} [or,
equivalently, \eqref{eq:Omega_phi_circ_accel} in the circular orbit case] is modified by a constant
multiplicative factor: the changes to the source term can be encapsulated in the changes in
$E$ and $L$ as given in Eq.~\eqref{eq:NonGeodesicEL}, while the new frequencies
can be obtained exactly as in the geodesic case, by rescaling Eq.~\eqref{eq:dt_dchi_geo} by an
appropriate multiplicative factor.

Specializing to a class of orbits which trace out the geodesic trajectory and requiring that the Fourier spectrum be sufficiently narrow restricts the orbits that can be efficiently computed in the frequency domain. This is true in the geodesic case as well, where previous studies have shown that the frequency domain approach is useful for eccentricities up to $e\sim0.5$-$0.7$ \cite{Barton:2008eb, Warburton-Barack:eccentric}. When we increase the rate at which the particle traces out its spatial trajectory we find, for both circular and eccentric orbits, that the number of spherical-harmonic $(\ell,m)$-modes needed in the mode-sum increases (this is not specific to a frequency-domain implementation and will also occur for a 1+1 time-domain implementation). For eccentric orbits in the frequency domain the mode frequency contains overtones of the radial frequency and we find the number of mode frequencies per $(\ell,m)$ increases as the orbital motion is speed up, making high eccentricity calculations even more challenging than in the geodesic case. Conversely, the number of frequency modes per $(\ell,m)$-mode drops as the orbital motion is slowed down. The breadth of the Fourier spectrum depends on the particular accelerated motion being considered. Given the range of possible accelerations we have not attempted to quantify the spectrum broadening in this work.

\subsection{Time-domain calculation}
\label{sec:TD}

The frequency-domain approach described in the previous section is best suited to bound,
periodic orbits in which a small number of frequencies contribute. In particular, this excludes
several cases which are of interest in exploring the history-dependence of the self-force. For
these cases we use a different, 1+1D time domain code. For the purposes of this paper we will
only briefly describe the code here; a much more detailed description (with code validation) will
be provided in a forthcoming work \cite{Diener_etal:dG_code}.

The code implements the explicit time evolution of the sourced scalar wave equation on a
Schwarzschild background spacetime,
\begin{equation}
	\nabla_{a}\nabla^{a}\Phi = S,
\end{equation}
where, as in Sec.\ref{sec:sing_and_mode_sum}, $\nabla_{a}$ is the Schwarzschild spacetime covariant derivative and $\Phi$ is the scalar field, while $S$ is the source term. We use the code in combination with the effective source
regularization scheme \cite{Barack:Golbourn:2007,Vega:Detweiler:2008}, which allows us to use an effective
source, $S := S_{\mathrm{eff}}$, to solve directly for the residual field,
$\Phi^R$, rather than retarded as in Eq.~\eqref{eq:Wave}. In particular, we use the worldtube approach of \cite{Barack:Golbourn:2007}, in which we evolve
the 
residual
field $\Phi^{\mathrm{R}}$ inside the world tube and the full retarded field
$\Phi^{\mathrm{ret}}$ outside.

Both the scalar field $\Phi^R$ and effective source $S_{\mathrm{eff}}$ are decomposed into
spherical-harmonic modes, leading to a set of uncoupled 1+1D wave equations, one for each
spherical-harmonic $(\ell, m)$ mode. For each 1+1D wave equation we use hyperboloidal coordinates
\cite{Zenginoglu:2010cq} in regions away from the orbit, matched to standard Schwarzschild
coordinates in the vicinity of the orbit. Our coordinates are thus based on Schwarzschild time $t$, and
the Tortoise radial coordinates, $r^* = r + 2M \log(\tfrac{r}{2M}-1)$ in the vicinity of the
worldline, and on hyperboloidal coordinates elsewhere. This allows us to avoid having to impose
artificial boundary conditions at finite values of $r^*_{\mathrm{min}}$ and $r^*_{\mathrm{max}}$;
instead we place the horizon at a finite hyperboloidal radius $\rho_{\mathrm{min}}$, and future null
infinity at finite hyperboloidal radius $\rho_{\mathrm{max}}$ (not to be confused with $\rho$ previously defined in Sec.~\ref{sec:sing_and_mode_sum}). Thus, at both physical outer
boundaries, no characteristic modes are entering the computational domain and boundary conditions
are handled automatically by the use of characteristic fluxes. The computational domain therefore
covers the whole exterior spacetime from the horizon to future null infinity. This coordinate setup
is the same as that used in \cite{Wardell_etal:SSF_via_Green_functions, Diener_etal:dG_code}; see there for more details.

We use a first order in space and time formulation of the equations, and evolve using the method of
lines. The discontinuous Galerkin (dG) method is used for spatial discretization (the radial
direction) and evolution in time is performed with a fourth order Runge-Kutta scheme. In the dG
scheme, the computational domain is split into a number of elements, and within each element the
solution is approximated using interpolating polynomials of order $n$, defined on $n+1$ discrete
nodes. At element boundaries the numerical approximation is thus multivalued (the boundary point is
represented in both the left and right elements, and the solution may have different values there),
but through the use of numerical fluxes we ensure that the discontinuities remain small during
evolution, and that they converge to zero with increasing discretization order. The notable
exception is at worldtube boundaries, where we transition from evolving the residual field to the
retarded field. In that case, the expected discontinuities are given analytically by the value of
the singular field on the worldtube boundaries, and the numerical fluxes are designed to drive the
numerical discontinuities towards these analytical values. In addition we use
the time-dependent coordinates of \cite{Field:2009kk} in the orbital region 
(between the hyperboloidal regions) in order to keep the worldline of the
particle fixed at a constant coordinate located at an element boundary. This is
necessary for eccentric orbits, but the time-dependent coordinates can be
turned off for the circular motion considered in this paper. At the
boundary between the orbital region and the hyperboloidal regions (alse kept
fixed at element boundaries) we have to take the coordinate transformation
into account when constructing the numerical fluxes. 

A key aspect of the time-domain calculation is the construction of the effective source. This is
obtained by constructing a \emph{puncture field}, defined to be the decomposition of the singular
field into spherical-harmonic modes. This mode decomposition must be done without evaluating on the
worldline (as is typically done in the case of mode-sum regularization), but may be done in an
approximate sense provided the approximation is exact on the worldline. To obtain expressions for a
puncture field, we thus follow exactly the methods described previously for constructing
regularization parameters, but with two notable modifications:
\begin{enumerate}
  \item We do not evaluate at $\Delta r = 0$, but instead keep the $\Delta r^0, \Delta r^1, \Delta r^2, \Delta r^3$ and $\Delta r^4$ terms in the series expansion about $\Delta r = 0$.
  \item We do not keep only the $m = 0$ mode, but also must keep $m = \pm 1, \pm 2$.
\end{enumerate}
The details of our approach are very much the same as described in \cite{Warburton:2013lea,Wardell:2015ada,Miller:2016hjv}, but now the coefficients in the small-$\Delta r$
expansion are somewhat more complicated as they incorporate information about the non-geodesic
nature of the worldline (through the acceleration, $a^a$, in a second-order puncture, and through
higher derivatives of the worldline if higher order punctures are used).

In this work, we chose to use a second-order puncture, keeping the leading two terms in the
expansion of the singular field, i.e., Eqs.~\eqref{eqn:phiSm1} and \eqref{eqn:phiS0}. This is analogous to using the first two mode-sum regularization
parameters, and allows us to compute the 
residual
field and self-force by evaluating a sum over
spherical-harmonic modes where the terms fall off as $1/\ell^2$. By choosing a second-order
puncture, the effective source we obtain is continuous but not differentiable at the location of
the particle, resulting in spherical-harmonic ($\ell$, $m$) modes of the 
residual
field that are
twice differentiable there. This would cause problems for a finite difference scheme (unless
extreme care is taken; see \cite{Thornburg:2016msc}). The advantage of the dG scheme is
that exponential convergence with dG element order is achieved not only for solutions that are
smooth everywhere, but also for solutions where non-smooth features can be localized at element
boundaries; an advantage that arises from numerical derivatives, calculated using data from only one element, never ``seeing''
the non-smoothness.

We shall explore circular orbits with non-uniform acceleration in Sec.~\ref{sec:TD_self_force}
below.

\section{Example applications}

In this section we calculate the self-force for specific accelerated worldlines. We consider circular orbits with uniform and non-uniform acceleration, and accelerated eccentric orbits that follow the same trajectory as geodesic orbits.

\subsection{Circular orbits: uniform acceleration}

\begin{figure}
	\centering
	\includegraphics[width=8.5cm]{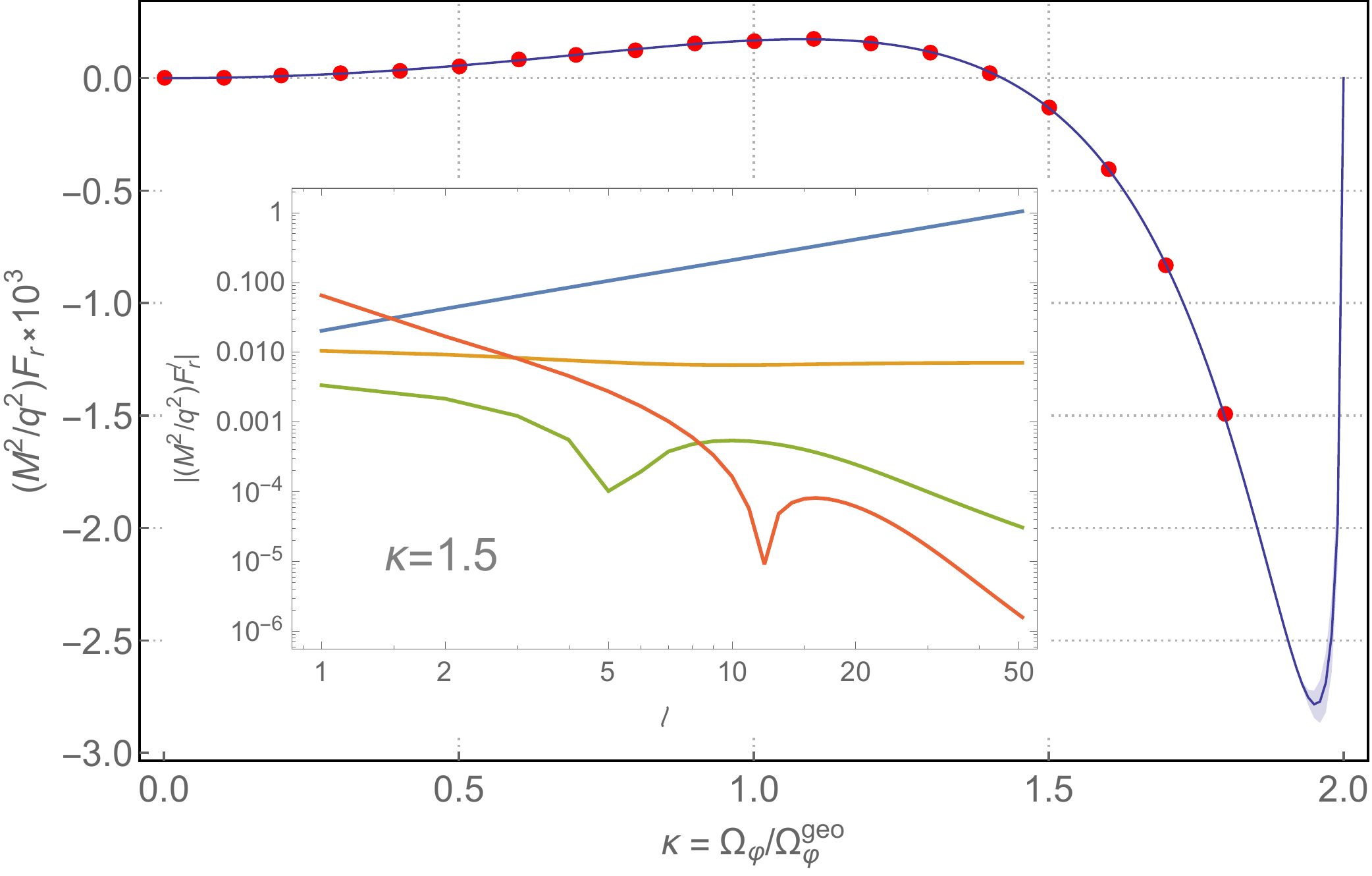}
	\caption{Covariant radial self-force, $F_r$, as a function of the orbital frequency for a circular orbit with radius $r_0=6M$. The (red) dots are our results computed using the mode-sum procedure described in this work. It is analytically known that the self-force vanishes for a static particle \cite{Burko:2000a,Wiseman:2000} and we take this as our data point at $\kappa=0$. The self-force changes sign around $\kappa \approx 1.40726$. This change in sign for uniformly accelerated circular orbits was first observed by Burko \cite{Burko:2000b}. The solid (blue) curve is computed using the Green function method described in \cite{Wardell_etal:SSF_via_Green_functions} (their error bars are shown as a shaded region about the curve). We find the results of the two methods agree for $\kappa\lesssim 1.8$. Above this value the mode-sum requires more than 100 $\ell$-modes to converge. Very high $\ell$-modes are difficult to compute accurately so we do not show results for these values of $\kappa$. (Inset) Behavior of the mode-sum on a log-log plot. At $\ell=50$, reading from top to bottom, the curves correspond to subtracting 0,1,2 and 3 leading-order non-zero regularization parameters after which the mode-sum grows/decays as $\ell,\ell^0, \ell^{-2}$ and $\ell^{-4}$, respectively, for large $\ell$.}\label{fig:circular_results}
\end{figure}

The description of generic accelerated motion along circular orbits is presented in Sec.~\ref{sec:circ_accel_description}. For uniform acceleration the orbital frequency, Eq.~\eqref{eq:Omega_phi_circ_accel}, is a constant multiple of the geodesic frequency, Eq.~\eqref{eq:Omega_varphi_geo}. We can thus parametrize uniformly accelerated circular orbits by their orbital radius and an acceleration factor, $\kappa$, defined such that
\begin{align}\label{eq:Omega_varphi_circ_accel_kappa}
	\Omega_\phi = \kappa \Omega_\phi^\text{geo}.
\end{align}
We compute the retarded field using a \textit{Mathematica} implementation similar to that in \cite{Warburton:2013lea} modified so that $\Omega_\phi$ is given by Eq.~\eqref{eq:Omega_varphi_circ_accel_kappa}. We then regularize the retarded field data using the parameters given in Sec.~\ref{sec:mode-sum_params}. Our main results for circular, uniformly accelerated orbits are presented in Fig.~\ref{fig:circular_results} and discussed in the figure caption.

\subsection{Bound accelerated eccentric orbits}\label{sec:results_ecc}

Our parametrization of accelerated motion for bound eccentric orbits presented in Sec.~\ref{sec:bound_eccentric_orbits_parametrization} has been chosen to 
exploit existing codes. In this section, we present results for accelerated orbits computed using a modified version of the frequency-domain code developed in \cite{Warburton-Barack:eccentric}.

\begin{figure}
	\includegraphics[width=8.5cm]{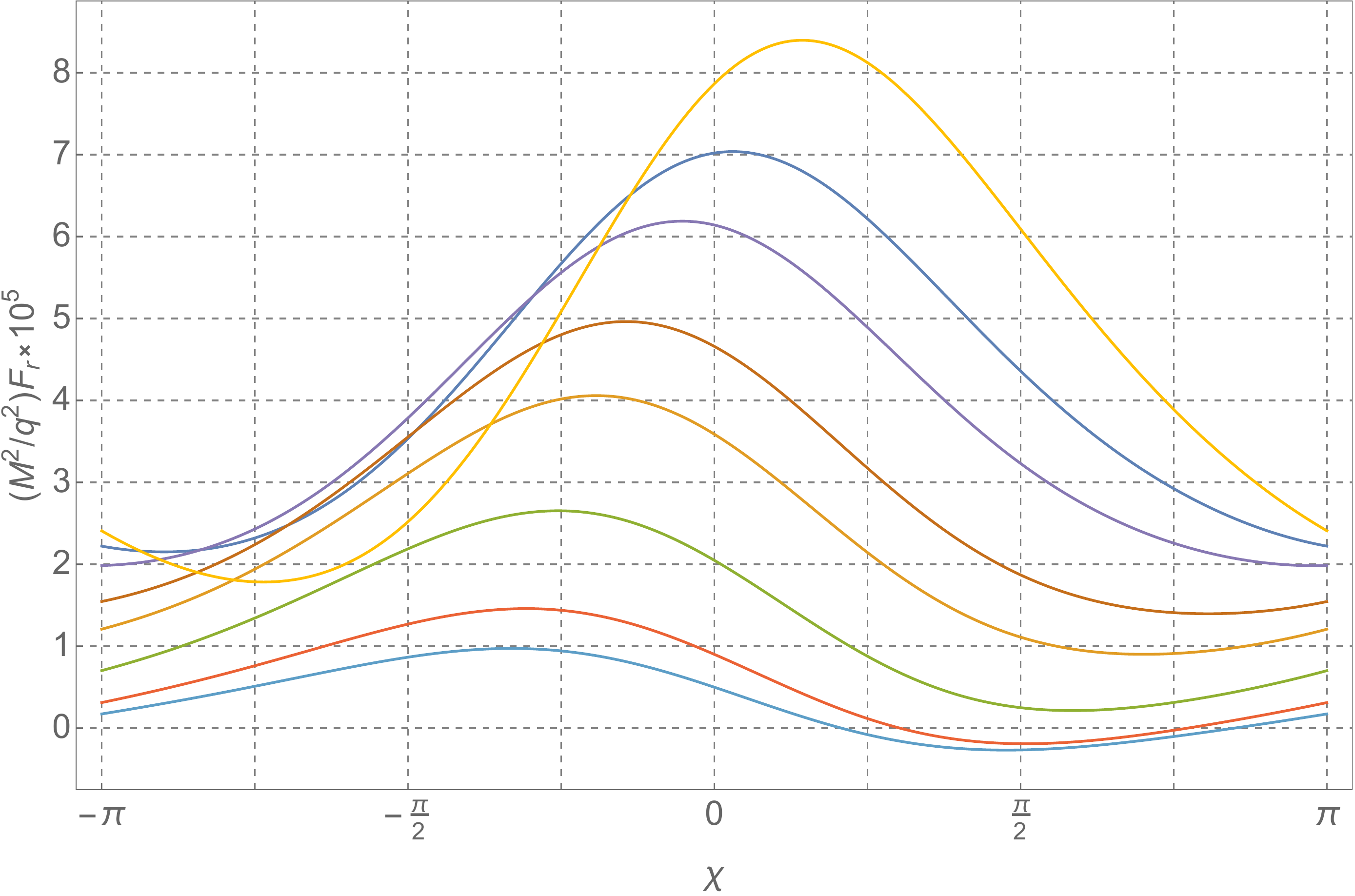}
	\caption{Covariant radial component of the SSF for a number of accelerated orbits with $(p,e)=(8,0.1)$. Each curve corresponds to an orbit with $dt/d\chi = \lambda\cdot dt/d\chi_\text{geo}$ where $\lambda$ is a constant. Reading from top to bottom at $\chi=0$ the curves correspond to $\lambda=\{0.9,1,1.1,1.3,1.5,2,3,4\}$. We observe as $\lambda$ increases (and the orbital motion slows) the magnitude to the self-force decreases. This is expected as in the limit $\lambda\rightarrow\infty$ the particle becomes static and the scalar-field self-force vanishes \cite{Burko:2000a,Wiseman:2000}.}\label{fig:ecc_accel}
\end{figure}

The description of the motion in Sec.~\ref{sec:bound_eccentric_orbits_parametrization} is uniquely determined once the relation between $t_0$ and $\chi$ is specified. We shall examine orbits whose acceleration is given by
\begin{align}
	\frac{dt_0}{d\chi} = \lambda \left(\frac{dt_0}{d\chi}\right)_\text{geo},
\end{align}
where $\lambda$ is a real constant and $\left(dt_0/d\chi \right)_\text{geo}$ is given by Eq.~\eqref{eq:dt_dchi_geo}. The spatial trajectory of the orbit is the same as the $\lambda=1$ geodesic. A value of $\lambda < 1$ ($\lambda > 1$) describes an orbit that completes one radial libration faster (slower) than the corresponding geodesic. We present sample results for the radial self-force $F_r$ for a number of different $\lambda$ values in Fig.~\ref{fig:ecc_accel}. As $\lambda$ increases the particle's motion slows and the magnitude of the self-force decreases. This is expected as in the limit $\lambda\rightarrow \infty$ the particle becomes static and the scalar-field self-force vanishes \cite{Burko:2000a,Wiseman:2000}. As $\lambda$ decreases the particle's speed increases, as does the magnitude of the radial self-force and we find contributions from the Fourier sum broaden. We also observe that the location, as a function of $\chi$, where the radial self-force is a maximum depends on $\lambda$.

\begin{figure}
	\includegraphics[width=8.5cm]{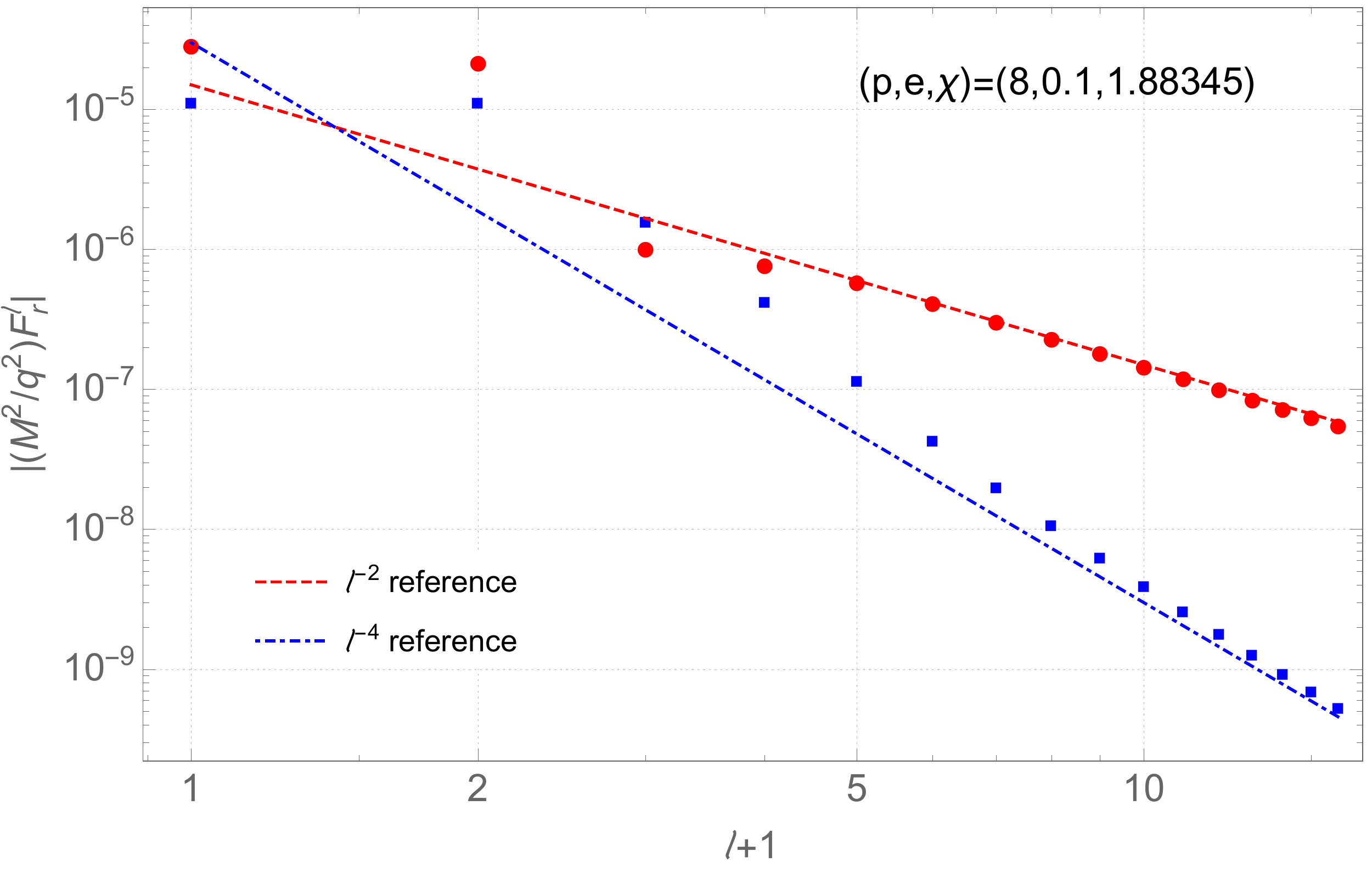}
	\caption{Convergence of the mode-sum for an accelerated eccentric orbit with $\lambda=2$ and orbital parameters $(p,e,\chi) = (8,0.1,1.88345)$. The (red) circular markers show the result of subtracting the leading two regularization parameters. For large $\ell$, the resulting modes approach the (red) dashed $\ell^{-2}$ reference line. Subtracting one further non-zero regularization parameter results in the (blue) square markers. For large $\ell$, these results approach the (blue) dot-dashed $\ell^{-4}$ reference line.}\label{fig:mode_sum_ecc}
\end{figure}

In Fig.~\ref{fig:mode_sum_ecc} we plot the contributions to the mode-sum at a particular point along an orbit with $\lambda = 2$. This figure demonstrates that the mode-sum converges at the expected rate and that employing higher-order regularization parameters correctly increases the rate of convergence of the mode-sum.

\subsection{Circular orbits: non-uniform acceleration}\label{sec:TD_self_force}

In this subsection we consider accelerated trajectories with sharper features. The Fourier spectrum associated with these wordlines is very broad which makes it infeasible to work with a frequency-domain calculation and so, unlike the previous two subsections, we employ a time-domain code. We will restrict our attention to circular orbits where we briefly accelerate the orbit before returning the particle to geodesic motion. Explicitly we accelerate the particle along its circular orbit via
\begin{align}\label{eq:Omega_accel}
	\Omega(t) = \Omega_\text{geo} + A e^{-(t-t_0)^2/\sigma^2},
\end{align}
where $A=0.05, t_0 = 800$ and $ \sigma = 3.6$.

\begin{figure}
	\includegraphics[width=8cm]{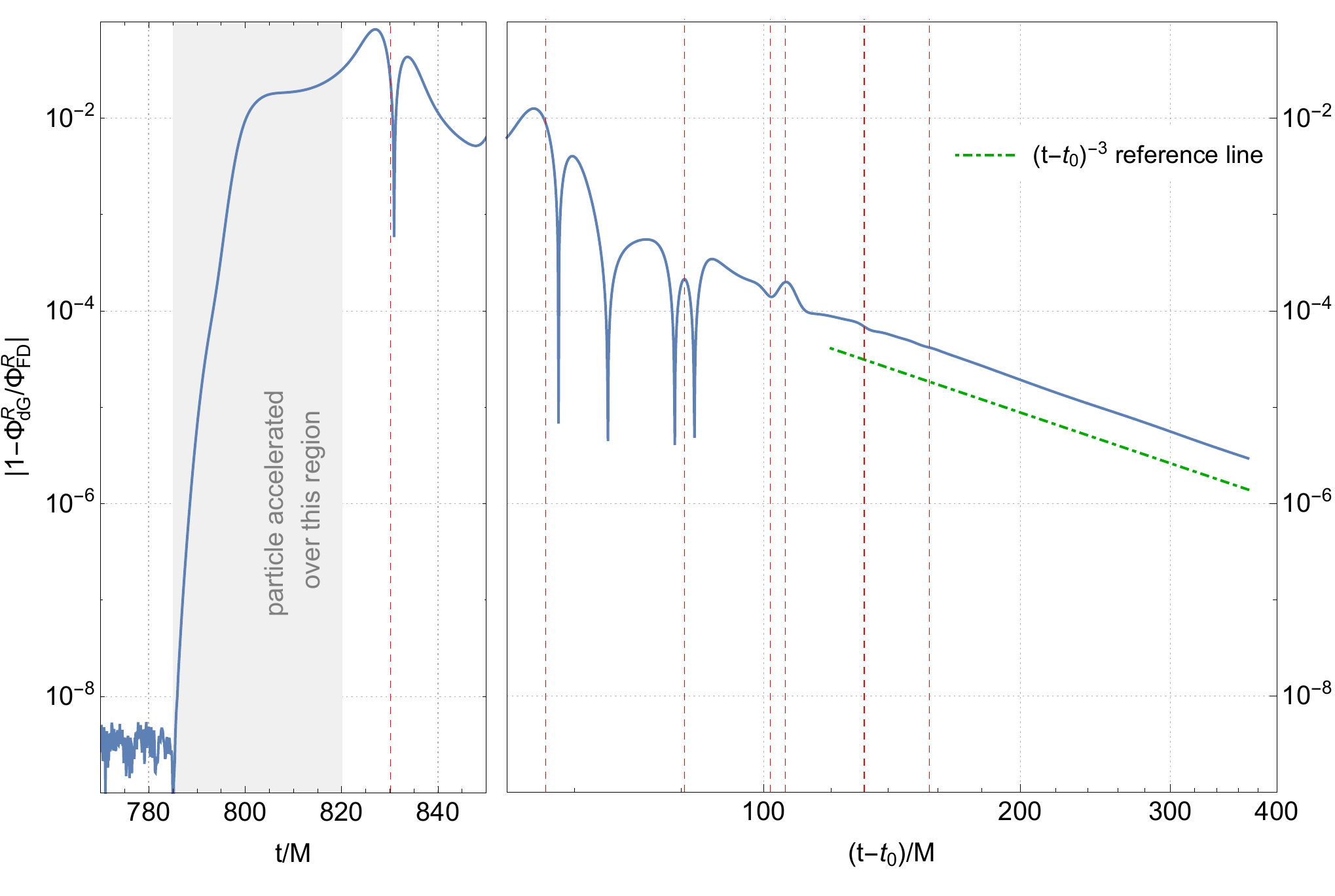}
	\caption{Difference in the scalar-field, $\Phi^R_\text{dG}$, for a particle with an accelerated worldline described by Eq.~\eqref{eq:Omega_accel} and the scalar field, $\Phi^R_\text{FD}$ for a particle moving along a geodesic with the same orbital radius. The acceleration is concentrated over the grey shaded region (being effectively zero elsewhere) with the maximum acceleration occurring at $t_0=800M$. Prior to the acceleration period the result of the dG and FD codes agree to a relative accuracy of $\sim10^{-9}$ The orbital radius is $r_0=6.7862M$. This orbital radius was chosen as the 2nd and 3rd null crossings (red, dashed, vertical lines) coalesce at a caustic -- see Fig.~\ref{fig:null_crossings}. From left to right the null crossings occur at $(t-t_0)/M \simeq \{30.2,55.5,55.5,80.8,101.7,106.0,131.2,156.4\}$, respectively. Considering the width (in time) of the acceleration phase, the null crossing times are consistent with the features in the curve. The apparent additional feature between the 3rd and 4th null crossing is not a real feature, but is simply a smooth zero crossing of the relative difference being plotted in a $\log$-$\log$ plot. At late time the field decays at the expected $t^{-3}$ rate for the $\ell=0$ mode (all the higher $\ell$-modes decay faster as $t^{-2\ell-3}$ \cite{1972PhRvD...5.2419P}). The (green) dot-dashed line shows a reference $t^{-3}$ curve.}\label{fig:circ_non_uniform}
\end{figure}

\begin{figure}
	\includegraphics[width=8cm]{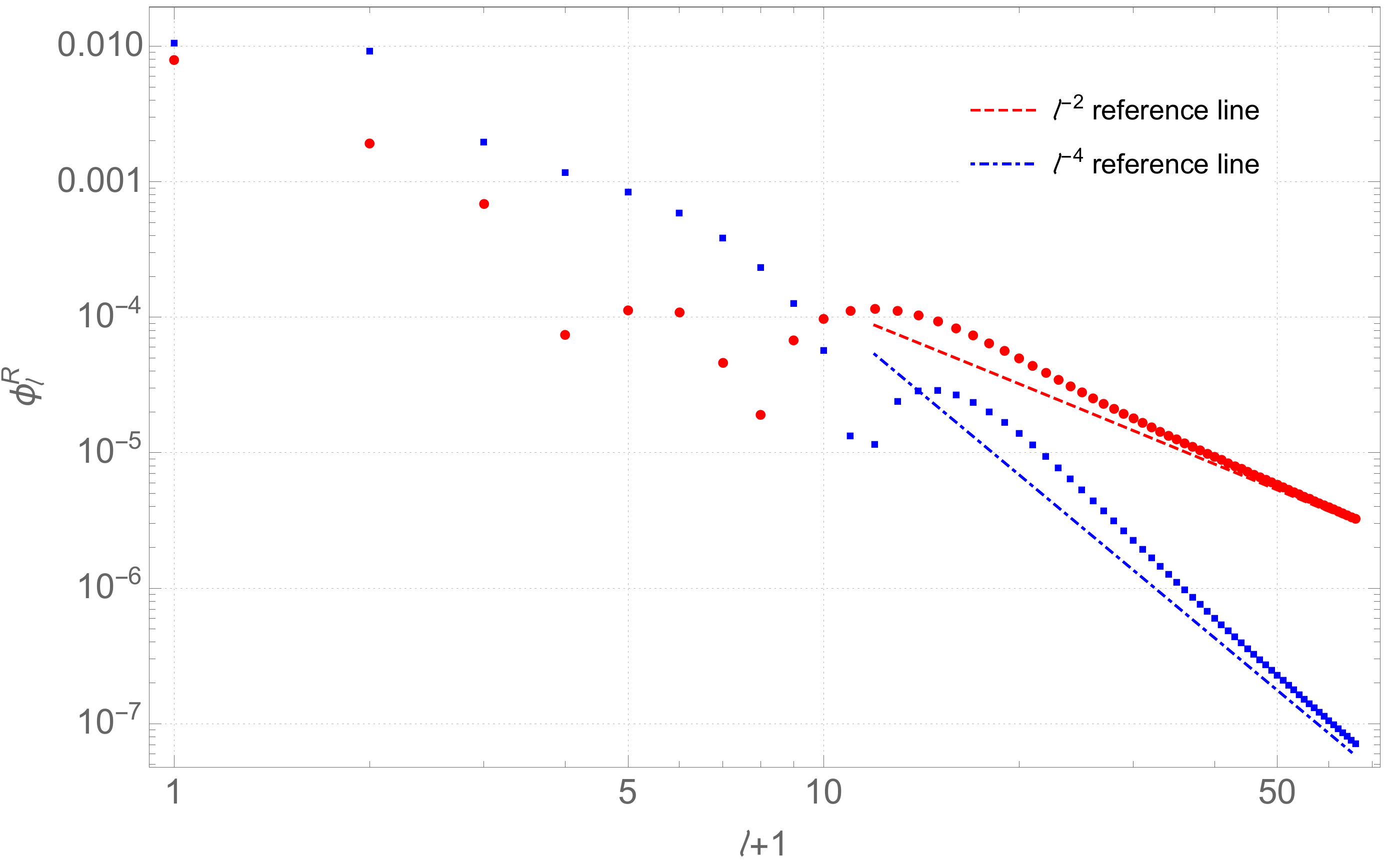}
	\caption{Convergence of the mode-sum for the regular scalar-field at the point of maximum acceleration ($t_0=800M$) for a particle with a worldline described by Eq.~\eqref{eq:Omega_accel}. The (red) round makers show the regular modes computed directly using the dG code. For large $\ell$ these modes approach the (red) dashed $\ell^{-2}$ reference curve. The (blue) square markers show the result of subtracting the $\Phi_{[2]}$ regularization parameter from the regular modes. For large $\ell$ these modes approach the (blue) dash-dotted $\ell^{-4}$ reference curve.}\label{fig:mode_sum_dG_accel}
\end{figure}

Our main results for this accelerated trajectory are given in Fig.~\ref{fig:circ_non_uniform}. For $t\lesssim785$, i.e., during the initial geodesic phase, we find our dG time-domain code gives a highly accurate result that agrees with the frequency-domain result to a relative accuracy of $\sim10^{-9}$. During the acceleration period the particle emits a burst of radiation which then scatters off the spacetime curvature and interacts with the particle at a later time. We observe the correct convergence rate of the mode-sum whilst the particle is accelerating --- see Fig.~\ref{fig:mode_sum_dG_accel}. We also observe features in the regular scalar field at subsequent times along the worldline connected to the period of acceleration by a null geodesic (these features are reminiscent of the structure of the Green function for the scalar field \cite{Casals:2013mpa}). Finally we observe the correct late-time behaviour of the scalar-field \cite{1972PhRvD...5.2419P}. All of these observations give us confidence that our dG time-domain code, the effective-source it uses and the mode-sum scheme are working correctly.

\begin{figure}
	\includegraphics[width=8cm]{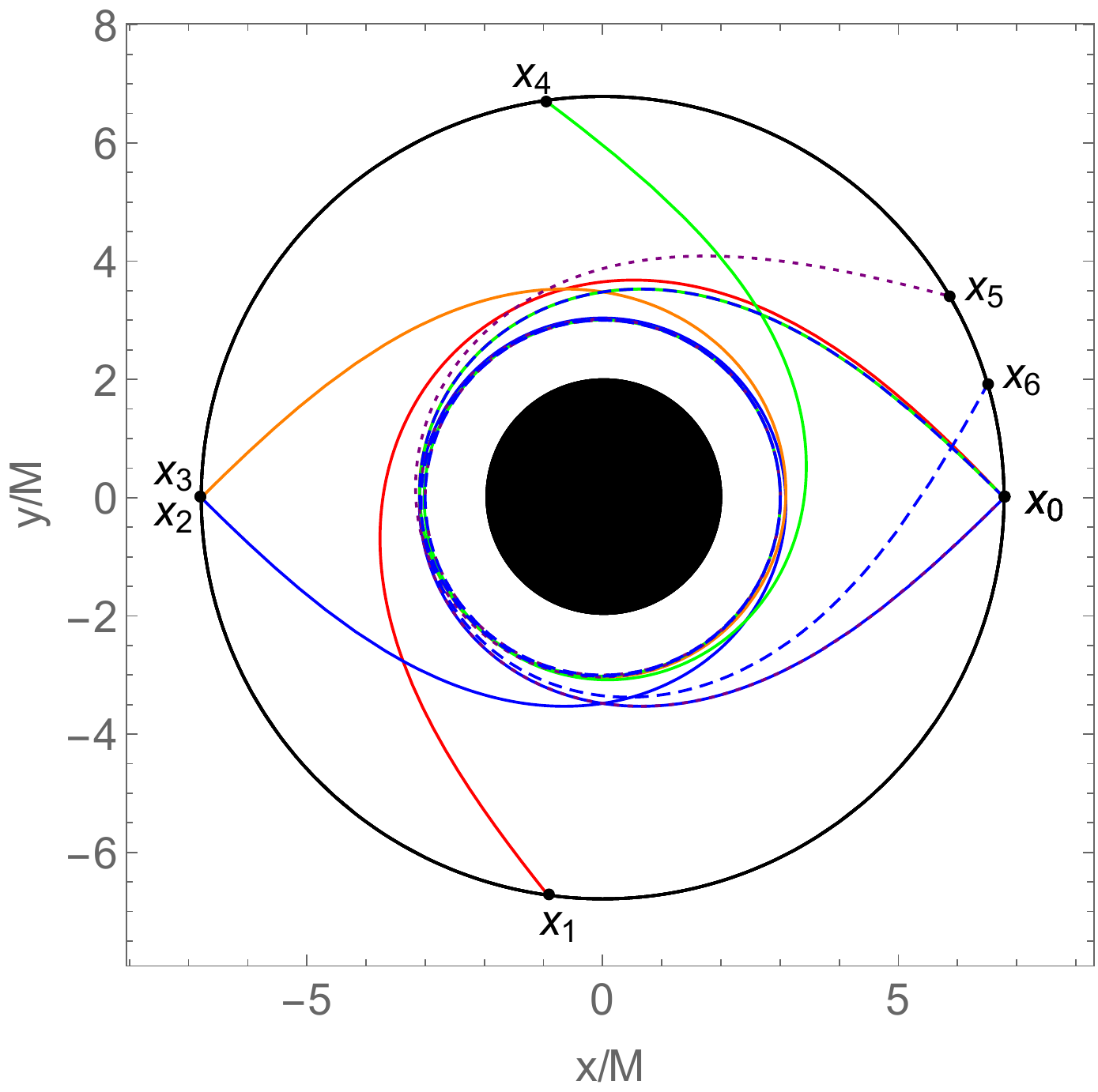}
	\caption{
	Null rays (colored curves) connecting the particle initially at $x_0 / M = (6.7862,0)$ at $t=0$ to subsequent points along a circular orbit (black curve) about a Schwarzschild black hole. The particle's orbital motion is in the clockwise direction. The orbital radius was chosen so that the second and third null connections coalesce into a caustic. Reading clockwise from $x_0$ the intersection of the null rays with the orbits occur at $t/M\simeq\{30.2,55.5,55.5,80.8,101.7,106.0\}$.}\label{fig:null_crossings}
\end{figure}

\section{Concluding remarks}
\label{sec:concluding_remarks}

In this work we have computed expansions of the Detweiler-Whiting singular field for a particle with a scalar field moving along an accelerated trajectory in Schwarzschild spacetime. Using these we have computed previously unknown regularization parameters which serve to increase the rate of convergence of the mode-sum, these are consistent with those previously derived for geodesic motion \cite{HOW:2012} and the special non-geodesic case of a static charge \cite{Casals:2012qq}. The singular-field expansion has also been used to compute a new effective-source which has been successfully employed as part of a new high-accuracy time-domain discontinuous-Galerkin code. Our new regularization parameters and effective-source have been tested by computing the self-force for a particle moving along a variety of accelerated trajectories, including circular motion (uniformly and non-uniformly accelerated) and bound eccentric motion.

The results of this work can find utility in a number of applications where accelerated motion arises in black hole perturbation theory. Examples include, self-consistent self-force calculations, where the field equations and the equations of motion for the particle are solved simultaneously as a coupled set; as well as local calculations of the regular field/metric for spinning bodies under the influence of e.g., the Mathisson-Papapetrou-Dixon force. In both cases the body's worldline is accelerated with respect to the background geometry.

It would also be possible to use the results of this work to explore the `history dependence' of the self-force. Similar to the results we presented for non-uniformly accelerated circular orbits, one could imagine accelerating short sections of the worldline for eccentric orbits and watching how quickly the field settles back down to the geodesic value. By accelerating different sections of the worldline, it might be possible to infer from where the dominant contribution to the geodesic self-force arises. In addition, by creating a burst of radiation during the acceleration period it might be possible to create `wiggles' in the self-force (believed to be related to ring-down), similar to those observed for highly eccentric orbits about a Kerr black hole \cite{Thornburg:2016msc}.

Finally, this work is naturally extended in a number of ways. First, the expansion of the singular
field and the regularization parameters should be extended to motion in Kerr spacetime
\cite{HOW:2014} as self-force calculations are now well established here
\cite{Warburton:Barack:2009, Warburton-Barack:eccentric,Warburton:2014bya,vandeMeent:2016pee}.
Second, the results could be extended to electromagnetic and gravitational perturbations
(\cite{Linz:2014} has done this for the leading-two regularization parameters). The former is
straightforward, but the latter requires additional care as accelerating the particle changes the
stress-energy tensor that sources the metric perturbation.

\section*{Acknowledgements}

AH gratefully acknowledges funding from the European Union's Horizon 2020 research and innovation programme under grant agreement 661705-GravityWaveWindow. BW and ACO gratefully acknowledge support from the Science Foundation Ireland under Grant No. 10/RFP/PHY2847. NW gratefully acknowledges support from a Royal Society - Science Foundation Ireland University Research Fellowship, Marie Curie International Outgoing Fellowship (PIOF-GA-2012-627781) and the Irish Research Council, which is funded under the National Development Plan for Ireland.
PD gratefully acknowledge support from the National Science Foundation under
Grant No. PHY-1307396 and thanks the Dublin group for their hospitality during
several visits.  These calculations were greatly assisted by the use of the software package Mathematica \cite{Mathematica}, in particular the open source tensor algebra package, xAct \cite{xTensor, xTensorOnline}.  We would also like to thank Roland Haas for enlightening discussions and Adam Day for his energizing encouragement.
\appendix

\section{Mode-sum regularization parameters for accelerated motion}\label{apdx:mode-sum_params}

Sec.~\ref{sec:mode-sum_params} presents the parameters needed to compute the regular self-force. This appendix gives the first subleading parameters that act to accelerate the convergence of the mode-sum. We give these higher order parameters explicitly here but also give all the parameters in electronic format online \cite{BlackHolePerturbationToolkit}.  Below we use the notation, $a^2 \equiv g_{\ab \bb} a^\ab a^\bb$, similarly, $a \cdot \dot{a} \equiv g_{\ab \bb} a^\ab \dot{a}^\bb$, while all $a^a$'s below are evaluated at $\xb$.

\begin{widetext}
\begin{equation}
F_{t}{}_{[2]} = \frac{F_{t\,\mathcal{E}}{}_{[2]} \mathcal{E} +\rb^2 F_{t\,\mathcal{K}}{}_{ [2]} \mathcal{K}}{6 L^3 \pi \rb^6 (\rb-2 M ) (L^2 + \rb^2)^{7/2}},
\end{equation} 
where
\begin{IEEEeqnarray*}{rCl}
F_{t\,\mathcal{E}}{}_{[2]} &=& 
3 E L^3 \rb^2 \ur (\rb-2 M) \big\{ 8 E^2 \rb^7 \left(L^2 -\rb^2\right) -\left(L^2+\rb^2\right)  \big[ 4 L^4 M \left( 9 L^2 +26 \rb^2 \right) + 2 M \rb^4 \left( 49 L^2 +23 \rb^2 \right)
\\
&& \qquad
 +\rb^5 \left( L^2 -7 \rb^2 \right) \big] \big\}
\\
&&
+a^2 L^2 \rb^5  (\rb-2 M) \left(L^2+\rb^2\right) \big\{E r^2 \left[a^{\phi } \rb \left(48 L^6+128 L^4 \rb^2+103 L^2 \rb^4+15 \rb^6 \right) + L \ur \left(8 L^4+23 L^2 \rb^2+23 \rb^4\right)\right]
\\
&& \quad
+ 2 a^t L  (\rb - 2 M) \left(L^2+\rb^2\right) \left(24 L^4+44 L^2 \rb^2+19 \rb^4\right)\big\}
\\
&&
+a^t L (\rb-2 M) \left(L^2+\rb^2\right) \big\{3 \rb^6 (\rb-2 M) \big[\rb^2 a^{\phi } \left(2 L^4+7 L^2 \rb^2-3 \rb^4\right) \left(2 L \ur - a^{\phi}  \rb^3 \right)
\\
&& \qquad
-4 a^r L^2 \left(L^2+\rb^2\right) \left(L^2 + 2\rb^2\right)\big]
-E^2 L^2 \rb^3 \left[2  L^4 M \left(8 L^2+17 \rb^2 \right)+2 M \rb^4 \left(L^2 +16 \rb^2  \right)+ 3 \rb^5 \left(L^2-7\rb^2 \right)\right]
\\
&&  \quad
+6 L^2  (\rb-2 M)  \left(L^2+r^2\right)\left[L^2 M \left(8 L^4+27 L^2 \rb^2+30 \rb^4\right) + \rb^6 (17 M-3 \rb)\right]\big\} 
\\
&&
+ 2 a^r E L^2 \rb^4 \left(L^2+\rb^2\right) \big\{3 \rb^5 a^{\phi }  (\rb-2 M)
   \left(2 L^4+7 L^2 \rb^2-3 \rb^4\right)
\\
&& \quad
+L \ur \left[L^4 M \left(8 L^2+17  \rb^2\right)+M \rb^4 \left(10 L^2-47 \rb^2\right) -3 \rb^5 \left( L^2 - 7 \rb^2 \right) \right]\big\}
\\
&&
+ a^\phi E \rb^3 (\rb-2 M) \big\{3 a^\phi \rb^8 \left[ 2 a^{\phi} \rb^{3} \left(L^2-\rb^2\right) \left(L^4+6 L^2 \rb^2+\rb^4\right) -  L  \ur \left(4 L^6+21 L^4 \rb^2-34 L^2 \rb^4-3 \rb^6\right) \right] 
\\
&& \quad
+ 9 E^2 L^2 \rb^7 \left(L^4-14 L^2 \rb^2+\rb^4\right)-L^2 \left(L^2+\rb^2\right) \big[16 L^6 M \left(2 L^2+5  \rb^2 \right)+ 2 M \rb^4 \left(  26 L^4 + 107 L^2 \rb^2 -15 \rb^4 \right)
\\
&& \qquad
+3 \rb^5 \left(2 L^4 -35 L^2 \rb^2 +3 \rb^4 \right) \big] \big\}
\\
&&
+ 4 a \cdot \dot{a} E L^3 \rb^6 (\rb-2 M) \left(L^2+\rb^2\right)^2 \left(24 L^4+44 L^2 \rb^2+19 \rb^4\right) - 8 \dot{a}^r E L^3 \rb^7  (\rb-2 M) \left(L^2+\rb^2\right)^2 \left(L^2+2 \rb^2\right)
\\
&&
- 4 \dot{a}^t L^2 \rb^6 (\rb-2 M)^2 \left(L^2+\rb^2\right)^2 \left[a^{\phi} \rb \left(8 L^4 +13 L^2 \rb^2+3 \rb^4\right)+2 L \ur \left(L^2+2 \rb^2\right)\right]
\\
&&
+ 4 \dot{a}^\phi L \rb^7  (\rb-2 M) \left(L^2+\rb^2\right)\big[E \rb^2 \left(L \ur-\rb^3 a^{\phi}\right) \left(2 L^4+7 L^2 \rb^2-3 \rb^4\right)
\\
&&\quad
- a^t L  (\rb - 2 M) \left(L^2+\rb^2\right) \left(8 L^4+13 L^2 \rb^2+3 \rb^4\right)\big]
\\
&&
- 2 \ddot{a}^t L^3 \rb^5 (\rb-2 M)^2 \left(L^2+\rb^2\right)^3  \left(8 L^2+7 \rb^2\right)
- 2 \ddot{a}^\phi E L^2 \rb^8 (\rb-2 M) \left(L^2+\rb^2\right)^2 \left(8 L^4+13 L^2 \rb^2+3 \rb^4\right),
\end{IEEEeqnarray*}
\begin{IEEEeqnarray*}{rCl}
F_{t\,\mathcal{K}}{}_{[2]} &=& 
-3 E L^3 \ur \rb^2 (\rb - 2 M) \left\{ E^2 \rb^5 \left(3 L^2 -5 \rb^2\right) 
-2 \left(L^2+\rb^2\right) \left[9 L^2 M \left(L^2+2 \rb^2 \right)+\rb^4 (13 M-2 \rb)\right]
\right\}
\\
&&
- a^2 L^2 \rb^5 (\rb-2 M) \left(L^2+\rb^2\right) \big\{ a^t L  (\rb - 2 M) \left(L^2+\rb^2\right) \left(24 L^2+23 \rb^2\right) + E \rb^2 \big[a^{\phi } \rb \left(24 L^4+43 L^2 \rb^2+15 \rb^4\right)
\\
&& \qquad
+4 L \ur \left(L^2+2 \rb^2\right)\big]\big\}
\\
&&
+ a^t L (\rb-2 M) \left(L^2+\rb^2\right) \big\{ 3 \rb^6 (\rb-2 M) \left[a^{\phi } \rb^2 \left(L^2-3 \rb^2\right) \left(a^{\phi } \rb^3 -2 L \ur \right)+2 a^r L^2 \left(L^2+\rb^2\right)\right] 
\\
&&\quad
+2 E^2 L^2 \rb^3 \left[ L^2 M \left(4 L^2+5 \rb^2\right)+\rb^4 (13 M-6\rb)\right]-3 L^2 (\rb-2 M) \left(L^2+\rb^2\right) \left[4 L^2 M (2 L^2+5 \rb^2)+3 \rb^4 (6 M-\rb)\right]\big\}
\\
&&
- 2 a^r E L^2 \rb^4 \left(L^2+\rb^2\right) \left\{3 a^{\phi } \rb^5  (\rb-2 M) \left(L^2-3 \rb^2\right)+L \ur \left[ L^2 M \left( 4 L^2 +5 \rb^2 \right) +\rb^4 (12 \rb-23 M)\right]\right\}
\\
&&
- a^\phi E \rb^3  (\rb - 2 M) \big\{ 3 a^\phi \rb^8 \left[a^\phi \rb^{3} \left(L^4-9 L^2 \rb^2-2 \rb^4\right) - L \ur \left(2 L^4-25 L^2 \rb^2-3 \rb^4\right) \right]
- 9 E^2 L^2 \rb^7 \left(7 L^2 - \rb^2 \right)
\\
&& \quad
- L^2 \left(L^2+\rb^2\right) \left[ 2 L^4 M \left(8 L^2 +13 \rb^2\right)+10 M \rb^4 \left(10 L^2 -3 \rb^2\right) -3 \rb^5 \left(17 L^2 -3 \rb^2\right) \right]
\big\}
\\
&&
-2 a \cdot \dot{a} E L^3 \rb^6  (\rb-2 M) \left(L^2+\rb^2\right)^2 \left(24 L^2+23 \rb^2\right)
\\
&&
+ 4 \dot{a}^t L^2 \rb^6  (\rb-2 M)^2 \left(L^2+\rb^2\right)^2 \left[a^{\phi } \rb  \left(4 L^2 +3 \rb^2 \right)+L \ur \right]
+4 \dot{a}^r E L^3 \rb^7 (\rb-2 M) \left(L^2+\rb^2\right)^2 
\\
&&
+4 \dot{a}^\phi L \rb^7 (\rb-2 M) \left(L^2+\rb^2\right) \left[E \rb^2 \left(L^2-3 \rb^2\right) \left(\rb^3 a^{\phi }-L \ur\right) + L a^t  (\rb - 2 M) \left(L^2+\rb^2\right) \left(4 L^2+3 \rb^2\right)\right]
\\
&&
+8 \ddot{a}^t L^3 \rb^5 (\rb-2 M)^2 \left(L^2+\rb^2\right)^3 
+ 2 \ddot{a}^\phi E L^2 \rb^8 (\rb-2 M) \left(L^2+\rb^2\right)^2 \left(4 L^2+3
   \rb^2\right),
\end{IEEEeqnarray*}

\begin{equation}
F_{r}{}_{[2]} = \frac{F_{r\,\mathcal{E}}{}_{[2]} \mathcal{E} + \rb^2 F_{r\,\mathcal{K}}{}_{ [2]} \mathcal{K}}{6 \pi  L^3 \rb^6 (\rb-2 M ) (L^2 + \rb^2)^{7/2}},
\end{equation} 
where
\begin{IEEEeqnarray*}{rCl}
F_{r\,\mathcal{E}}{}_{[2]} &=& 
24 E^4 L^3 \rb^{10} \left(\rb^2-L^2\right)
+12 E^2 L^3 \rb^3 \left(L^2+\rb^2\right) \left[L^4 M \left(9 L^2+26 \rb^2\right) + M \rb^4 \left(23 L^2 + 14 \rb^2 \right) + \rb^5 \left( L^2 -3 \rb^2 \right) \right]
\\
&&
-3 L^3  (\rb - 2 M)\left(L^2+\rb^2\right)^2 \left[2 L^4 M \left(14 L^2+ 41 \rb^2 \right) + 2 M \rb^4 \left(41 L^2+ 16 \rb^2 \right) - \rb^5 \left(L^2 + 3 \rb^2\right)\right]
\\
&&
- a^2 L^2 \rb^5 \left(L^2+\rb^2\right) \big[a^{\phi } \ur \rb^4 \left(48 L^6+128 L^4 \rb^2+103 L^2 \rb^4+15 \rb^6\right)+2 a^r L \rb^2\left(L^2+\rb^2\right) \left(24 L^4+44 L^2 \rb^2+19 \rb^4\right)
\\
&& \quad
- 3 L  (\rb - 2 M) \left(L^2+\rb^2\right) \left(8 L^4+17 L^2 \rb^2+11 \rb^4\right)+E^2 L \rb^3 \left(8 L^4+23 L^2 \rb^2+23 \rb^4\right)\big]
\\
&&
+ 3 a^r L \rb^2 \left(L^2+\rb^2\right) \big\{
4 a^r L^2 \rb^6 \left(L^2+\rb^2\right) \left(L^2+2 \rb^2\right)
+ a^{\phi }  \rb^8 \big[a^{\phi } \rb^{3} \left(2 L^4+7 L^2 \rb^2-3 \rb^4\right)
\\
&& \qquad
- 4 L \ur  \left(2 L^4+7 L^2 \rb^2-3 \rb^4\right) \big]
+3 L^2 E^2 \rb^7 \left(L^2-7 \rb^2\right)
-2 L^2 \left(L^2+\rb^2\right) \big[ L^4 M \left(8 L^2+27 \rb^2 \right) 
\\
&& \qquad
+ M \rb^4 \left( 28 L^2 + 27 \rb^2 \right) + \rb^5 \left( L^2 - 8 \rb^2 \right) 
\big]
\big\}
\\
&&
-3 a^\phi \rb^4 \big\{
a^\phi \rb^5 \big[
2 a^\phi \rb^{6} \ur  \left(L^2-\rb^2 \right) \left(L^4+6 L^2 \rb^2+\rb^4\right)
- E^2 L \rb^3 \left(4 L^6+21 L^4 \rb^2-34 L^2 \rb^4-3 \rb^6\right)
\\
&& \qquad
+ L  (\rb - 2 M) \left(L^2+\rb^2\right) \left(4 L^6+21 L^4 \rb^2-27 L^2 \rb^4-4 \rb^6\right) 
\big]
+ 3 E^2 L^2 \rb^7 \ur \left(L^4-14 L^2 \rb^2+\rb^4\right)
\\
&& \quad
-L^2 \ur \left(L^2+\rb^2\right) \left[2 L^6 M \left(8 L^2 +19 \rb^2\right)+2 M \rb^4 \left(10 L^4 + 20 L^2 \rb^2 - 3 \rb^4 \right)+ \rb^5 \left( 2 L^4 -21 L^2 \rb^2 + \rb^4 \right)\right]
\big\}
\\
&&
-4 a \cdot \dot{a} L^3 \rb^7 \ur \left(L^2+\rb^2\right)^2 \left(24 L^4+44 L^2 \rb^2+19 \rb^4\right)
\\
&&
+4 \dot{a}^r L^2 \rb^8 \left(L^2+\rb^2\right)^2 \left[a^{\phi } \rb \left(8 L^4+13 L^2 \rb^2+3 \rb^4\right)+4 L \ur \left(L^2+2 \rb^2\right)\right]
\\
&&
+ 4 \dot{a}^\phi L \rb^7 \left(L^2+\rb^2\right) \big\{a^{\phi } \ur \rb^6\left(2 L^4+7 L^2 \rb^2-3 \rb^4\right)+ a^r L \rb^2 \left(L^2+\rb^2\right) \left(8 L^4+13 L^2 \rb^2+3 \rb^4\right)
\\
&& \quad
- L \left[2 (\rb-2M) \left(L^2+\rb^2\right) \left(L^4+2 \rb^4\right)+E^2 \rb^3 \left(2 L^4+7 L^2 \rb^2-3 \rb^4\right)\right]\big\}
\\
&&
+2 \ddot{a}^r L^3 \rb^7 \left(L^2+\rb^2\right)^3 \left(8 L^2+7 \rb^2\right)
+ 2 \ddot{a}^\phi L^2 \rb^9 \ur \left(L^2+\rb^2\right)^2 \left(8 L^4+13 L^2 \rb^2+3 \rb^4\right),
\end{IEEEeqnarray*}
\begin{IEEEeqnarray*}{rCl}
F_{r\,\mathcal{K}}{}_{[2]} &=& 
3 E^4 L^3 \rb^8 \left(3 L^2-5 \rb^2\right)
-3 E^2 L^3 \rb^3 \left(L^2+\rb^2\right) \left[2 M \left( 9 L^4 +17 L^2 \rb^2 +16 \rb^4\right)+ \rb^3 \left(L^2 - 7 \rb^2\right)\right]
\\
&&
+3 L^3 (\rb -2 M) \left(L^2+\rb^2\right)^2  \left[14 L^2 M \left(L^2+2 \rb^2 \right)-\rb^4 (\rb-16 M)\right]
\\
&&
+ a^2 L^2 \rb^5 \left(L^2+\rb^2\right) \big[
a^r L \rb^2 \left(L^2+\rb^2\right) \left(24 L^2+23 \rb^2\right)
+ a^{\phi } \rb^4 \ur \left(24 L^4+43 L^2 \rb^2+15 \rb^4\right)
\\
&& \quad
+4 E^2 L \rb^3 \left(L^2+2 \rb^2\right)
-3 L (\rb-2 M) \left(L^2+\rb^2\right) \left(4 L^2+5 \rb^2\right)
\big]
\\
&&
- 3 a^r L \rb^2 \left(L^2+\rb^2\right) \big\{
2 a^r L^2 \rb^6 \left(L^2+\rb^2\right)
+ a^\phi \rb^8 \big[
a^{\phi} \rb^{3} \left(L^2 -3 \rb^2\right)-4 L \ur \left(L^2-3 \rb^2\right)
\big] -12 E^2 L^2 \rb^7
\\
&& \quad
-L^2 \left(L^2+\rb^2\right) \left[4 L^2 M \left(2 L^2+5 \rb^2 \right)+3 \rb^4 (10 M -3 \rb)\right]
\big\}
\\
&&
+ 3 a^\phi \rb^4 \big \{
a^\phi \rb^5 \big[a^\phi\rb^{6} \ur \left(L^4-9 L^2 r^2-2 r^4\right)
- E^2 L \rb^3 \left(2 L^4-25 L^2\rb^2-3 \rb^4\right)
\\
&& \qquad
+2 L (\rb-2 M) \left(L^2+\rb^2\right) \left(L^4-11 L^2 \rb^2-2 \rb^4\right)
\big]
- 3 E^2 L^2 \rb^7 \ur \left(7 L^2 - \rb^2\right)
\\
&& \quad
-L^2 \ur \left(L^2+\rb^2\right) \left[4 L^4 M\left(2 L^2 +3 \rb^2 \right) +M \rb^4  \left(22 L^2 - 6 \rb^2 \right)- \rb^5 \left( 11 L^2 -\rb^2\right)\right]
\big\}
\\
&&
+2 a \cdot \dot{a} L^3 \rb^7 \ur \left(L^2+\rb^2\right)^2 \left(24 L^2+23 \rb^2\right)
-4 \dot{a}^r L^2 \rb^8 \left(L^2+\rb^2\right)^2 \left[ a^{\phi} \rb \left(4 L^2 +3 \rb^2 \right)+2 L \ur \right]
\\
&&
-4 \dot{a}^\phi L \rb^7 \left(L^2+r^2\right) \big[
a^r L \rb^2 \left(L^2+\rb^2\right) \left(4 L^2+3 \rb^2\right)
+a^{\phi } \rb^{6} \ur \left(L^2-3 \rb^2\right)
- E^2 L \rb^{3} \left(L^2-3 \rb^2\right)
\\
&& \quad
- L  (\rb-2M) \left(L^2+\rb^2\right) \left(L^2+4 \rb^2\right)
\big]
\\
&&
-8 \ddot{a}^r L^3 r^7 \left(L^2+\rb^2\right)^3
-2 \ddot{a}^\phi L^2 \rb^9 \ur \left(L^2+\rb^2\right)^2 \left(4 L^2+3 \rb^2\right),
\end{IEEEeqnarray*}

\begin{equation}
F_{\theta}{}_{[2]} = 0,
\end{equation}

\begin{equation}
F_{\phi}{}_{[2]} = \frac{F_{\phi\,\mathcal{E}}{}_{[2]} \mathcal{E} + \rb^2 F_{\phi\,\mathcal{K}}{}_{ [2]} \mathcal{K}}{6 \pi  L^4 \rb^4  (L^2 + \rb^2)^{5/2}},
\end{equation} 
where
\begin{IEEEeqnarray*}{rCl}
F_{\phi\,\mathcal{E}}{}_{[2]} &=& 
3 E^2 L^3 \rb^7 \ur \left(7 L^2 - \rb^2\right)
+3 L^3 \ur \left[ 2 L^6 M \left(14 L^2 + 43 \rb^2 \right) + 2 L^2 M \rb^4 \left( 46 L^2 + 17 \rb^2\right)- \rb^5 \left( L^4 - \rb^4\right) \right]
\\
&&
-3 a^2 L^2 \rb^5 \left(L^2+\rb^2\right) \left[a^{\phi } \rb \left(32 L^6+56 L^4 \rb^2+21 L^2 \rb^4-\rb^6\right)+L \ur \left(8 L^4+13 L^2 \rb^2+3 \rb^4\right)\right]
\\
&&
+18 a^r L^2 \rb^6 \left(L^4-\rb^4\right) \left(\rb^3 a^{\phi} - L \ur \right)
+ 3 a^\phi \rb^3 \big\{ 
a^\phi \rb^{8} \left( a^\phi \rb^3 - 3 L \ur \right) \left(3 L^4-7 L^2 \rb^2-2 \rb^4\right)
\\
&& \quad
+3 E^2 L^2 \rb^5 \left(2 L^4-7 L^2 \rb^2-\rb^4\right)
-L^2 \left(L^2+\rb^2\right) \big[16  L^4 M \left( L^2 + \rb^2 \right)  +2 M \rb^4 \left( 19 L^2 + \rb^2 \right)
\\
&& \qquad
+3 \rb^3 \left( 2 L^4 -5 L^2 \rb^2 - \rb^4 \right)
\big]
\big\}
\\
&&
-6 a \cdot \dot{a} L^3 r^4 \left(L^2+\rb^2\right)^2 \left(16 L^4+16 L^2 \rb^2+\rb^4\right)
\\
&&
+12 \dot{a}^r L^3 \rb^5 \left(L^2+\rb^2\right)^2 \left(2 L^2+\rb^2\right)
+4 \dot{a}^\phi L \rb^8 \left[2 a^{\phi } \left(L^2+\rb^2\right) \left(8 L^6+14 L^4 \rb^2+4 L^2 \rb^4+\rb^6\right)+3 L \rb \ur \left(L^4-\rb^4\right)\right]
\\
&&
+2 \ddot{a}^\phi L^2 \rb^6 \left(L^2+\rb^2\right)^2 \left(16 L^4+12 L^2 \rb^2-\rb^4\right),
\end{IEEEeqnarray*}
\begin{IEEEeqnarray*}{rCl}
F_{\phi\,\mathcal{K}}{}_{[2]} &=& 
-3 E^2 L^3 \rb^5 \ur \left(3 L^2 - \rb^2\right)
-3 L^3 \ur \left(L^2+\rb^2\right) \left(14 L^4 M+16 L^2 M \rb^2+\rb^5\right)
\\
&&
+3 a^2 L^2 \rb^5 \left(L^2+\rb^2\right) \left[ a^{\phi } \rb \left(16 L^4 +14 L^2 \rb^2 - \rb^4 \right)+ L \ur \left(4 L^2 + 3 \rb^2 \right) \right]
+ 18 a^r L^2 \rb^6 \left(L^2+\rb^2\right) \left(a^{\phi} \rb^3 - L \ur \right)
\\
&&
+ 3 a^\phi \rb^3 \big\{
2 a^\phi \rb^8 \left( a^\phi \rb^3 - 3 L \ur \right) \left(3 L^2+\rb^2\right)
+ 3 E^2 L^2 \rb^5 \left(5 L^2+\rb^2\right)+L^2 \left(L^2+\rb^2\right) \big[2 M \left( 4 L^4 + 14 L^2 \rb^2 + \rb^4 \right)
\\
&& \qquad
-3 \rb^3 \left( 4 L^2 + \rb^2 \right)\big]
\big\}
\\
&&
+ 6 a \cdot \dot{a} L^3 \rb^4 \left(L^2+\rb^2\right)^2 \left(8 L^2+\rb^2\right)
-12 \dot{a}^r L^3 \rb^5 \left(L^2+\rb^2\right)^2
\\
&&
-4 \dot{a}^\phi L \rb^8 \left(L^2+\rb^2\right) \left[a^{\phi } \left(8 L^4+7 L^2 \rb^2+2 \rb^4\right) - 3 L \rb \ur \right]
-2 \ddot{a}^\phi L^2 \rb^6 \left(8 L^2 - \rb^2\right) \left(L^2+r^2\right)^2.
\end{IEEEeqnarray*}
 
For practical use it is often easier to have parametric forms for the 4-acceleration $a^a$, the 4-jerk $\dot{a}^a=D_Ua^a$, and the 4-jounce $\ddot{a}^a=D^2_Ua^a$.  Below, overdots, $(\dot{})$, symbolize one or more partial differentiation with respect to proper time, while a superscript in parentheses represents higher orders of differentiation, e.g., $\rb^{(n)} \equiv \frac{\partial^n \rb}{\partial \tau^n}$ :
\begin{align*}
a^t &= \dut +\frac{2 E M \ur}{(\rb-2 M)^2},
\\
a^r &=\dur+\frac{L^2 (3 M-\rb)}{\rb^4}+\frac{M}{\rb^2},
\\
a^\theta &=0,
\\
a^\phi &=\dup+\frac{2 L \ur}{\rb^3},
\\
\dot{a}^t &= \ddut + \frac{3 M \ur}{\rb \left(\rb - 2 M\right)} \dut + \frac{3 E M}{\left(\rb - 2 M\right)^2} \dur - \frac{E M \left[  M \left( 3 L^2 + 5 \rb^2 \right) - \rb \left( 3 L^2 + 4 \rb^2 \right)\right]}{\rb^4 \left(\rb - 2 M\right)^2} - \frac{2 E^3 M \left(2 \rb - 3 M \right)}{\rb \left(\rb - 2 M\right)^3},
\\
\dot{a}^r &=\ddur-\frac{3 L \left(\rb - 3 M \right)}{\rb^2} \dup+\frac{2 M \left(3 L^2-\rb^2\right)-3 L^2 \rb}{\rb^5} \ur,
\\
\dot{a}^\theta &= 0,
\\
\dot{a}^\phi &= \ddup + \frac{3 L}{\rb^3} \dur +\frac{ 3 \ur}{\rb}  \dup +\frac{L M \left(3 L^2+\rb^2 \right) - L^3 \rb}{\rb^7},
\\
\ddot{a}^t &= \dddut + \frac{4 M \ur}{\rb \left(\rb - 2 M\right)} \ddut + \frac{4 E M}{\left(\rb - 2 M\right)^2} \ddur + \left[ 6 \dur - \frac{15 E^2}{\rb}+\frac{2\left(5 \rb - 7 M \right)}{\rb^2} + \frac{3 L^2 \left(3 \rb - 4 M\right)}{\rb^4}\right] \frac{M \dut}{\rb \left(\rb- 2 M\right)}
\\
&\quad
- \frac{3 E M \left(3 \rb - 2 M\right)}{\rb \left(\rb - 2 M\right)^3} \ur \dur + \frac{2 E^3 M \left(30 M^2 - 23 M \rb +6 \rb^2 \right)}{ \rb^2 \left(\rb - 2 M\right)^4} \ur - \frac{E  M \left( 34 M^2 - 35 M \rb + 12 \rb^2 \right)}{\rb^3 \left(\rb - 2 M\right)^3} \ur
\\
&\quad
- \frac{3 E L^2 M \left( 6 M^2 - 7 M \rb + 3 \rb^2 \right)}{\rb^5 \left(\rb - 2 M\right)^3} \ur,
\\
\ddot{a}^r &= \dddur - \frac{4 L \left(\rb - 3 M\right)}{ \rb^2} \ddup - \left[ 3 \rb - 10 M + \frac{L^2 M \left( \rb - 2 M\right)}{E^2 \rb^3}\right] \dup{}^2 + \frac{M \left(L^2 +\rb^2 \right)}{E^2 \rb^4} \dur{}^2
\\
&\quad
- \left[\frac{2 M}{E^2} \dur + \frac{12 \rb - 25 M}{\rb} +  \frac{2 M \left( M \rb^2 - L^2 M + L^2 \rb\right)}{E^2 \rb^4} \right] \frac{L \ur \dup}{\rb^2}
\\
&\quad
- \left[ L^2 \left(6 \rb - 11 M\right) + 2 M \rb^2 - \frac{2 M \left( L^2 + M \rb \right)}{E^2} - \frac{2 L^4 M \left(\rb - M\right)}{E^2 \rb^3} \right] \frac{\dur}{\rb^5} - \frac{2 E^2 M \left(L^2 - 3 \rb^2\right)}{\rb^6}
\\
&\quad
+\frac{M \left(L^2 + \rb^2\right) \left(M \rb^2 - L^2 M +L^2 \rb \right)^2}{E^2 \rb^{12}} + \frac{L^4 \left(M^2 - 4 M \rb +\rb^2 \right)}{ \rb^9} -\frac{L^2 M \left( 5 \rb - 7 M \right)}{\rb^7} - \frac{6 M \left(\rb - 2 M \right)}{\rb^5},
\\
\ddot{a}^\theta & = 0,
\\
\ddot{a}^\phi &=\dddup+ \frac{4 L}{\rb^3} \ddur+\frac{4}{\rb}\ur \ddup+\left[6 \dur +\frac{M}{\rb^2}- \frac{6 L^2 \left(\rb - 3 M \right)}{\rb^4}\right] \frac{\dup}{\rb} -\left[ L^2 \left(2 \rb - 3 M\right) + 2 M \rb^2 \right] \frac{2 L \ur}{\rb^8}.
\end{align*}
\end{widetext}

\bibliographystyle{apsrev4-1}
\bibliography{references}

\begin{thebibliography}{125}%
\makeatletter
\providecommand \@ifxundefined [1]{%
 \@ifx{#1\undefined}
}%
\providecommand \@ifnum [1]{%
 \ifnum #1\expandafter \@firstoftwo
 \else \expandafter \@secondoftwo
 \fi
}%
\providecommand \@ifx [1]{%
 \ifx #1\expandafter \@firstoftwo
 \else \expandafter \@secondoftwo
 \fi
}%
\providecommand \natexlab [1]{#1}%
\providecommand \enquote  [1]{``#1''}%
\providecommand \bibnamefont  [1]{#1}%
\providecommand \bibfnamefont [1]{#1}%
\providecommand \citenamefont [1]{#1}%
\providecommand \href@noop [0]{\@secondoftwo}%
\providecommand \href [0]{\begingroup \@sanitize@url \@href}%
\providecommand \@href[1]{\@@startlink{#1}\@@href}%
\providecommand \@@href[1]{\endgroup#1\@@endlink}%
\providecommand \@sanitize@url [0]{\catcode `\\12\catcode `\$12\catcode
  `\&12\catcode `\#12\catcode `\^12\catcode `\_12\catcode `\%12\relax}%
\providecommand \@@startlink[1]{}%
\providecommand \@@endlink[0]{}%
\providecommand \url  [0]{\begingroup\@sanitize@url \@url }%
\providecommand \@url [1]{\endgroup\@href {#1}{\urlprefix }}%
\providecommand \urlprefix  [0]{URL }%
\providecommand \Eprint [0]{\href }%
\providecommand \doibase [0]{http://dx.doi.org/}%
\providecommand \selectlanguage [0]{\@gobble}%
\providecommand \bibinfo  [0]{\@secondoftwo}%
\providecommand \bibfield  [0]{\@secondoftwo}%
\providecommand \translation [1]{[#1]}%
\providecommand \BibitemOpen [0]{}%
\providecommand \bibitemStop [0]{}%
\providecommand \bibitemNoStop [0]{.\EOS\space}%
\providecommand \EOS [0]{\spacefactor3000\relax}%
\providecommand \BibitemShut  [1]{\csname bibitem#1\endcsname}%
\let\auto@bib@innerbib\@empty
\bibitem [{\citenamefont {Abbott}\ \emph
  {et~al.}(2016{\natexlab{a}})\citenamefont {Abbott} \emph
  {et~al.}}]{Abbott:2016blz}%
  \BibitemOpen
  \bibfield  {author} {\bibinfo {author} {\bibfnamefont {B.~P.}\ \bibnamefont
  {Abbott}} \emph {et~al.} (\bibinfo {collaboration} {LIGO Scientific,
  Virgo}),\ }\href {\doibase 10.1103/PhysRevLett.116.061102} {\bibfield
  {journal} {\bibinfo  {journal} {Phys. Rev. Lett.}\ }\textbf {\bibinfo
  {volume} {116}},\ \bibinfo {pages} {061102} (\bibinfo {year}
  {2016}{\natexlab{a}})},\ \Eprint {http://arxiv.org/abs/1602.03837}
  {arXiv:1602.03837 [gr-qc]} \BibitemShut {NoStop}%
\bibitem [{\citenamefont {Abbott}\ \emph
  {et~al.}(2016{\natexlab{b}})\citenamefont {Abbott} \emph
  {et~al.}}]{Abbott:2016nmj}%
  \BibitemOpen
  \bibfield  {author} {\bibinfo {author} {\bibfnamefont {B.~P.}\ \bibnamefont
  {Abbott}} \emph {et~al.} (\bibinfo {collaboration} {LIGO Scientific,
  Virgo}),\ }\href {\doibase 10.1103/PhysRevLett.116.241103} {\bibfield
  {journal} {\bibinfo  {journal} {Phys. Rev. Lett.}\ }\textbf {\bibinfo
  {volume} {116}},\ \bibinfo {pages} {241103} (\bibinfo {year}
  {2016}{\natexlab{b}})},\ \Eprint {http://arxiv.org/abs/1606.04855}
  {arXiv:1606.04855} \BibitemShut {NoStop}%
\bibitem [{\citenamefont {Abbott}\ \emph
  {et~al.}(2017{\natexlab{a}})\citenamefont {Abbott} \emph
  {et~al.}}]{Abbott:2017vtc}%
  \BibitemOpen
  \bibfield  {author} {\bibinfo {author} {\bibfnamefont {B.~P.}\ \bibnamefont
  {Abbott}} \emph {et~al.} (\bibinfo {collaboration} {LIGO Scientific,
  VIRGO}),\ }\href {\doibase 10.1103/PhysRevLett.118.221101} {\bibfield
  {journal} {\bibinfo  {journal} {Phys. Rev. Lett.}\ }\textbf {\bibinfo
  {volume} {118}},\ \bibinfo {pages} {221101} (\bibinfo {year}
  {2017}{\natexlab{a}})},\ \Eprint {http://arxiv.org/abs/1706.01812}
  {arXiv:1706.01812} \BibitemShut {NoStop}%
\bibitem [{\citenamefont {Abbott}\ \emph
  {et~al.}(2017{\natexlab{b}})\citenamefont {Abbott} \emph
  {et~al.}}]{Abbott:2017oio}%
  \BibitemOpen
  \bibfield  {author} {\bibinfo {author} {\bibfnamefont {B.~P.}\ \bibnamefont
  {Abbott}} \emph {et~al.} (\bibinfo {collaboration} {LIGO Scientific,
  Virgo}),\ }\href {\doibase 10.1103/PhysRevLett.119.141101} {\bibfield
  {journal} {\bibinfo  {journal} {Phys. Rev. Lett.}\ }\textbf {\bibinfo
  {volume} {119}},\ \bibinfo {pages} {141101} (\bibinfo {year}
  {2017}{\natexlab{b}})},\ \Eprint {http://arxiv.org/abs/1709.09660}
  {arXiv:1709.09660 [gr-qc]} \BibitemShut {NoStop}%
\bibitem [{\citenamefont {Abbott}\ \emph
  {et~al.}(2017{\natexlab{c}})\citenamefont {Abbott} \emph
  {et~al.}}]{TheLIGOScientific:2017qsa}%
  \BibitemOpen
  \bibfield  {author} {\bibinfo {author} {\bibfnamefont {B.~P.}\ \bibnamefont
  {Abbott}} \emph {et~al.} (\bibinfo {collaboration} {LIGO Scientific,
  Virgo}),\ }\href {\doibase 10.1103/PhysRevLett.119.161101} {\bibfield
  {journal} {\bibinfo  {journal} {Phys. Rev. Lett.}\ }\textbf {\bibinfo
  {volume} {119}},\ \bibinfo {pages} {161101} (\bibinfo {year}
  {2017}{\natexlab{c}})},\ \Eprint {http://arxiv.org/abs/1710.05832}
  {arXiv:1710.05832} \BibitemShut {NoStop}%
\bibitem [{\citenamefont {Abbott}\ \emph
  {et~al.}(2017{\natexlab{d}})\citenamefont {Abbott} \emph
  {et~al.}}]{Abbott:2017gyy}%
  \BibitemOpen
  \bibfield  {author} {\bibinfo {author} {\bibfnamefont {B.~P.}\ \bibnamefont
  {Abbott}} \emph {et~al.} (\bibinfo {collaboration} {LIGO Scientific,
  Virgo}),\ }\href {\doibase 10.3847/2041-8213/aa9f0c} {\bibfield  {journal}
  {\bibinfo  {journal} {Astrophys. J.}\ }\textbf {\bibinfo {volume} {851}},\
  \bibinfo {pages} {L35} (\bibinfo {year} {2017}{\natexlab{d}})},\ \Eprint
  {http://arxiv.org/abs/1711.05578} {arXiv:1711.05578 [astro-ph.HE]}
  \BibitemShut {NoStop}%
\bibitem [{\citenamefont {Abbott}\ \emph
  {et~al.}(2017{\natexlab{e}})\citenamefont {Abbott} \emph
  {et~al.}}]{GBM:2017lvd}%
  \BibitemOpen
  \bibfield  {author} {\bibinfo {author} {\bibfnamefont {B.~P.}\ \bibnamefont
  {Abbott}} \emph {et~al.} (\bibinfo {collaboration} {LIGO Scientific, Virgo,
  GROND, SALT Group, OzGrav, DFN, INTEGRAL, Insight-Hxmt, MAXI Team, Fermi-LAT,
  J-GEM, RATIR, IceCube, CAASTRO, LWA, ePESSTO, GRAWITA, RIMAS, SKA South
  Africa/MeerKAT, H.E.S.S., 1M2H Team, IKI-GW Follow-up, Fermi GBM, Pi of Sky,
  DWF (Deeper Wider Faster Program), Dark Energy Survey, MASTER, AstroSat
  Cadmium Zinc Telluride Imager Team, Swift, Pierre Auger, ASKAP, VINROUGE,
  JAGWAR, Chandra Team at McGill University, TTU-NRAO, GROWTH, AGILE Team, MWA,
  ATCA, AST3, TOROS, Pan-STARRS, NuSTAR, ATLAS Telescopes, BOOTES, CaltechNRAO,
  High Time Resolution Universe Survey, Nordic Optical Telescope, Las Cumbres
  Observatory Group, TZAC Consortium, LOFAR, IPN, DLT40, Texas Tech University,
  HAWC, ANTARES, KU, Dark Energy Camera GW-EM, CALET, Euro VLBI Team, ALMA}),\
  }\href {\doibase 10.3847/2041-8213/aa91c9} {\bibfield  {journal} {\bibinfo
  {journal} {Astrophys. J.}\ }\textbf {\bibinfo {volume} {848}},\ \bibinfo
  {pages} {L12} (\bibinfo {year} {2017}{\natexlab{e}})},\ \Eprint
  {http://arxiv.org/abs/1710.05833} {arXiv:1710.05833 [astro-ph.HE]}
  \BibitemShut {NoStop}%
\bibitem [{\citenamefont {Armano}\ \emph {et~al.}(2016)\citenamefont {Armano}
  \emph {et~al.}}]{Armano:2016bkm}%
  \BibitemOpen
  \bibfield  {author} {\bibinfo {author} {\bibfnamefont {M.}~\bibnamefont
  {Armano}} \emph {et~al.},\ }\href {\doibase 10.1103/PhysRevLett.116.231101}
  {\bibfield  {journal} {\bibinfo  {journal} {Phys. Rev. Lett.}\ }\textbf
  {\bibinfo {volume} {116}},\ \bibinfo {pages} {231101} (\bibinfo {year}
  {2016})}\BibitemShut {NoStop}%
\bibitem [{\citenamefont {Armano}\ \emph {et~al.}(2018)\citenamefont {Armano}
  \emph {et~al.}}]{Armano:2018}%
  \BibitemOpen
  \bibfield  {author} {\bibinfo {author} {\bibfnamefont {M.}~\bibnamefont
  {Armano}} \emph {et~al.},\ }\href {\doibase 10.1103/PhysRevLett.120.061101}
  {\bibfield  {journal} {\bibinfo  {journal} {Phys. Rev. Lett.}\ }\textbf
  {\bibinfo {volume} {120}},\ \bibinfo {pages} {061101} (\bibinfo {year}
  {2018})}\BibitemShut {NoStop}%
\bibitem [{\citenamefont {Amaro-Seoane}\ and\ \citenamefont
  {Santamaria}(2010)}]{AmaroSeoane:2009ui}%
  \BibitemOpen
  \bibfield  {author} {\bibinfo {author} {\bibfnamefont {P.}~\bibnamefont
  {Amaro-Seoane}}\ and\ \bibinfo {author} {\bibfnamefont {L.}~\bibnamefont
  {Santamaria}},\ }\href {\doibase 10.1088/0004-637X/722/2/1197} {\bibfield
  {journal} {\bibinfo  {journal} {Astrophys. J.}\ }\textbf {\bibinfo {volume}
  {722}},\ \bibinfo {pages} {1197} (\bibinfo {year} {2010})},\ \Eprint
  {http://arxiv.org/abs/0910.0254} {arXiv:0910.0254} \BibitemShut {NoStop}%
\bibitem [{\citenamefont {Kocsis}\ and\ \citenamefont
  {Levin}(2012)}]{Kocsis:2011jy}%
  \BibitemOpen
  \bibfield  {author} {\bibinfo {author} {\bibfnamefont {B.}~\bibnamefont
  {Kocsis}}\ and\ \bibinfo {author} {\bibfnamefont {J.}~\bibnamefont {Levin}},\
  }\href {\doibase 10.1103/PhysRevD.85.123005} {\bibfield  {journal} {\bibinfo
  {journal} {Phys. Rev.}\ }\textbf {\bibinfo {volume} {D85}},\ \bibinfo {pages}
  {123005} (\bibinfo {year} {2012})},\ \Eprint {http://arxiv.org/abs/1109.4170}
  {arXiv:1109.4170} \BibitemShut {NoStop}%
\bibitem [{\citenamefont {Sesana}(2016)}]{Sesana:2016ljz}%
  \BibitemOpen
  \bibfield  {author} {\bibinfo {author} {\bibfnamefont {A.}~\bibnamefont
  {Sesana}},\ }\href {\doibase 10.1103/PhysRevLett.116.231102} {\bibfield
  {journal} {\bibinfo  {journal} {Phys. Rev. Lett.}\ }\textbf {\bibinfo
  {volume} {116}},\ \bibinfo {pages} {231102} (\bibinfo {year} {2016})},\
  \Eprint {http://arxiv.org/abs/1602.06951} {arXiv:1602.06951} \BibitemShut
  {NoStop}%
\bibitem [{\citenamefont {Colpi}\ and\ \citenamefont
  {Sesana}(2017)}]{Colpi:2016fup}%
  \BibitemOpen
  \bibfield  {author} {\bibinfo {author} {\bibfnamefont {M.}~\bibnamefont
  {Colpi}}\ and\ \bibinfo {author} {\bibfnamefont {A.}~\bibnamefont {Sesana}},\
  }in\ \href {\doibase 10.1142/9789813141766_0002} {\emph {\bibinfo {booktitle}
  {An Overview of Gravitational Waves: Theory, Sources and Detection}}},\
  \bibinfo {editor} {edited by\ \bibinfo {editor} {\bibfnamefont
  {G.}~\bibnamefont {Auger}}}\ (\bibinfo {year} {2017})\ pp.\ \bibinfo {pages}
  {43--140},\ \Eprint {http://arxiv.org/abs/1610.05309} {arXiv:1610.05309}
  \BibitemShut {NoStop}%
\bibitem [{\citenamefont {Sesana}(2017)}]{Sesana:2017vsj}%
  \BibitemOpen
  \bibfield  {author} {\bibinfo {author} {\bibfnamefont {A.}~\bibnamefont
  {Sesana}},\ }\bibfield  {booktitle} {\emph {\bibinfo {booktitle}
  {{Proceedings, 11th International LISA Symposium: Zurich, Switzerland,
  September 5-9, 2016}}},\ }\href {\doibase 10.1088/1742-6596/840/1/012018}
  {\bibfield  {journal} {\bibinfo  {journal} {J. Phys. Conf. Ser.}\ }\textbf
  {\bibinfo {volume} {840}},\ \bibinfo {pages} {012018} (\bibinfo {year}
  {2017})},\ \Eprint {http://arxiv.org/abs/1702.04356} {arXiv:1702.04356}
  \BibitemShut {NoStop}%
\bibitem [{\citenamefont {Vitale}(2016)}]{Vitale:2016rfr}%
  \BibitemOpen
  \bibfield  {author} {\bibinfo {author} {\bibfnamefont {S.}~\bibnamefont
  {Vitale}},\ }\href {\doibase 10.1103/PhysRevLett.117.051102} {\bibfield
  {journal} {\bibinfo  {journal} {Phys. Rev. Lett.}\ }\textbf {\bibinfo
  {volume} {117}},\ \bibinfo {pages} {051102} (\bibinfo {year} {2016})},\
  \Eprint {http://arxiv.org/abs/1605.01037} {arXiv:1605.01037} \BibitemShut
  {NoStop}%
\bibitem [{\citenamefont {Bonvin}\ \emph {et~al.}(2017)\citenamefont {Bonvin},
  \citenamefont {Caprini}, \citenamefont {Sturani},\ and\ \citenamefont
  {Tamanini}}]{Bonvin:2016qxr}%
  \BibitemOpen
  \bibfield  {author} {\bibinfo {author} {\bibfnamefont {C.}~\bibnamefont
  {Bonvin}}, \bibinfo {author} {\bibfnamefont {C.}~\bibnamefont {Caprini}},
  \bibinfo {author} {\bibfnamefont {R.}~\bibnamefont {Sturani}}, \ and\
  \bibinfo {author} {\bibfnamefont {N.}~\bibnamefont {Tamanini}},\ }\href
  {\doibase 10.1103/PhysRevD.95.044029} {\bibfield  {journal} {\bibinfo
  {journal} {Phys. Rev.}\ }\textbf {\bibinfo {volume} {D95}},\ \bibinfo {pages}
  {044029} (\bibinfo {year} {2017})},\ \Eprint
  {http://arxiv.org/abs/1609.08093} {arXiv:1609.08093} \BibitemShut {NoStop}%
\bibitem [{\citenamefont {Sesana}\ \emph {et~al.}(2009)\citenamefont {Sesana},
  \citenamefont {Gair}, \citenamefont {Mandel},\ and\ \citenamefont
  {Vecchio}}]{Sesana:2009wg}%
  \BibitemOpen
  \bibfield  {author} {\bibinfo {author} {\bibfnamefont {A.}~\bibnamefont
  {Sesana}}, \bibinfo {author} {\bibfnamefont {J.}~\bibnamefont {Gair}},
  \bibinfo {author} {\bibfnamefont {I.}~\bibnamefont {Mandel}}, \ and\ \bibinfo
  {author} {\bibfnamefont {A.}~\bibnamefont {Vecchio}},\ }\href {\doibase
  10.1088/0004-637X/698/2/L129} {\bibfield  {journal} {\bibinfo  {journal}
  {Astrophys. J.}\ }\textbf {\bibinfo {volume} {698}},\ \bibinfo {pages} {L129}
  (\bibinfo {year} {2009})},\ \Eprint {http://arxiv.org/abs/0903.4177}
  {arXiv:0903.4177} \BibitemShut {NoStop}%
\bibitem [{\citenamefont {Nishizawa}\ \emph {et~al.}(2016)\citenamefont
  {Nishizawa}, \citenamefont {Berti}, \citenamefont {Klein},\ and\
  \citenamefont {Sesana}}]{Nishizawa:2016jji}%
  \BibitemOpen
  \bibfield  {author} {\bibinfo {author} {\bibfnamefont {A.}~\bibnamefont
  {Nishizawa}}, \bibinfo {author} {\bibfnamefont {E.}~\bibnamefont {Berti}},
  \bibinfo {author} {\bibfnamefont {A.}~\bibnamefont {Klein}}, \ and\ \bibinfo
  {author} {\bibfnamefont {A.}~\bibnamefont {Sesana}},\ }\href {\doibase
  10.1103/PhysRevD.94.064020} {\bibfield  {journal} {\bibinfo  {journal} {Phys.
  Rev.}\ }\textbf {\bibinfo {volume} {D94}},\ \bibinfo {pages} {064020}
  (\bibinfo {year} {2016})},\ \Eprint {http://arxiv.org/abs/1605.01341}
  {arXiv:1605.01341} \BibitemShut {NoStop}%
\bibitem [{\citenamefont {Nishizawa}\ \emph {et~al.}(2017)\citenamefont
  {Nishizawa}, \citenamefont {Sesana}, \citenamefont {Berti},\ and\
  \citenamefont {Klein}}]{Nishizawa:2016eza}%
  \BibitemOpen
  \bibfield  {author} {\bibinfo {author} {\bibfnamefont {A.}~\bibnamefont
  {Nishizawa}}, \bibinfo {author} {\bibfnamefont {A.}~\bibnamefont {Sesana}},
  \bibinfo {author} {\bibfnamefont {E.}~\bibnamefont {Berti}}, \ and\ \bibinfo
  {author} {\bibfnamefont {A.}~\bibnamefont {Klein}},\ }\href {\doibase
  10.1093/mnras/stw2993} {\bibfield  {journal} {\bibinfo  {journal} {Mon. Not.
  Roy. Astron. Soc.}\ }\textbf {\bibinfo {volume} {465}},\ \bibinfo {pages}
  {4375} (\bibinfo {year} {2017})},\ \Eprint {http://arxiv.org/abs/1606.09295}
  {arXiv:1606.09295} \BibitemShut {NoStop}%
\bibitem [{\citenamefont {Breivik}\ \emph {et~al.}(2016)\citenamefont
  {Breivik}, \citenamefont {Rodriguez}, \citenamefont {Larson}, \citenamefont
  {Kalogera},\ and\ \citenamefont {Rasio}}]{Breivik:2016ddj}%
  \BibitemOpen
  \bibfield  {author} {\bibinfo {author} {\bibfnamefont {K.}~\bibnamefont
  {Breivik}}, \bibinfo {author} {\bibfnamefont {C.~L.}\ \bibnamefont
  {Rodriguez}}, \bibinfo {author} {\bibfnamefont {S.~L.}\ \bibnamefont
  {Larson}}, \bibinfo {author} {\bibfnamefont {V.}~\bibnamefont {Kalogera}}, \
  and\ \bibinfo {author} {\bibfnamefont {F.~A.}\ \bibnamefont {Rasio}},\ }\href
  {\doibase 10.3847/2041-8205/830/1/L18} {\bibfield  {journal} {\bibinfo
  {journal} {Astrophys. J.}\ }\textbf {\bibinfo {volume} {830}},\ \bibinfo
  {pages} {L18} (\bibinfo {year} {2016})},\ \Eprint
  {http://arxiv.org/abs/1606.09558} {arXiv:1606.09558} \BibitemShut {NoStop}%
\bibitem [{\citenamefont {Inayoshi}\ \emph {et~al.}(2017)\citenamefont
  {Inayoshi}, \citenamefont {Tamanini}, \citenamefont {Caprini},\ and\
  \citenamefont {Haiman}}]{Inayoshi:2017hgw}%
  \BibitemOpen
  \bibfield  {author} {\bibinfo {author} {\bibfnamefont {K.}~\bibnamefont
  {Inayoshi}}, \bibinfo {author} {\bibfnamefont {N.}~\bibnamefont {Tamanini}},
  \bibinfo {author} {\bibfnamefont {C.}~\bibnamefont {Caprini}}, \ and\
  \bibinfo {author} {\bibfnamefont {Z.}~\bibnamefont {Haiman}},\ }\href
  {\doibase 10.1103/PhysRevD.96.063014} {\bibfield  {journal} {\bibinfo
  {journal} {Phys. Rev.}\ }\textbf {\bibinfo {volume} {D96}},\ \bibinfo {pages}
  {063014} (\bibinfo {year} {2017})},\ \Eprint
  {http://arxiv.org/abs/1702.06529} {arXiv:1702.06529} \BibitemShut {NoStop}%
\bibitem [{\citenamefont {Barausse}\ \emph {et~al.}(2016)\citenamefont
  {Barausse}, \citenamefont {Yunes},\ and\ \citenamefont
  {Chamberlain}}]{Barausse:2016eii}%
  \BibitemOpen
  \bibfield  {author} {\bibinfo {author} {\bibfnamefont {E.}~\bibnamefont
  {Barausse}}, \bibinfo {author} {\bibfnamefont {N.}~\bibnamefont {Yunes}}, \
  and\ \bibinfo {author} {\bibfnamefont {K.}~\bibnamefont {Chamberlain}},\
  }\href {\doibase 10.1103/PhysRevLett.116.241104} {\bibfield  {journal}
  {\bibinfo  {journal} {Phys. Rev. Lett.}\ }\textbf {\bibinfo {volume} {116}},\
  \bibinfo {pages} {241104} (\bibinfo {year} {2016})},\ \Eprint
  {http://arxiv.org/abs/1603.04075} {arXiv:1603.04075} \BibitemShut {NoStop}%
\bibitem [{\citenamefont {{Amaro-Seoane}}\ \emph {et~al.}(2017)\citenamefont
  {{Amaro-Seoane}} \emph {et~al.}}]{LISA}%
  \BibitemOpen
  \bibfield  {author} {\bibinfo {author} {\bibfnamefont {P.}~\bibnamefont
  {{Amaro-Seoane}}} \emph {et~al.},\ }\href@noop {} {\bibfield  {journal}
  {\bibinfo  {journal} {ArXiv e-prints}\ } (\bibinfo {year} {2017})},\ \Eprint
  {http://arxiv.org/abs/1702.00786} {arXiv:1702.00786} \BibitemShut {NoStop}%
\bibitem [{\citenamefont {Glampedakis}\ and\ \citenamefont
  {Babak}(2006)}]{Glampedakis:2005cf}%
  \BibitemOpen
  \bibfield  {author} {\bibinfo {author} {\bibfnamefont {K.}~\bibnamefont
  {Glampedakis}}\ and\ \bibinfo {author} {\bibfnamefont {S.}~\bibnamefont
  {Babak}},\ }\href {\doibase 10.1088/0264-9381/23/12/013} {\bibfield
  {journal} {\bibinfo  {journal} {Class. Quant. Grav.}\ }\textbf {\bibinfo
  {volume} {23}},\ \bibinfo {pages} {4167} (\bibinfo {year} {2006})},\ \Eprint
  {http://arxiv.org/abs/gr-qc/0510057} {arXiv:gr-qc/0510057} \BibitemShut
  {NoStop}%
\bibitem [{\citenamefont {Vigeland}\ and\ \citenamefont
  {Hughes}(2010)}]{Vigeland:2009pr}%
  \BibitemOpen
  \bibfield  {author} {\bibinfo {author} {\bibfnamefont {S.~J.}\ \bibnamefont
  {Vigeland}}\ and\ \bibinfo {author} {\bibfnamefont {S.~A.}\ \bibnamefont
  {Hughes}},\ }\href {\doibase 10.1103/PhysRevD.81.024030} {\bibfield
  {journal} {\bibinfo  {journal} {Phys. Rev.}\ }\textbf {\bibinfo {volume}
  {D81}},\ \bibinfo {pages} {024030} (\bibinfo {year} {2010})},\ \Eprint
  {http://arxiv.org/abs/0911.1756} {arXiv:0911.1756} \BibitemShut {NoStop}%
\bibitem [{\citenamefont {Moore}\ \emph {et~al.}(2017)\citenamefont {Moore},
  \citenamefont {Chua},\ and\ \citenamefont {Gair}}]{Moore:2017lxy}%
  \BibitemOpen
  \bibfield  {author} {\bibinfo {author} {\bibfnamefont {C.~J.}\ \bibnamefont
  {Moore}}, \bibinfo {author} {\bibfnamefont {A.~J.~K.}\ \bibnamefont {Chua}},
  \ and\ \bibinfo {author} {\bibfnamefont {J.~R.}\ \bibnamefont {Gair}},\
  }\href {\doibase 10.1088/1361-6382/aa85fa} {\bibfield  {journal} {\bibinfo
  {journal} {Class. Quant. Grav.}\ }\textbf {\bibinfo {volume} {34}},\ \bibinfo
  {pages} {195009} (\bibinfo {year} {2017})},\ \Eprint
  {http://arxiv.org/abs/1707.00712} {arXiv:1707.00712} \BibitemShut {NoStop}%
\bibitem [{\citenamefont {Barausse}\ \emph {et~al.}(2014)\citenamefont
  {Barausse}, \citenamefont {Cardoso},\ and\ \citenamefont
  {Pani}}]{Barausse:2014tra}%
  \BibitemOpen
  \bibfield  {author} {\bibinfo {author} {\bibfnamefont {E.}~\bibnamefont
  {Barausse}}, \bibinfo {author} {\bibfnamefont {V.}~\bibnamefont {Cardoso}}, \
  and\ \bibinfo {author} {\bibfnamefont {P.}~\bibnamefont {Pani}},\ }\href
  {\doibase 10.1103/PhysRevD.89.104059} {\bibfield  {journal} {\bibinfo
  {journal} {Phys. Rev.}\ }\textbf {\bibinfo {volume} {D89}},\ \bibinfo {pages}
  {104059} (\bibinfo {year} {2014})},\ \Eprint {http://arxiv.org/abs/1404.7149}
  {arXiv:1404.7149} \BibitemShut {NoStop}%
\bibitem [{\citenamefont {Babak}\ \emph {et~al.}(2017)\citenamefont {Babak},
  \citenamefont {Gair}, \citenamefont {Sesana}, \citenamefont {Barausse},
  \citenamefont {Sopuerta}, \citenamefont {Berry}, \citenamefont {Berti},
  \citenamefont {Amaro-Seoane}, \citenamefont {Petiteau},\ and\ \citenamefont
  {Klein}}]{Babak:2017tow}%
  \BibitemOpen
  \bibfield  {author} {\bibinfo {author} {\bibfnamefont {S.}~\bibnamefont
  {Babak}}, \bibinfo {author} {\bibfnamefont {J.}~\bibnamefont {Gair}},
  \bibinfo {author} {\bibfnamefont {A.}~\bibnamefont {Sesana}}, \bibinfo
  {author} {\bibfnamefont {E.}~\bibnamefont {Barausse}}, \bibinfo {author}
  {\bibfnamefont {C.~F.}\ \bibnamefont {Sopuerta}}, \bibinfo {author}
  {\bibfnamefont {C.~P.~L.}\ \bibnamefont {Berry}}, \bibinfo {author}
  {\bibfnamefont {E.}~\bibnamefont {Berti}}, \bibinfo {author} {\bibfnamefont
  {P.}~\bibnamefont {Amaro-Seoane}}, \bibinfo {author} {\bibfnamefont
  {A.}~\bibnamefont {Petiteau}}, \ and\ \bibinfo {author} {\bibfnamefont
  {A.}~\bibnamefont {Klein}},\ }\href {\doibase 10.1103/PhysRevD.95.103012}
  {\bibfield  {journal} {\bibinfo  {journal} {Phys. Rev.}\ }\textbf {\bibinfo
  {volume} {D95}},\ \bibinfo {pages} {103012} (\bibinfo {year} {2017})},\
  \Eprint {http://arxiv.org/abs/1703.09722} {arXiv:1703.09722} \BibitemShut
  {NoStop}%
\bibitem [{\citenamefont {Detweiler}(2008)}]{Detweiler:2008ft}%
  \BibitemOpen
  \bibfield  {author} {\bibinfo {author} {\bibfnamefont {S.~L.}\ \bibnamefont
  {Detweiler}},\ }\href {\doibase 10.1103/PhysRevD.77.124026} {\bibfield
  {journal} {\bibinfo  {journal} {Phys. Rev.}\ }\textbf {\bibinfo {volume}
  {D77}},\ \bibinfo {pages} {124026} (\bibinfo {year} {2008})},\ \Eprint
  {http://arxiv.org/abs/0804.3529} {arXiv:0804.3529} \BibitemShut {NoStop}%
\bibitem [{\citenamefont {Blanchet}\ \emph
  {et~al.}(2010{\natexlab{a}})\citenamefont {Blanchet}, \citenamefont
  {Detweiler}, \citenamefont {Le~Tiec},\ and\ \citenamefont
  {Whiting}}]{Blanchet:2009sd}%
  \BibitemOpen
  \bibfield  {author} {\bibinfo {author} {\bibfnamefont {L.}~\bibnamefont
  {Blanchet}}, \bibinfo {author} {\bibfnamefont {S.~L.}\ \bibnamefont
  {Detweiler}}, \bibinfo {author} {\bibfnamefont {A.}~\bibnamefont {Le~Tiec}},
  \ and\ \bibinfo {author} {\bibfnamefont {B.~F.}\ \bibnamefont {Whiting}},\
  }\href {\doibase 10.1103/PhysRevD.81.064004} {\bibfield  {journal} {\bibinfo
  {journal} {Phys. Rev.}\ }\textbf {\bibinfo {volume} {D81}},\ \bibinfo {pages}
  {064004} (\bibinfo {year} {2010}{\natexlab{a}})},\ \Eprint
  {http://arxiv.org/abs/0910.0207} {arXiv:0910.0207} \BibitemShut {NoStop}%
\bibitem [{\citenamefont {Blanchet}\ \emph
  {et~al.}(2010{\natexlab{b}})\citenamefont {Blanchet}, \citenamefont
  {Detweiler}, \citenamefont {Le~Tiec},\ and\ \citenamefont
  {Whiting}}]{Blanchet:2010zd}%
  \BibitemOpen
  \bibfield  {author} {\bibinfo {author} {\bibfnamefont {L.}~\bibnamefont
  {Blanchet}}, \bibinfo {author} {\bibfnamefont {S.~L.}\ \bibnamefont
  {Detweiler}}, \bibinfo {author} {\bibfnamefont {A.}~\bibnamefont {Le~Tiec}},
  \ and\ \bibinfo {author} {\bibfnamefont {B.~F.}\ \bibnamefont {Whiting}},\
  }\href {\doibase 10.1103/PhysRevD.81.084033} {\bibfield  {journal} {\bibinfo
  {journal} {Phys. Rev.}\ }\textbf {\bibinfo {volume} {D81}},\ \bibinfo {pages}
  {084033} (\bibinfo {year} {2010}{\natexlab{b}})},\ \Eprint
  {http://arxiv.org/abs/1002.0726} {arXiv:1002.0726} \BibitemShut {NoStop}%
\bibitem [{\citenamefont {Blanchet}\ \emph {et~al.}(2011)\citenamefont
  {Blanchet}, \citenamefont {Detweiler}, \citenamefont {Le~Tiec},\ and\
  \citenamefont {Whiting}}]{Blanchet:2011aha}%
  \BibitemOpen
  \bibfield  {author} {\bibinfo {author} {\bibfnamefont {L.}~\bibnamefont
  {Blanchet}}, \bibinfo {author} {\bibfnamefont {S.}~\bibnamefont {Detweiler}},
  \bibinfo {author} {\bibfnamefont {A.}~\bibnamefont {Le~Tiec}}, \ and\
  \bibinfo {author} {\bibfnamefont {B.~F.}\ \bibnamefont {Whiting}},\
  }\bibfield  {booktitle} {\emph {\bibinfo {booktitle} {{Mass and motion in
  general relativity. Proceedings, School on Mass, Orleans, France, June 23-25,
  2008}}},\ }\href {\doibase 10.1007/978-90-481-3015-3_15} {\bibfield
  {journal} {\bibinfo  {journal} {Fundam. Theor. Phys.}\ }\textbf {\bibinfo
  {volume} {162}},\ \bibinfo {pages} {415} (\bibinfo {year} {2011})},\ \bibinfo
  {note} {[,415(2010)]},\ \Eprint {http://arxiv.org/abs/1007.2614}
  {arXiv:1007.2614} \BibitemShut {NoStop}%
\bibitem [{\citenamefont {Le~Tiec}\ \emph {et~al.}(2012)\citenamefont
  {Le~Tiec}, \citenamefont {Barausse},\ and\ \citenamefont
  {Buonanno}}]{LeTiec:2011dp}%
  \BibitemOpen
  \bibfield  {author} {\bibinfo {author} {\bibfnamefont {A.}~\bibnamefont
  {Le~Tiec}}, \bibinfo {author} {\bibfnamefont {E.}~\bibnamefont {Barausse}}, \
  and\ \bibinfo {author} {\bibfnamefont {A.}~\bibnamefont {Buonanno}},\ }\href
  {\doibase 10.1103/PhysRevLett.108.131103} {\bibfield  {journal} {\bibinfo
  {journal} {Phys. Rev. Lett.}\ }\textbf {\bibinfo {volume} {108}},\ \bibinfo
  {pages} {131103} (\bibinfo {year} {2012})},\ \Eprint
  {http://arxiv.org/abs/1111.5609} {arXiv:1111.5609} \BibitemShut {NoStop}%
\bibitem [{\citenamefont {Le~Tiec}\ \emph {et~al.}(2013)\citenamefont {Le~Tiec}
  \emph {et~al.}}]{Tiec:2013twa}%
  \BibitemOpen
  \bibfield  {author} {\bibinfo {author} {\bibfnamefont {A.}~\bibnamefont
  {Le~Tiec}} \emph {et~al.},\ }\href {\doibase 10.1103/PhysRevD.88.124027}
  {\bibfield  {journal} {\bibinfo  {journal} {Phys. Rev.}\ }\textbf {\bibinfo
  {volume} {D88}},\ \bibinfo {pages} {124027} (\bibinfo {year} {2013})},\
  \Eprint {http://arxiv.org/abs/1309.0541} {arXiv:1309.0541} \BibitemShut
  {NoStop}%
\bibitem [{\citenamefont {Dolan}\ \emph {et~al.}(2014)\citenamefont {Dolan},
  \citenamefont {Warburton}, \citenamefont {Harte}, \citenamefont {Le~Tiec},
  \citenamefont {Wardell},\ and\ \citenamefont {Barack}}]{Dolan:2013roa}%
  \BibitemOpen
  \bibfield  {author} {\bibinfo {author} {\bibfnamefont {S.~R.}\ \bibnamefont
  {Dolan}}, \bibinfo {author} {\bibfnamefont {N.}~\bibnamefont {Warburton}},
  \bibinfo {author} {\bibfnamefont {A.~I.}\ \bibnamefont {Harte}}, \bibinfo
  {author} {\bibfnamefont {A.}~\bibnamefont {Le~Tiec}}, \bibinfo {author}
  {\bibfnamefont {B.}~\bibnamefont {Wardell}}, \ and\ \bibinfo {author}
  {\bibfnamefont {L.}~\bibnamefont {Barack}},\ }\href {\doibase
  10.1103/PhysRevD.89.064011} {\bibfield  {journal} {\bibinfo  {journal} {Phys.
  Rev.}\ }\textbf {\bibinfo {volume} {D89}},\ \bibinfo {pages} {064011}
  (\bibinfo {year} {2014})},\ \Eprint {http://arxiv.org/abs/1312.0775}
  {arXiv:1312.0775} \BibitemShut {NoStop}%
\bibitem [{\citenamefont {Isoyama}\ \emph {et~al.}(2014)\citenamefont
  {Isoyama}, \citenamefont {Barack}, \citenamefont {Dolan}, \citenamefont
  {Le~Tiec}, \citenamefont {Nakano}, \citenamefont {Shah}, \citenamefont
  {Tanaka},\ and\ \citenamefont {Warburton}}]{Isoyama:2014mja}%
  \BibitemOpen
  \bibfield  {author} {\bibinfo {author} {\bibfnamefont {S.}~\bibnamefont
  {Isoyama}}, \bibinfo {author} {\bibfnamefont {L.}~\bibnamefont {Barack}},
  \bibinfo {author} {\bibfnamefont {S.~R.}\ \bibnamefont {Dolan}}, \bibinfo
  {author} {\bibfnamefont {A.}~\bibnamefont {Le~Tiec}}, \bibinfo {author}
  {\bibfnamefont {H.}~\bibnamefont {Nakano}}, \bibinfo {author} {\bibfnamefont
  {A.~G.}\ \bibnamefont {Shah}}, \bibinfo {author} {\bibfnamefont
  {T.}~\bibnamefont {Tanaka}}, \ and\ \bibinfo {author} {\bibfnamefont
  {N.}~\bibnamefont {Warburton}},\ }\href {\doibase
  10.1103/PhysRevLett.113.161101} {\bibfield  {journal} {\bibinfo  {journal}
  {Phys. Rev. Lett.}\ }\textbf {\bibinfo {volume} {113}},\ \bibinfo {pages}
  {161101} (\bibinfo {year} {2014})},\ \Eprint {http://arxiv.org/abs/1404.6133}
  {arXiv:1404.6133} \BibitemShut {NoStop}%
\bibitem [{\citenamefont {Akcay}\ \emph {et~al.}(2015)\citenamefont {Akcay},
  \citenamefont {Le~Tiec}, \citenamefont {Barack}, \citenamefont {Sago},\ and\
  \citenamefont {Warburton}}]{Akcay:2015pza}%
  \BibitemOpen
  \bibfield  {author} {\bibinfo {author} {\bibfnamefont {S.}~\bibnamefont
  {Akcay}}, \bibinfo {author} {\bibfnamefont {A.}~\bibnamefont {Le~Tiec}},
  \bibinfo {author} {\bibfnamefont {L.}~\bibnamefont {Barack}}, \bibinfo
  {author} {\bibfnamefont {N.}~\bibnamefont {Sago}}, \ and\ \bibinfo {author}
  {\bibfnamefont {N.}~\bibnamefont {Warburton}},\ }\href {\doibase
  10.1103/PhysRevD.91.124014} {\bibfield  {journal} {\bibinfo  {journal} {Phys.
  Rev.}\ }\textbf {\bibinfo {volume} {D91}},\ \bibinfo {pages} {124014}
  (\bibinfo {year} {2015})},\ \Eprint {http://arxiv.org/abs/1503.01374}
  {arXiv:1503.01374} \BibitemShut {NoStop}%
\bibitem [{\citenamefont {Akcay}\ \emph {et~al.}(2017)\citenamefont {Akcay},
  \citenamefont {Dempsey},\ and\ \citenamefont {Dolan}}]{Akcay:2016dku}%
  \BibitemOpen
  \bibfield  {author} {\bibinfo {author} {\bibfnamefont {S.}~\bibnamefont
  {Akcay}}, \bibinfo {author} {\bibfnamefont {D.}~\bibnamefont {Dempsey}}, \
  and\ \bibinfo {author} {\bibfnamefont {S.}~\bibnamefont {Dolan}},\ }\href
  {\doibase 10.1088/1361-6382/aa61d6} {\bibfield  {journal} {\bibinfo
  {journal} {Class. Quant. Grav.}\ }\textbf {\bibinfo {volume} {34}},\ \bibinfo
  {pages} {084001} (\bibinfo {year} {2017})},\ \Eprint
  {http://arxiv.org/abs/1608.04811} {arXiv:1608.04811} \BibitemShut {NoStop}%
\bibitem [{\citenamefont {Zimmerman}\ \emph {et~al.}()\citenamefont
  {Zimmerman}, \citenamefont {Lewis}, \citenamefont {Ossokine}, \citenamefont
  {Pound},\ and\ \citenamefont {Pfeiffer}}]{Zimmerman:2018}%
  \BibitemOpen
  \bibfield  {author} {\bibinfo {author} {\bibfnamefont {A.}~\bibnamefont
  {Zimmerman}}, \bibinfo {author} {\bibfnamefont {A.~G.}\ \bibnamefont
  {Lewis}}, \bibinfo {author} {\bibfnamefont {S.}~\bibnamefont {Ossokine}},
  \bibinfo {author} {\bibfnamefont {A.}~\bibnamefont {Pound}}, \ and\ \bibinfo
  {author} {\bibfnamefont {H.~P.}\ \bibnamefont {Pfeiffer}},\ }\href@noop {} {\
  }\bibinfo {note} {In preparation}\BibitemShut {NoStop}%
\bibitem [{\citenamefont {Le~Tiec}(2014)}]{Tiec:2014lba}%
  \BibitemOpen
  \bibfield  {author} {\bibinfo {author} {\bibfnamefont {A.}~\bibnamefont
  {Le~Tiec}},\ }\href {\doibase 10.1142/S0218271814300225} {\bibfield
  {journal} {\bibinfo  {journal} {Int. J. Mod. Phys.}\ }\textbf {\bibinfo
  {volume} {D23}},\ \bibinfo {pages} {1430022} (\bibinfo {year} {2014})},\
  \Eprint {http://arxiv.org/abs/1408.5505} {arXiv:1408.5505} \BibitemShut
  {NoStop}%
\bibitem [{\citenamefont {Damour}(2010)}]{Damour:2009sm}%
  \BibitemOpen
  \bibfield  {author} {\bibinfo {author} {\bibfnamefont {T.}~\bibnamefont
  {Damour}},\ }\href {\doibase 10.1103/PhysRevD.81.024017} {\bibfield
  {journal} {\bibinfo  {journal} {Phys. Rev.}\ }\textbf {\bibinfo {volume}
  {D81}},\ \bibinfo {pages} {024017} (\bibinfo {year} {2010})},\ \Eprint
  {http://arxiv.org/abs/0910.5533} {arXiv:0910.5533} \BibitemShut {NoStop}%
\bibitem [{\citenamefont {Barack}\ \emph {et~al.}(2010)\citenamefont {Barack},
  \citenamefont {Damour},\ and\ \citenamefont {Sago}}]{Barack:2010ny}%
  \BibitemOpen
  \bibfield  {author} {\bibinfo {author} {\bibfnamefont {L.}~\bibnamefont
  {Barack}}, \bibinfo {author} {\bibfnamefont {T.}~\bibnamefont {Damour}}, \
  and\ \bibinfo {author} {\bibfnamefont {N.}~\bibnamefont {Sago}},\ }\href
  {\doibase 10.1103/PhysRevD.82.084036} {\bibfield  {journal} {\bibinfo
  {journal} {Phys. Rev.}\ }\textbf {\bibinfo {volume} {D82}},\ \bibinfo {pages}
  {084036} (\bibinfo {year} {2010})},\ \Eprint {http://arxiv.org/abs/1008.0935}
  {arXiv:1008.0935} \BibitemShut {NoStop}%
\bibitem [{\citenamefont {Akcay}\ \emph {et~al.}(2012)\citenamefont {Akcay},
  \citenamefont {Barack}, \citenamefont {Damour},\ and\ \citenamefont
  {Sago}}]{Akcay:2012ea}%
  \BibitemOpen
  \bibfield  {author} {\bibinfo {author} {\bibfnamefont {S.}~\bibnamefont
  {Akcay}}, \bibinfo {author} {\bibfnamefont {L.}~\bibnamefont {Barack}},
  \bibinfo {author} {\bibfnamefont {T.}~\bibnamefont {Damour}}, \ and\ \bibinfo
  {author} {\bibfnamefont {N.}~\bibnamefont {Sago}},\ }\href {\doibase
  10.1103/PhysRevD.86.104041} {\bibfield  {journal} {\bibinfo  {journal} {Phys.
  Rev.}\ }\textbf {\bibinfo {volume} {D86}},\ \bibinfo {pages} {104041}
  (\bibinfo {year} {2012})},\ \Eprint {http://arxiv.org/abs/1209.0964}
  {arXiv:1209.0964} \BibitemShut {NoStop}%
\bibitem [{\citenamefont {Mino}\ \emph {et~al.}(1997)\citenamefont {Mino},
  \citenamefont {Sasaki},\ and\ \citenamefont
  {Tanaka}}]{Mino:Sasaki:Tanaka:1996}%
  \BibitemOpen
  \bibfield  {author} {\bibinfo {author} {\bibfnamefont {Y.}~\bibnamefont
  {Mino}}, \bibinfo {author} {\bibfnamefont {M.}~\bibnamefont {Sasaki}}, \ and\
  \bibinfo {author} {\bibfnamefont {T.}~\bibnamefont {Tanaka}},\ }\href
  {\doibase 10.1103/PhysRevD.55.3457} {\bibfield  {journal} {\bibinfo
  {journal} {Phys. Rev.}\ }\textbf {\bibinfo {volume} {D55}},\ \bibinfo {pages}
  {3457} (\bibinfo {year} {1997})},\ \Eprint
  {http://arxiv.org/abs/gr-qc/9606018} {arXiv:gr-qc/9606018} \BibitemShut
  {NoStop}%
\bibitem [{\citenamefont {Quinn}\ and\ \citenamefont
  {Wald}(1997)}]{Quinn:Wald:1997}%
  \BibitemOpen
  \bibfield  {author} {\bibinfo {author} {\bibfnamefont {T.~C.}\ \bibnamefont
  {Quinn}}\ and\ \bibinfo {author} {\bibfnamefont {R.~M.}\ \bibnamefont
  {Wald}},\ }\href {\doibase 10.1103/PhysRevD.56.3381} {\bibfield  {journal}
  {\bibinfo  {journal} {Phys. Rev.}\ }\textbf {\bibinfo {volume} {D56}},\
  \bibinfo {pages} {3381} (\bibinfo {year} {1997})},\ \Eprint
  {http://arxiv.org/abs/gr-qc/9610053} {arXiv:gr-qc/9610053} \BibitemShut
  {NoStop}%
\bibitem [{\citenamefont {Detweiler}\ and\ \citenamefont
  {Whiting}(2003)}]{Detweiler-Whiting-2003}%
  \BibitemOpen
  \bibfield  {author} {\bibinfo {author} {\bibfnamefont {S.}~\bibnamefont
  {Detweiler}}\ and\ \bibinfo {author} {\bibfnamefont {B.~F.}\ \bibnamefont
  {Whiting}},\ }\href@noop {} {\bibfield  {journal} {\bibinfo  {journal} {Phys.
  Rev. D}\ }\textbf {\bibinfo {volume} {67}},\ \bibinfo {pages} {024025}
  (\bibinfo {year} {2003})}\BibitemShut {NoStop}%
\bibitem [{\citenamefont {Gralla}\ and\ \citenamefont
  {Wald}(2008)}]{Gralla:Wald:2008}%
  \BibitemOpen
  \bibfield  {author} {\bibinfo {author} {\bibfnamefont {S.~E.}\ \bibnamefont
  {Gralla}}\ and\ \bibinfo {author} {\bibfnamefont {R.~M.}\ \bibnamefont
  {Wald}},\ }\href@noop {} {\bibfield  {journal} {\bibinfo  {journal} {Class.
  Quantum Grav.}\ }\textbf {\bibinfo {volume} {25}},\ \bibinfo {pages} {205009}
  (\bibinfo {year} {2008})},\ \Eprint {http://arxiv.org/abs/arXiv:0806.3293}
  {arXiv:0806.3293} \BibitemShut {NoStop}%
\bibitem [{\citenamefont {Poisson}\ \emph {et~al.}(2011)\citenamefont
  {Poisson}, \citenamefont {Pound},\ and\ \citenamefont {Vega}}]{Poisson:2003}%
  \BibitemOpen
  \bibfield  {author} {\bibinfo {author} {\bibfnamefont {E.}~\bibnamefont
  {Poisson}}, \bibinfo {author} {\bibfnamefont {A.}~\bibnamefont {Pound}}, \
  and\ \bibinfo {author} {\bibfnamefont {I.}~\bibnamefont {Vega}},\ }\href@noop
  {} {\bibfield  {journal} {\bibinfo  {journal} {Living Rev. Relativity}\
  }\textbf {\bibinfo {volume} {14}},\ \bibinfo {pages} {7} (\bibinfo {year}
  {2011})},\ \Eprint {http://arxiv.org/abs/arXiv:1102.0529v3}
  {arXiv:1102.0529v3} \BibitemShut {NoStop}%
\bibitem [{\citenamefont {Barack}\ and\ \citenamefont
  {Ori}(2000{\natexlab{a}})}]{Barack:1999wf}%
  \BibitemOpen
  \bibfield  {author} {\bibinfo {author} {\bibfnamefont {L.}~\bibnamefont
  {Barack}}\ and\ \bibinfo {author} {\bibfnamefont {A.}~\bibnamefont {Ori}},\
  }\href {\doibase 10.1103/PhysRevD.61.061502} {\bibfield  {journal} {\bibinfo
  {journal} {Phys.Rev.}\ }\textbf {\bibinfo {volume} {D61}},\ \bibinfo {pages}
  {061502} (\bibinfo {year} {2000}{\natexlab{a}})},\ \Eprint
  {http://arxiv.org/abs/gr-qc/9912010} {arXiv:gr-qc/9912010} \BibitemShut
  {NoStop}%
\bibitem [{\citenamefont {Vega}\ and\ \citenamefont
  {Detweiler}(2008)}]{Vega:Detweiler:2008}%
  \BibitemOpen
  \bibfield  {author} {\bibinfo {author} {\bibfnamefont {I.}~\bibnamefont
  {Vega}}\ and\ \bibinfo {author} {\bibfnamefont {S.}~\bibnamefont
  {Detweiler}},\ }\href@noop {} {\bibfield  {journal} {\bibinfo  {journal}
  {Phys. Rev. D}\ }\textbf {\bibinfo {volume} {77}},\ \bibinfo {pages} {084008}
  (\bibinfo {year} {2008})}\BibitemShut {NoStop}%
\bibitem [{\citenamefont {Barack}\ and\ \citenamefont
  {Golbourn}(2007)}]{Barack:Golbourn:2007}%
  \BibitemOpen
  \bibfield  {author} {\bibinfo {author} {\bibfnamefont {L.}~\bibnamefont
  {Barack}}\ and\ \bibinfo {author} {\bibfnamefont {D.~A.}\ \bibnamefont
  {Golbourn}},\ }\href {\doibase 10.1103/PhysRevD.76.044020} {\bibfield
  {journal} {\bibinfo  {journal} {Phys. Rev.}\ }\textbf {\bibinfo {volume}
  {D76}},\ \bibinfo {pages} {044020} (\bibinfo {year} {2007})},\ \Eprint
  {http://arxiv.org/abs/arXiv:0705.3620} {arXiv:arXiv:0705.3620} \BibitemShut
  {NoStop}%
\bibitem [{\citenamefont {Barack}\ and\ \citenamefont
  {Sago}(2007)}]{Barack:Sago:2007}%
  \BibitemOpen
  \bibfield  {author} {\bibinfo {author} {\bibfnamefont {L.}~\bibnamefont
  {Barack}}\ and\ \bibinfo {author} {\bibfnamefont {N.}~\bibnamefont {Sago}},\
  }\href {\doibase 10.1103/PhysRevD.75.064021} {\bibfield  {journal} {\bibinfo
  {journal} {Phys. Rev.}\ }\textbf {\bibinfo {volume} {D75}},\ \bibinfo {pages}
  {064021} (\bibinfo {year} {2007})},\ \Eprint
  {http://arxiv.org/abs/gr-qc/0701069} {arXiv:gr-qc/0701069} \BibitemShut
  {NoStop}%
\bibitem [{\citenamefont {Barack}\ and\ \citenamefont
  {Sago}(2010)}]{Barack:Sago:2010}%
  \BibitemOpen
  \bibfield  {author} {\bibinfo {author} {\bibfnamefont {L.}~\bibnamefont
  {Barack}}\ and\ \bibinfo {author} {\bibfnamefont {N.}~\bibnamefont {Sago}},\
  }\href {\doibase 10.1103/PhysRevD.81.084021} {\bibfield  {journal} {\bibinfo
  {journal} {Phys. Rev.}\ }\textbf {\bibinfo {volume} {D81}},\ \bibinfo {pages}
  {084021} (\bibinfo {year} {2010})},\ \Eprint
  {http://arxiv.org/abs/arXiv:1002.2386} {arXiv:arXiv:1002.2386} \BibitemShut
  {NoStop}%
\bibitem [{\citenamefont {Akcay}\ \emph {et~al.}(2013)\citenamefont {Akcay},
  \citenamefont {Warburton},\ and\ \citenamefont {Barack}}]{Akcay:2013wfa}%
  \BibitemOpen
  \bibfield  {author} {\bibinfo {author} {\bibfnamefont {S.}~\bibnamefont
  {Akcay}}, \bibinfo {author} {\bibfnamefont {N.}~\bibnamefont {Warburton}}, \
  and\ \bibinfo {author} {\bibfnamefont {L.}~\bibnamefont {Barack}},\ }\href
  {\doibase 10.1103/PhysRevD.88.104009} {\bibfield  {journal} {\bibinfo
  {journal} {Phys. Rev.}\ }\textbf {\bibinfo {volume} {D88}},\ \bibinfo {pages}
  {104009} (\bibinfo {year} {2013})},\ \Eprint {http://arxiv.org/abs/1308.5223}
  {arXiv:1308.5223} \BibitemShut {NoStop}%
\bibitem [{\citenamefont {Osburn}\ \emph {et~al.}(2014)\citenamefont {Osburn},
  \citenamefont {Forseth}, \citenamefont {Evans},\ and\ \citenamefont
  {Hopper}}]{Osburn:2014hoa}%
  \BibitemOpen
  \bibfield  {author} {\bibinfo {author} {\bibfnamefont {T.}~\bibnamefont
  {Osburn}}, \bibinfo {author} {\bibfnamefont {E.}~\bibnamefont {Forseth}},
  \bibinfo {author} {\bibfnamefont {C.~R.}\ \bibnamefont {Evans}}, \ and\
  \bibinfo {author} {\bibfnamefont {S.}~\bibnamefont {Hopper}},\ }\href
  {\doibase 10.1103/PhysRevD.90.104031} {\bibfield  {journal} {\bibinfo
  {journal} {Phys. Rev.}\ }\textbf {\bibinfo {volume} {D90}},\ \bibinfo {pages}
  {104031} (\bibinfo {year} {2014})},\ \Eprint {http://arxiv.org/abs/1409.4419}
  {arXiv:1409.4419} \BibitemShut {NoStop}%
\bibitem [{\citenamefont {Merlin}\ and\ \citenamefont
  {Shah}(2015)}]{Merlin:2014qda}%
  \BibitemOpen
  \bibfield  {author} {\bibinfo {author} {\bibfnamefont {C.}~\bibnamefont
  {Merlin}}\ and\ \bibinfo {author} {\bibfnamefont {A.~G.}\ \bibnamefont
  {Shah}},\ }\href {\doibase 10.1103/PhysRevD.91.024005} {\bibfield  {journal}
  {\bibinfo  {journal} {Phys. Rev.}\ }\textbf {\bibinfo {volume} {D91}},\
  \bibinfo {pages} {024005} (\bibinfo {year} {2015})},\ \Eprint
  {http://arxiv.org/abs/1410.2998} {arXiv:1410.2998} \BibitemShut {NoStop}%
\bibitem [{\citenamefont {van~de Meent}(2016)}]{vandeMeent:2016pee}%
  \BibitemOpen
  \bibfield  {author} {\bibinfo {author} {\bibfnamefont {M.}~\bibnamefont
  {van~de Meent}},\ }\href {\doibase 10.1103/PhysRevD.94.044034} {\bibfield
  {journal} {\bibinfo  {journal} {Phys. Rev.}\ }\textbf {\bibinfo {volume}
  {D94}},\ \bibinfo {pages} {044034} (\bibinfo {year} {2016})},\ \Eprint
  {http://arxiv.org/abs/1606.06297} {arXiv:1606.06297} \BibitemShut {NoStop}%
\bibitem [{\citenamefont {van~de Meent}(2018)}]{vandeMeent:2017bcc}%
  \BibitemOpen
  \bibfield  {author} {\bibinfo {author} {\bibfnamefont {M.}~\bibnamefont
  {van~de Meent}},\ }\href {\doibase 10.1103/PhysRevD.97.104033} {\bibfield
  {journal} {\bibinfo  {journal} {Phys. Rev.}\ }\textbf {\bibinfo {volume}
  {D97}},\ \bibinfo {pages} {104033} (\bibinfo {year} {2018})},\ \Eprint
  {http://arxiv.org/abs/1711.09607} {arXiv:1711.09607 [gr-qc]} \BibitemShut
  {NoStop}%
\bibitem [{\citenamefont {Warburton}\ \emph {et~al.}(2012)\citenamefont
  {Warburton}, \citenamefont {Akcay}, \citenamefont {Barack}, \citenamefont
  {Gair},\ and\ \citenamefont {Sago}}]{Warburton:2011fk}%
  \BibitemOpen
  \bibfield  {author} {\bibinfo {author} {\bibfnamefont {N.}~\bibnamefont
  {Warburton}}, \bibinfo {author} {\bibfnamefont {S.}~\bibnamefont {Akcay}},
  \bibinfo {author} {\bibfnamefont {L.}~\bibnamefont {Barack}}, \bibinfo
  {author} {\bibfnamefont {J.~R.}\ \bibnamefont {Gair}}, \ and\ \bibinfo
  {author} {\bibfnamefont {N.}~\bibnamefont {Sago}},\ }\href {\doibase
  10.1103/PhysRevD.85.061501} {\bibfield  {journal} {\bibinfo  {journal}
  {Phys.Rev.}\ }\textbf {\bibinfo {volume} {D85}},\ \bibinfo {pages} {061501}
  (\bibinfo {year} {2012})},\ \Eprint {http://arxiv.org/abs/arXiv:1111.6908}
  {arXiv:arXiv:1111.6908} \BibitemShut {NoStop}%
\bibitem [{\citenamefont {Osburn}\ \emph {et~al.}(2016)\citenamefont {Osburn},
  \citenamefont {Warburton},\ and\ \citenamefont {Evans}}]{Osburn:2015duj}%
  \BibitemOpen
  \bibfield  {author} {\bibinfo {author} {\bibfnamefont {T.}~\bibnamefont
  {Osburn}}, \bibinfo {author} {\bibfnamefont {N.}~\bibnamefont {Warburton}}, \
  and\ \bibinfo {author} {\bibfnamefont {C.~R.}\ \bibnamefont {Evans}},\ }\href
  {\doibase 10.1103/PhysRevD.93.064024} {\bibfield  {journal} {\bibinfo
  {journal} {Phys. Rev.}\ }\textbf {\bibinfo {volume} {D93}},\ \bibinfo {pages}
  {064024} (\bibinfo {year} {2016})},\ \Eprint
  {http://arxiv.org/abs/1511.01498} {arXiv:1511.01498} \BibitemShut {NoStop}%
\bibitem [{\citenamefont {Warburton}\ \emph {et~al.}(2017)\citenamefont
  {Warburton}, \citenamefont {Osburn},\ and\ \citenamefont
  {Evans}}]{Warburton:2017sxk}%
  \BibitemOpen
  \bibfield  {author} {\bibinfo {author} {\bibfnamefont {N.}~\bibnamefont
  {Warburton}}, \bibinfo {author} {\bibfnamefont {T.}~\bibnamefont {Osburn}}, \
  and\ \bibinfo {author} {\bibfnamefont {C.~R.}\ \bibnamefont {Evans}},\ }\href
  {\doibase 10.1103/PhysRevD.96.084057} {\bibfield  {journal} {\bibinfo
  {journal} {Phys. Rev.}\ }\textbf {\bibinfo {volume} {D96}},\ \bibinfo {pages}
  {084057} (\bibinfo {year} {2017})},\ \Eprint
  {http://arxiv.org/abs/1708.03720} {arXiv:1708.03720 [gr-qc]} \BibitemShut
  {NoStop}%
\bibitem [{\citenamefont {Diener}\ \emph {et~al.}(2012)\citenamefont {Diener},
  \citenamefont {Vega}, \citenamefont {Wardell},\ and\ \citenamefont
  {Detweiler}}]{Diener:2011cc}%
  \BibitemOpen
  \bibfield  {author} {\bibinfo {author} {\bibfnamefont {P.}~\bibnamefont
  {Diener}}, \bibinfo {author} {\bibfnamefont {I.}~\bibnamefont {Vega}},
  \bibinfo {author} {\bibfnamefont {B.}~\bibnamefont {Wardell}}, \ and\
  \bibinfo {author} {\bibfnamefont {S.}~\bibnamefont {Detweiler}},\ }\href
  {\doibase 10.1103/PhysRevLett.108.191102} {\bibfield  {journal} {\bibinfo
  {journal} {Phys. Rev. Lett.}\ }\textbf {\bibinfo {volume} {108}},\ \bibinfo
  {pages} {191102} (\bibinfo {year} {2012})},\ \Eprint
  {http://arxiv.org/abs/1112.4821} {arXiv:1112.4821} \BibitemShut {NoStop}%
\bibitem [{\citenamefont {Mathisson}(2010)}]{Mathisson2010}%
  \BibitemOpen
  \bibfield  {author} {\bibinfo {author} {\bibfnamefont {M.}~\bibnamefont
  {Mathisson}},\ }\href {\doibase 10.1007/s10714-010-0939-y} {\bibfield
  {journal} {\bibinfo  {journal} {General Relativity and Gravitation}\ }\textbf
  {\bibinfo {volume} {42}},\ \bibinfo {pages} {1011} (\bibinfo {year}
  {2010})}\BibitemShut {NoStop}%
\bibitem [{\citenamefont {Papapetrou}(1951)}]{Papa51}%
  \BibitemOpen
  \bibfield  {author} {\bibinfo {author} {\bibfnamefont {A.}~\bibnamefont
  {Papapetrou}},\ }\href@noop {} {\bibfield  {journal} {\bibinfo  {journal}
  {Proc. R. Soc. Lond.}\ }\textbf {\bibinfo {volume} {A209}},\ \bibinfo {pages}
  {248} (\bibinfo {year} {1951})}\BibitemShut {NoStop}%
\bibitem [{\citenamefont {Dixon}(1970)}]{Dixo70}%
  \BibitemOpen
  \bibfield  {author} {\bibinfo {author} {\bibfnamefont {W.~G.}\ \bibnamefont
  {Dixon}},\ }\href {\doibase 10.1098/rspa.1970.0020} {\bibfield  {journal}
  {\bibinfo  {journal} {Royal Society of London Proceedings Series A}\ }\textbf
  {\bibinfo {volume} {314}},\ \bibinfo {pages} {499} (\bibinfo {year}
  {1970})}\BibitemShut {NoStop}%
\bibitem [{\citenamefont {Kyrian}\ and\ \citenamefont
  {Semerak}(2007)}]{Kyrian:2007}%
  \BibitemOpen
  \bibfield  {author} {\bibinfo {author} {\bibfnamefont {K.}~\bibnamefont
  {Kyrian}}\ and\ \bibinfo {author} {\bibfnamefont {O.}~\bibnamefont
  {Semerak}},\ }\href {\doibase 10.1111/j.1365-2966.2007.12502.x} {\bibfield
  {journal} {\bibinfo  {journal} {Mon. Not. Roy. Astron. Soc.}\ }\textbf
  {\bibinfo {volume} {382}},\ \bibinfo {pages} {1922} (\bibinfo {year}
  {2007})}\BibitemShut {NoStop}%
\bibitem [{\citenamefont {Harms}\ \emph
  {et~al.}(2016{\natexlab{a}})\citenamefont {Harms}, \citenamefont
  {Lukes-Gerakopoulos}, \citenamefont {Bernuzzi},\ and\ \citenamefont
  {Nagar}}]{Harms:2015ixa}%
  \BibitemOpen
  \bibfield  {author} {\bibinfo {author} {\bibfnamefont {E.}~\bibnamefont
  {Harms}}, \bibinfo {author} {\bibfnamefont {G.}~\bibnamefont
  {Lukes-Gerakopoulos}}, \bibinfo {author} {\bibfnamefont {S.}~\bibnamefont
  {Bernuzzi}}, \ and\ \bibinfo {author} {\bibfnamefont {A.}~\bibnamefont
  {Nagar}},\ }\href {\doibase 10.1103/PhysRevD.93.044015} {\bibfield  {journal}
  {\bibinfo  {journal} {Phys. Rev.}\ }\textbf {\bibinfo {volume} {D93}},\
  \bibinfo {pages} {044015} (\bibinfo {year} {2016}{\natexlab{a}})},\ \Eprint
  {http://arxiv.org/abs/1510.05548} {arXiv:1510.05548} \BibitemShut {NoStop}%
\bibitem [{\citenamefont {Harms}\ \emph
  {et~al.}(2016{\natexlab{b}})\citenamefont {Harms}, \citenamefont
  {Lukes-Gerakopoulos}, \citenamefont {Bernuzzi},\ and\ \citenamefont
  {Nagar}}]{Harms:2016ctx}%
  \BibitemOpen
  \bibfield  {author} {\bibinfo {author} {\bibfnamefont {E.}~\bibnamefont
  {Harms}}, \bibinfo {author} {\bibfnamefont {G.}~\bibnamefont
  {Lukes-Gerakopoulos}}, \bibinfo {author} {\bibfnamefont {S.}~\bibnamefont
  {Bernuzzi}}, \ and\ \bibinfo {author} {\bibfnamefont {A.}~\bibnamefont
  {Nagar}},\ }\href {\doibase 10.1103/PhysRevD.94.104010} {\bibfield  {journal}
  {\bibinfo  {journal} {Phys. Rev.}\ }\textbf {\bibinfo {volume} {D94}},\
  \bibinfo {pages} {104010} (\bibinfo {year} {2016}{\natexlab{b}})},\ \Eprint
  {http://arxiv.org/abs/1609.00356} {arXiv:1609.00356} \BibitemShut {NoStop}%
\bibitem [{\citenamefont {Han}(2010)}]{Han:2010tp}%
  \BibitemOpen
  \bibfield  {author} {\bibinfo {author} {\bibfnamefont {W.-B.}\ \bibnamefont
  {Han}},\ }\href {\doibase 10.1103/PhysRevD.82.084013} {\bibfield  {journal}
  {\bibinfo  {journal} {Phys. Rev.}\ }\textbf {\bibinfo {volume} {D82}},\
  \bibinfo {pages} {084013} (\bibinfo {year} {2010})},\ \Eprint
  {http://arxiv.org/abs/1008.3324} {arXiv:1008.3324} \BibitemShut {NoStop}%
\bibitem [{\citenamefont {Harte}(2012)}]{Harte:2011ku}%
  \BibitemOpen
  \bibfield  {author} {\bibinfo {author} {\bibfnamefont {A.~I.}\ \bibnamefont
  {Harte}},\ }\href {\doibase 10.1088/0264-9381/29/5/055012} {\bibfield
  {journal} {\bibinfo  {journal} {Class.Quant.Grav.}\ }\textbf {\bibinfo
  {volume} {29}},\ \bibinfo {pages} {055012} (\bibinfo {year} {2012})},\
  \Eprint {http://arxiv.org/abs/arXiv:1103.0543} {arXiv:arXiv:1103.0543}
  \BibitemShut {NoStop}%
\bibitem [{\citenamefont {Burko}(2004)}]{Burko:2004}%
  \BibitemOpen
  \bibfield  {author} {\bibinfo {author} {\bibfnamefont {L.~M.}\ \bibnamefont
  {Burko}},\ }\href {\doibase 10.1103/PhysRevD.69.044011} {\bibfield  {journal}
  {\bibinfo  {journal} {Phys. Rev. D}\ }\textbf {\bibinfo {volume} {69}},\
  \bibinfo {pages} {044011} (\bibinfo {year} {2004})}\BibitemShut {NoStop}%
\bibitem [{\citenamefont {Huerta}\ and\ \citenamefont
  {Gair}(2011)}]{Huerta:2011kt}%
  \BibitemOpen
  \bibfield  {author} {\bibinfo {author} {\bibfnamefont {E.~A.}\ \bibnamefont
  {Huerta}}\ and\ \bibinfo {author} {\bibfnamefont {J.~R.}\ \bibnamefont
  {Gair}},\ }\href {\doibase 10.1103/PhysRevD.84.064023} {\bibfield  {journal}
  {\bibinfo  {journal} {Phys. Rev.}\ }\textbf {\bibinfo {volume} {D84}},\
  \bibinfo {pages} {064023} (\bibinfo {year} {2011})},\ \Eprint
  {http://arxiv.org/abs/1105.3567} {arXiv:1105.3567} \BibitemShut {NoStop}%
\bibitem [{\citenamefont {Steinhoff}\ and\ \citenamefont
  {Puetzfeld}(2012)}]{Steinhoff:2012}%
  \BibitemOpen
  \bibfield  {author} {\bibinfo {author} {\bibfnamefont {J.}~\bibnamefont
  {Steinhoff}}\ and\ \bibinfo {author} {\bibfnamefont {D.}~\bibnamefont
  {Puetzfeld}},\ }\href {\doibase 10.1103/PhysRevD.86.044033} {\bibfield
  {journal} {\bibinfo  {journal} {Phys. Rev.}\ }\textbf {\bibinfo {volume}
  {D86}},\ \bibinfo {pages} {044033} (\bibinfo {year} {2012})},\ \Eprint
  {http://arxiv.org/abs/1205.3926} {arXiv:1205.3926} \BibitemShut {NoStop}%
\bibitem [{\citenamefont {Burko}\ and\ \citenamefont
  {Khanna}(2015)}]{Burko:2015}%
  \BibitemOpen
  \bibfield  {author} {\bibinfo {author} {\bibfnamefont {L.~M.}\ \bibnamefont
  {Burko}}\ and\ \bibinfo {author} {\bibfnamefont {G.}~\bibnamefont {Khanna}},\
  }\href {\doibase 10.1103/PhysRevD.91.104017} {\bibfield  {journal} {\bibinfo
  {journal} {Phys. Rev.}\ }\textbf {\bibinfo {volume} {D91}},\ \bibinfo {pages}
  {104017} (\bibinfo {year} {2015})},\ \Eprint
  {http://arxiv.org/abs/1503.05097} {arXiv:1503.05097} \BibitemShut {NoStop}%
\bibitem [{\citenamefont {Penrose}(1969)}]{Penrose:1969pc}%
  \BibitemOpen
  \bibfield  {author} {\bibinfo {author} {\bibfnamefont {R.}~\bibnamefont
  {Penrose}},\ }\href@noop {} {\bibfield  {journal} {\bibinfo  {journal} {Riv.
  Nuovo Cim.}\ }\textbf {\bibinfo {volume} {1}},\ \bibinfo {pages} {252}
  (\bibinfo {year} {1969})},\ \bibinfo {note} {[Gen. Rel.
  Grav.34,1141(2002)]}\BibitemShut {NoStop}%
\bibitem [{\citenamefont {{Wald}}(1974)}]{Wald:1974}%
  \BibitemOpen
  \bibfield  {author} {\bibinfo {author} {\bibfnamefont {R.}~\bibnamefont
  {{Wald}}},\ }\href {\doibase 10.1016/0003-4916(74)90125-0} {\bibfield
  {journal} {\bibinfo  {journal} {Annals of Physics}\ }\textbf {\bibinfo
  {volume} {82}},\ \bibinfo {pages} {548} (\bibinfo {year} {1974})}\BibitemShut
  {NoStop}%
\bibitem [{\citenamefont {Hubeny}(1999)}]{Hubeny:1998}%
  \BibitemOpen
  \bibfield  {author} {\bibinfo {author} {\bibfnamefont {V.~E.}\ \bibnamefont
  {Hubeny}},\ }\href {\doibase 10.1103/PhysRevD.59.064013} {\bibfield
  {journal} {\bibinfo  {journal} {Phys. Rev.}\ }\textbf {\bibinfo {volume}
  {D59}},\ \bibinfo {pages} {064013} (\bibinfo {year} {1999})},\ \Eprint
  {http://arxiv.org/abs/gr-qc/9808043} {arXiv:gr-qc/9808043} \BibitemShut
  {NoStop}%
\bibitem [{\citenamefont {Jacobson}\ and\ \citenamefont
  {Sotiriou}(2009)}]{Jacobson:2009}%
  \BibitemOpen
  \bibfield  {author} {\bibinfo {author} {\bibfnamefont {T.}~\bibnamefont
  {Jacobson}}\ and\ \bibinfo {author} {\bibfnamefont {T.~P.}\ \bibnamefont
  {Sotiriou}},\ }\href {\doibase 10.1103/PhysRevLett.103.141101} {\bibfield
  {journal} {\bibinfo  {journal} {Phys. Rev. Lett.}\ }\textbf {\bibinfo
  {volume} {103}},\ \bibinfo {pages} {141101} (\bibinfo {year} {2009})},\
  \bibinfo {note} {[Erratum: Phys. Rev. Lett.103,209903(2009)]},\ \Eprint
  {http://arxiv.org/abs/0907.4146} {arXiv:0907.4146} \BibitemShut {NoStop}%
\bibitem [{\citenamefont {Barausse}\ \emph {et~al.}(2010)\citenamefont
  {Barausse}, \citenamefont {Cardoso},\ and\ \citenamefont
  {Khanna}}]{Barausse:2010}%
  \BibitemOpen
  \bibfield  {author} {\bibinfo {author} {\bibfnamefont {E.}~\bibnamefont
  {Barausse}}, \bibinfo {author} {\bibfnamefont {V.}~\bibnamefont {Cardoso}}, \
  and\ \bibinfo {author} {\bibfnamefont {G.}~\bibnamefont {Khanna}},\ }\href
  {\doibase 10.1103/PhysRevLett.105.261102} {\bibfield  {journal} {\bibinfo
  {journal} {Phys. Rev. Lett.}\ }\textbf {\bibinfo {volume} {105}},\ \bibinfo
  {pages} {261102} (\bibinfo {year} {2010})},\ \Eprint
  {http://arxiv.org/abs/1008.5159} {arXiv:1008.5159} \BibitemShut {NoStop}%
\bibitem [{\citenamefont {Barausse}\ \emph {et~al.}(2011)\citenamefont
  {Barausse}, \citenamefont {Cardoso},\ and\ \citenamefont
  {Khanna}}]{Barausse:2011}%
  \BibitemOpen
  \bibfield  {author} {\bibinfo {author} {\bibfnamefont {E.}~\bibnamefont
  {Barausse}}, \bibinfo {author} {\bibfnamefont {V.}~\bibnamefont {Cardoso}}, \
  and\ \bibinfo {author} {\bibfnamefont {G.}~\bibnamefont {Khanna}},\ }\href
  {\doibase 10.1103/PhysRevD.84.104006} {\bibfield  {journal} {\bibinfo
  {journal} {Phys. Rev.}\ }\textbf {\bibinfo {volume} {D84}},\ \bibinfo {pages}
  {104006} (\bibinfo {year} {2011})},\ \Eprint {http://arxiv.org/abs/1106.1692}
  {arXiv:1106.1692} \BibitemShut {NoStop}%
\bibitem [{\citenamefont {Colleoni}\ and\ \citenamefont
  {Barack}(2015)}]{Colleoni:2015}%
  \BibitemOpen
  \bibfield  {author} {\bibinfo {author} {\bibfnamefont {M.}~\bibnamefont
  {Colleoni}}\ and\ \bibinfo {author} {\bibfnamefont {L.}~\bibnamefont
  {Barack}},\ }\href {\doibase 10.1103/PhysRevD.91.104024} {\bibfield
  {journal} {\bibinfo  {journal} {Phys. Rev.}\ }\textbf {\bibinfo {volume}
  {D91}},\ \bibinfo {pages} {104024} (\bibinfo {year} {2015})},\ \Eprint
  {http://arxiv.org/abs/1501.07330} {arXiv:1501.07330} \BibitemShut {NoStop}%
\bibitem [{\citenamefont {Colleoni}\ \emph {et~al.}(2015)\citenamefont
  {Colleoni}, \citenamefont {Barack}, \citenamefont {Shah},\ and\ \citenamefont
  {van~de Meent}}]{Colleoni:2015b}%
  \BibitemOpen
  \bibfield  {author} {\bibinfo {author} {\bibfnamefont {M.}~\bibnamefont
  {Colleoni}}, \bibinfo {author} {\bibfnamefont {L.}~\bibnamefont {Barack}},
  \bibinfo {author} {\bibfnamefont {A.~G.}\ \bibnamefont {Shah}}, \ and\
  \bibinfo {author} {\bibfnamefont {M.}~\bibnamefont {van~de Meent}},\ }\href
  {\doibase 10.1103/PhysRevD.92.084044} {\bibfield  {journal} {\bibinfo
  {journal} {Phys. Rev.}\ }\textbf {\bibinfo {volume} {D92}},\ \bibinfo {pages}
  {084044} (\bibinfo {year} {2015})},\ \Eprint
  {http://arxiv.org/abs/1508.04031} {arXiv:1508.04031} \BibitemShut {NoStop}%
\bibitem [{\citenamefont {Isoyama}\ \emph {et~al.}(2011)\citenamefont
  {Isoyama}, \citenamefont {Sago},\ and\ \citenamefont
  {Tanaka}}]{Isoyama:2011}%
  \BibitemOpen
  \bibfield  {author} {\bibinfo {author} {\bibfnamefont {S.}~\bibnamefont
  {Isoyama}}, \bibinfo {author} {\bibfnamefont {N.}~\bibnamefont {Sago}}, \
  and\ \bibinfo {author} {\bibfnamefont {T.}~\bibnamefont {Tanaka}},\ }\href
  {\doibase 10.1103/PhysRevD.84.124024} {\bibfield  {journal} {\bibinfo
  {journal} {Phys. Rev.}\ }\textbf {\bibinfo {volume} {D84}},\ \bibinfo {pages}
  {124024} (\bibinfo {year} {2011})},\ \Eprint {http://arxiv.org/abs/1108.6207}
  {arXiv:1108.6207} \BibitemShut {NoStop}%
\bibitem [{\citenamefont {Zimmerman}\ \emph {et~al.}(2013)\citenamefont
  {Zimmerman}, \citenamefont {Vega}, \citenamefont {Poisson},\ and\
  \citenamefont {Haas}}]{Zimmerman:2012}%
  \BibitemOpen
  \bibfield  {author} {\bibinfo {author} {\bibfnamefont {P.}~\bibnamefont
  {Zimmerman}}, \bibinfo {author} {\bibfnamefont {I.}~\bibnamefont {Vega}},
  \bibinfo {author} {\bibfnamefont {E.}~\bibnamefont {Poisson}}, \ and\
  \bibinfo {author} {\bibfnamefont {R.}~\bibnamefont {Haas}},\ }\href {\doibase
  10.1103/PhysRevD.87.041501} {\bibfield  {journal} {\bibinfo  {journal} {Phys.
  Rev.}\ }\textbf {\bibinfo {volume} {D87}},\ \bibinfo {pages} {041501}
  (\bibinfo {year} {2013})},\ \Eprint {http://arxiv.org/abs/1211.3889}
  {arXiv:1211.3889} \BibitemShut {NoStop}%
\bibitem [{\citenamefont {Sorce}\ and\ \citenamefont
  {Wald}(2017)}]{Sorce:2017dst}%
  \BibitemOpen
  \bibfield  {author} {\bibinfo {author} {\bibfnamefont {J.}~\bibnamefont
  {Sorce}}\ and\ \bibinfo {author} {\bibfnamefont {R.~M.}\ \bibnamefont
  {Wald}},\ }\href {\doibase 10.1103/PhysRevD.96.104014} {\bibfield  {journal}
  {\bibinfo  {journal} {Phys. Rev.}\ }\textbf {\bibinfo {volume} {D96}},\
  \bibinfo {pages} {104014} (\bibinfo {year} {2017})},\ \Eprint
  {http://arxiv.org/abs/1707.05862} {arXiv:1707.05862 [gr-qc]} \BibitemShut
  {NoStop}%
\bibitem [{\citenamefont {Heffernan}\ \emph {et~al.}(2012)\citenamefont
  {Heffernan}, \citenamefont {Ottewill},\ and\ \citenamefont
  {Wardell}}]{HOW:2012}%
  \BibitemOpen
  \bibfield  {author} {\bibinfo {author} {\bibfnamefont {A.}~\bibnamefont
  {Heffernan}}, \bibinfo {author} {\bibfnamefont {A.}~\bibnamefont {Ottewill}},
  \ and\ \bibinfo {author} {\bibfnamefont {B.}~\bibnamefont {Wardell}},\ }\href
  {\doibase 10.1103/PhysRevD.86.104023} {\bibfield  {journal} {\bibinfo
  {journal} {Phys.Rev.}\ }\textbf {\bibinfo {volume} {D86}},\ \bibinfo {pages}
  {104023} (\bibinfo {year} {2012})},\ \Eprint {http://arxiv.org/abs/1204.0794}
  {arXiv:1204.0794} \BibitemShut {NoStop}%
\bibitem [{\citenamefont {Heffernan}\ \emph {et~al.}(2014)\citenamefont
  {Heffernan}, \citenamefont {Ottewill},\ and\ \citenamefont
  {Wardell}}]{HOW:2014}%
  \BibitemOpen
  \bibfield  {author} {\bibinfo {author} {\bibfnamefont {A.}~\bibnamefont
  {Heffernan}}, \bibinfo {author} {\bibfnamefont {A.}~\bibnamefont {Ottewill}},
  \ and\ \bibinfo {author} {\bibfnamefont {B.}~\bibnamefont {Wardell}},\ }\href
  {\doibase 10.1103/PhysRevD.89.024030} {\bibfield  {journal} {\bibinfo
  {journal} {Phys.Rev.}\ }\textbf {\bibinfo {volume} {D89}},\ \bibinfo {pages}
  {024030} (\bibinfo {year} {2014})},\ \Eprint {http://arxiv.org/abs/1211.6446}
  {arXiv:1211.6446} \BibitemShut {NoStop}%
\bibitem [{\citenamefont {Heffernan}(2012)}]{Heffernan:2014}%
  \BibitemOpen
  \bibfield  {author} {\bibinfo {author} {\bibfnamefont {A.}~\bibnamefont
  {Heffernan}},\ }\emph {\bibinfo {title} {{The Self-Force Problem: Local
  Behaviour of the Detweiler-Whiting Singular Field}}},\ \href
  {https://inspirehep.net/record/1287046/files/arXiv:1403.6177.pdf} {Ph.D.
  thesis},\ \bibinfo  {school} {University Coll., Dublin} (\bibinfo {year}
  {2012}),\ \Eprint {http://arxiv.org/abs/1403.6177} {arXiv:1403.6177}
  \BibitemShut {NoStop}%
\bibitem [{\citenamefont {Barack}\ and\ \citenamefont
  {Ori}(2000{\natexlab{b}})}]{Barack:Ori:2000}%
  \BibitemOpen
  \bibfield  {author} {\bibinfo {author} {\bibfnamefont {L.}~\bibnamefont
  {Barack}}\ and\ \bibinfo {author} {\bibfnamefont {A.}~\bibnamefont {Ori}},\
  }\href@noop {} {\bibfield  {journal} {\bibinfo  {journal} {Phys. Rev. D}\
  }\textbf {\bibinfo {volume} {61}},\ \bibinfo {pages} {061502} (\bibinfo
  {year} {2000}{\natexlab{b}})}\BibitemShut {NoStop}%
\bibitem [{\citenamefont {Barack}(2001)}]{Barack:2001}%
  \BibitemOpen
  \bibfield  {author} {\bibinfo {author} {\bibfnamefont {L.}~\bibnamefont
  {Barack}},\ }\href {\doibase 10.1103/PhysRevD.64.084021} {\bibfield
  {journal} {\bibinfo  {journal} {Phys.Rev.}\ }\textbf {\bibinfo {volume}
  {D64}},\ \bibinfo {pages} {084021} (\bibinfo {year} {2001})},\ \Eprint
  {http://arxiv.org/abs/gr-qc/0105040} {arXiv:gr-qc/0105040} \BibitemShut
  {NoStop}%
\bibitem [{\citenamefont {Barack}\ and\ \citenamefont
  {Ori}(2002)}]{Barack:2002mha}%
  \BibitemOpen
  \bibfield  {author} {\bibinfo {author} {\bibfnamefont {L.}~\bibnamefont
  {Barack}}\ and\ \bibinfo {author} {\bibfnamefont {A.}~\bibnamefont {Ori}},\
  }\href {\doibase 10.1103/PhysRevD.66.084022} {\bibfield  {journal} {\bibinfo
  {journal} {Phys.Rev.}\ }\textbf {\bibinfo {volume} {D66}},\ \bibinfo {pages}
  {084022} (\bibinfo {year} {2002})},\ \Eprint
  {http://arxiv.org/abs/gr-qc/0204093} {arXiv:gr-qc/0204093} \BibitemShut
  {NoStop}%
\bibitem [{\citenamefont {Barack}\ and\ \citenamefont
  {Ori}(2003)}]{Barack:2002bt}%
  \BibitemOpen
  \bibfield  {author} {\bibinfo {author} {\bibfnamefont {L.}~\bibnamefont
  {Barack}}\ and\ \bibinfo {author} {\bibfnamefont {A.}~\bibnamefont {Ori}},\
  }\href {\doibase 10.1103/PhysRevD.67.024029} {\bibfield  {journal} {\bibinfo
  {journal} {Phys.Rev.}\ }\textbf {\bibinfo {volume} {D67}},\ \bibinfo {pages}
  {024029} (\bibinfo {year} {2003})},\ \Eprint
  {http://arxiv.org/abs/gr-qc/0209072} {arXiv:gr-qc/0209072} \BibitemShut
  {NoStop}%
\bibitem [{\citenamefont {Detweiler}\ \emph {et~al.}(2003)\citenamefont
  {Detweiler}, \citenamefont {Messaritaki},\ and\ \citenamefont
  {Whiting}}]{Detweiler:2002gi}%
  \BibitemOpen
  \bibfield  {author} {\bibinfo {author} {\bibfnamefont {S.}~\bibnamefont
  {Detweiler}}, \bibinfo {author} {\bibfnamefont {E.}~\bibnamefont
  {Messaritaki}}, \ and\ \bibinfo {author} {\bibfnamefont {B.~F.}\ \bibnamefont
  {Whiting}},\ }\href {\doibase 10.1103/PhysRevD.67.104016} {\bibfield
  {journal} {\bibinfo  {journal} {Phys. Rev.}\ }\textbf {\bibinfo {volume}
  {D67}},\ \bibinfo {pages} {104016} (\bibinfo {year} {2003})},\ \Eprint
  {http://arxiv.org/abs/gr-qc/0205079} {arXiv:gr-qc/0205079} \BibitemShut
  {NoStop}%
\bibitem [{\citenamefont {Haas}\ and\ \citenamefont
  {Poisson}(2006)}]{Haas:2006ne}%
  \BibitemOpen
  \bibfield  {author} {\bibinfo {author} {\bibfnamefont {R.}~\bibnamefont
  {Haas}}\ and\ \bibinfo {author} {\bibfnamefont {E.}~\bibnamefont {Poisson}},\
  }\href {\doibase 10.1103/PhysRevD.74.044009} {\bibfield  {journal} {\bibinfo
  {journal} {Phys. Rev.}\ }\textbf {\bibinfo {volume} {D74}},\ \bibinfo {pages}
  {044009} (\bibinfo {year} {2006})},\ \Eprint
  {http://arxiv.org/abs/gr-qc/0605077} {arXiv:gr-qc/0605077} \BibitemShut
  {NoStop}%
\bibitem [{\citenamefont {Casals}\ \emph {et~al.}(2012)\citenamefont {Casals},
  \citenamefont {Poisson},\ and\ \citenamefont {Vega}}]{Casals:2012qq}%
  \BibitemOpen
  \bibfield  {author} {\bibinfo {author} {\bibfnamefont {M.}~\bibnamefont
  {Casals}}, \bibinfo {author} {\bibfnamefont {E.}~\bibnamefont {Poisson}}, \
  and\ \bibinfo {author} {\bibfnamefont {I.}~\bibnamefont {Vega}},\ }\href
  {\doibase 10.1103/PhysRevD.86.064033} {\bibfield  {journal} {\bibinfo
  {journal} {Phys. Rev.}\ }\textbf {\bibinfo {volume} {D86}},\ \bibinfo {pages}
  {064033} (\bibinfo {year} {2012})},\ \Eprint {http://arxiv.org/abs/1206.3772}
  {arXiv:1206.3772} \BibitemShut {NoStop}%
\bibitem [{\citenamefont {Linz}\ \emph {et~al.}(2014)\citenamefont {Linz},
  \citenamefont {Friedman},\ and\ \citenamefont {Wiseman}}]{Linz:2014}%
  \BibitemOpen
  \bibfield  {author} {\bibinfo {author} {\bibfnamefont {T.~M.}\ \bibnamefont
  {Linz}}, \bibinfo {author} {\bibfnamefont {J.~L.}\ \bibnamefont {Friedman}},
  \ and\ \bibinfo {author} {\bibfnamefont {A.~G.}\ \bibnamefont {Wiseman}},\
  }\href {\doibase 10.1103/PhysRevD.90.024064} {\bibfield  {journal} {\bibinfo
  {journal} {Phys. Rev.}\ }\textbf {\bibinfo {volume} {D90}},\ \bibinfo {pages}
  {024064} (\bibinfo {year} {2014})},\ \Eprint {http://arxiv.org/abs/1404.7039}
  {arXiv:1404.7039} \BibitemShut {NoStop}%
\bibitem [{\citenamefont {Quinn}(2000)}]{Quinn:2000}%
  \BibitemOpen
  \bibfield  {author} {\bibinfo {author} {\bibfnamefont {T.~C.}\ \bibnamefont
  {Quinn}},\ }\href {\doibase 10.1103/PhysRevD.62.064029} {\bibfield  {journal}
  {\bibinfo  {journal} {Phys. Rev.}\ }\textbf {\bibinfo {volume} {D62}},\
  \bibinfo {pages} {064029} (\bibinfo {year} {2000})},\ \Eprint
  {http://arxiv.org/abs/gr-qc/0005030} {arXiv:gr-qc/0005030} \BibitemShut
  {NoStop}%
\bibitem [{\citenamefont {D\'ecanini}\ and\ \citenamefont
  {Folacci}(2006)}]{Decanini:Folacci:2005a}%
  \BibitemOpen
  \bibfield  {author} {\bibinfo {author} {\bibfnamefont {Y.}~\bibnamefont
  {D\'ecanini}}\ and\ \bibinfo {author} {\bibfnamefont {A.}~\bibnamefont
  {Folacci}},\ }\href {\doibase 10.1103/PhysRevD.73.044027} {\bibfield
  {journal} {\bibinfo  {journal} {Phys. Rev.}\ }\textbf {\bibinfo {volume}
  {D73}},\ \bibinfo {pages} {044027} (\bibinfo {year} {2006})},\ \Eprint
  {http://arxiv.org/abs/gr-qc/0511115} {arXiv:gr-qc/0511115} \BibitemShut
  {NoStop}%
\bibitem [{\citenamefont {Mino}\ and\ \citenamefont
  {Sasaki}(2002)}]{Mino:2001mq}%
  \BibitemOpen
  \bibfield  {author} {\bibinfo {author} {\bibfnamefont {N.}~\bibnamefont
  {Mino}}\ and\ \bibinfo {author} {\bibnamefont {Sasaki}},\ }\href {\doibase
  10.1143/PTP.108.1039} {\bibfield  {journal} {\bibinfo  {journal} {Prog.
  Theor. Phys.}\ }\textbf {\bibinfo {volume} {108}},\ \bibinfo {pages} {1039}
  (\bibinfo {year} {2002})},\ \Eprint {http://arxiv.org/abs/gr-qc/0111074}
  {arXiv:gr-qc/0111074} \BibitemShut {NoStop}%
\bibitem [{\citenamefont {Diener}\ \emph {et~al.}({\natexlab{a}})\citenamefont
  {Diener}, \citenamefont {Warburton},\ and\ \citenamefont
  {Wardell}}]{InspiralComparison}%
  \BibitemOpen
  \bibfield  {author} {\bibinfo {author} {\bibfnamefont {P.}~\bibnamefont
  {Diener}}, \bibinfo {author} {\bibfnamefont {N.}~\bibnamefont {Warburton}}, \
  and\ \bibinfo {author} {\bibfnamefont {B.}~\bibnamefont {Wardell}},\
  }\href@noop {} {\enquote {\bibinfo {title} {{Comparison of Self-consistent
  and Geodesic Evolution of a Scalar Charge in a Schwarzschild Space-time}},}\
  } ({\natexlab{a}}),\ \bibinfo {note} {in preparation}\BibitemShut {NoStop}%
\bibitem [{\citenamefont {{Darwin}}(1961)}]{Darwin-1961}%
  \BibitemOpen
  \bibfield  {author} {\bibinfo {author} {\bibfnamefont {C.}~\bibnamefont
  {{Darwin}}},\ }\href {\doibase 10.1098/rspa.1961.0142} {\bibfield  {journal}
  {\bibinfo  {journal} {Royal Society of London Proceedings Series A}\ }\textbf
  {\bibinfo {volume} {263}},\ \bibinfo {pages} {39} (\bibinfo {year}
  {1961})}\BibitemShut {NoStop}%
\bibitem [{\citenamefont {{Cutler}}\ \emph {et~al.}(1994)\citenamefont
  {{Cutler}}, \citenamefont {{Kennefick}},\ and\ \citenamefont
  {{Poisson}}}]{Cutler-Kennefick-Poisson}%
  \BibitemOpen
  \bibfield  {author} {\bibinfo {author} {\bibfnamefont {C.}~\bibnamefont
  {{Cutler}}}, \bibinfo {author} {\bibfnamefont {D.}~\bibnamefont
  {{Kennefick}}}, \ and\ \bibinfo {author} {\bibfnamefont {E.}~\bibnamefont
  {{Poisson}}},\ }\href {\doibase 10.1103/PhysRevD.50.3816} {\bibfield
  {journal} {\bibinfo  {journal} {\prd}\ }\textbf {\bibinfo {volume} {50}},\
  \bibinfo {pages} {3816} (\bibinfo {year} {1994})}\BibitemShut {NoStop}%
\bibitem [{\citenamefont {{Warburton}}\ and\ \citenamefont
  {{Barack}}(2011)}]{Warburton-Barack:eccentric}%
  \BibitemOpen
  \bibfield  {author} {\bibinfo {author} {\bibfnamefont {N.}~\bibnamefont
  {{Warburton}}}\ and\ \bibinfo {author} {\bibfnamefont {L.}~\bibnamefont
  {{Barack}}},\ }\href {\doibase 10.1103/PhysRevD.83.124038} {\bibfield
  {journal} {\bibinfo  {journal} {\prd}\ }\textbf {\bibinfo {volume} {83}},\
  \bibinfo {eid} {124038} (\bibinfo {year} {2011})},\ \Eprint
  {http://arxiv.org/abs/1103.0287} {arXiv:1103.0287} \BibitemShut {NoStop}%
\bibitem [{\citenamefont {Diener}\ \emph {et~al.}({\natexlab{b}})\citenamefont
  {Diener}, \citenamefont {Warburton},\ and\ \citenamefont
  {Wardell}}]{Diener_etal:dG_code}%
  \BibitemOpen
  \bibfield  {author} {\bibinfo {author} {\bibfnamefont {P.}~\bibnamefont
  {Diener}}, \bibinfo {author} {\bibfnamefont {N.}~\bibnamefont {Warburton}}, \
  and\ \bibinfo {author} {\bibfnamefont {B.}~\bibnamefont {Wardell}},\
  }\href@noop {} {\enquote {\bibinfo {title} {{A Discontinuous Galerkin Code
  for High-Precision Time-Domain Self-force Calculations}},}\ }
  ({\natexlab{b}}),\ \bibinfo {note} {in preparation}\BibitemShut {NoStop}%
\bibitem [{\citenamefont {Hopper}(2018)}]{Hopper:2017iyq}%
  \BibitemOpen
  \bibfield  {author} {\bibinfo {author} {\bibfnamefont {S.}~\bibnamefont
  {Hopper}},\ }\href {\doibase 10.1103/PhysRevD.97.064007} {\bibfield
  {journal} {\bibinfo  {journal} {Phys. Rev.}\ }\textbf {\bibinfo {volume}
  {D97}},\ \bibinfo {pages} {064007} (\bibinfo {year} {2018})},\ \Eprint
  {http://arxiv.org/abs/1706.05455} {arXiv:1706.05455 [gr-qc]} \BibitemShut
  {NoStop}%
\bibitem [{\citenamefont {Colleoni}\ \emph {et~al.}(2018)\citenamefont
  {Colleoni}, \citenamefont {Isoyama}, \citenamefont {Sago},\ and\
  \citenamefont {Barack}}]{Colleoni:Soichiro:Sago:Barack:2018}%
  \BibitemOpen
  \bibfield  {author} {\bibinfo {author} {\bibfnamefont {M.}~\bibnamefont
  {Colleoni}}, \bibinfo {author} {\bibfnamefont {S.}~\bibnamefont {Isoyama}},
  \bibinfo {author} {\bibfnamefont {N.}~\bibnamefont {Sago}}, \ and\ \bibinfo
  {author} {\bibfnamefont {L.}~\bibnamefont {Barack}},\ }\href@noop {}
  {\enquote {\bibinfo {title} {In preparation},}\ } (\bibinfo {year} {2018}),\
  \bibinfo {note} {in preparation}\BibitemShut {NoStop}%
\bibitem [{\citenamefont {Barton}\ \emph {et~al.}(2008)\citenamefont {Barton},
  \citenamefont {Lazar}, \citenamefont {Kennefick}, \citenamefont {Khanna},\
  and\ \citenamefont {Burko}}]{Barton:2008eb}%
  \BibitemOpen
  \bibfield  {author} {\bibinfo {author} {\bibfnamefont {J.~L.}\ \bibnamefont
  {Barton}}, \bibinfo {author} {\bibfnamefont {D.~J.}\ \bibnamefont {Lazar}},
  \bibinfo {author} {\bibfnamefont {D.~J.}\ \bibnamefont {Kennefick}}, \bibinfo
  {author} {\bibfnamefont {G.}~\bibnamefont {Khanna}}, \ and\ \bibinfo {author}
  {\bibfnamefont {L.~M.}\ \bibnamefont {Burko}},\ }\href {\doibase
  10.1103/PhysRevD.78.064042} {\bibfield  {journal} {\bibinfo  {journal} {Phys.
  Rev.}\ }\textbf {\bibinfo {volume} {D78}},\ \bibinfo {pages} {064042}
  (\bibinfo {year} {2008})},\ \Eprint {http://arxiv.org/abs/0804.1075}
  {arXiv:0804.1075 [gr-qc]} \BibitemShut {NoStop}%
\bibitem [{\citenamefont {Zenginoglu}(2011)}]{Zenginoglu:2010cq}%
  \BibitemOpen
  \bibfield  {author} {\bibinfo {author} {\bibfnamefont {A.}~\bibnamefont
  {Zenginoglu}},\ }\href {\doibase 10.1016/j.jcp.2010.12.016} {\bibfield
  {journal} {\bibinfo  {journal} {J. Comput. Phys.}\ }\textbf {\bibinfo
  {volume} {230}},\ \bibinfo {pages} {2286} (\bibinfo {year} {2011})},\ \Eprint
  {http://arxiv.org/abs/1008.3809} {arXiv:1008.3809} \BibitemShut {NoStop}%
\bibitem [{\citenamefont {Wardell}\ \emph {et~al.}(2014)\citenamefont
  {Wardell}, \citenamefont {Galley}, \citenamefont {Zenginoğlu}, \citenamefont
  {Casals}, \citenamefont {Dolan},\ and\ \citenamefont
  {Ottewill}}]{Wardell_etal:SSF_via_Green_functions}%
  \BibitemOpen
  \bibfield  {author} {\bibinfo {author} {\bibfnamefont {B.}~\bibnamefont
  {Wardell}}, \bibinfo {author} {\bibfnamefont {C.~R.}\ \bibnamefont {Galley}},
  \bibinfo {author} {\bibfnamefont {A.}~\bibnamefont {Zenginoğlu}}, \bibinfo
  {author} {\bibfnamefont {M.}~\bibnamefont {Casals}}, \bibinfo {author}
  {\bibfnamefont {S.~R.}\ \bibnamefont {Dolan}}, \ and\ \bibinfo {author}
  {\bibfnamefont {A.~C.}\ \bibnamefont {Ottewill}},\ }\href {\doibase
  10.1103/PhysRevD.89.084021} {\bibfield  {journal} {\bibinfo  {journal} {Phys.
  Rev.}\ }\textbf {\bibinfo {volume} {D89}},\ \bibinfo {pages} {084021}
  (\bibinfo {year} {2014})},\ \Eprint {http://arxiv.org/abs/1401.1506}
  {arXiv:1401.1506} \BibitemShut {NoStop}%
\bibitem [{\citenamefont {Field}\ \emph {et~al.}(2009)\citenamefont {Field},
  \citenamefont {Hesthaven},\ and\ \citenamefont {Lau}}]{Field:2009kk}%
  \BibitemOpen
  \bibfield  {author} {\bibinfo {author} {\bibfnamefont {S.~E.}\ \bibnamefont
  {Field}}, \bibinfo {author} {\bibfnamefont {J.~S.}\ \bibnamefont
  {Hesthaven}}, \ and\ \bibinfo {author} {\bibfnamefont {S.~R.}\ \bibnamefont
  {Lau}},\ }\href {\doibase 10.1088/0264-9381/26/16/165010} {\bibfield
  {journal} {\bibinfo  {journal} {Class. Quant. Grav.}\ }\textbf {\bibinfo
  {volume} {26}},\ \bibinfo {pages} {165010} (\bibinfo {year} {2009})},\
  \Eprint {http://arxiv.org/abs/0902.1287} {arXiv:0902.1287} \BibitemShut
  {NoStop}%
\bibitem [{\citenamefont {Warburton}\ and\ \citenamefont
  {Wardell}(2014)}]{Warburton:2013lea}%
  \BibitemOpen
  \bibfield  {author} {\bibinfo {author} {\bibfnamefont {N.}~\bibnamefont
  {Warburton}}\ and\ \bibinfo {author} {\bibfnamefont {B.}~\bibnamefont
  {Wardell}},\ }\href {\doibase 10.1103/PhysRevD.89.044046} {\bibfield
  {journal} {\bibinfo  {journal} {Phys. Rev.}\ }\textbf {\bibinfo {volume}
  {D89}},\ \bibinfo {pages} {044046} (\bibinfo {year} {2014})},\ \Eprint
  {http://arxiv.org/abs/1311.3104} {arXiv:1311.3104} \BibitemShut {NoStop}%
\bibitem [{\citenamefont {Wardell}\ and\ \citenamefont
  {Warburton}(2015)}]{Wardell:2015ada}%
  \BibitemOpen
  \bibfield  {author} {\bibinfo {author} {\bibfnamefont {B.}~\bibnamefont
  {Wardell}}\ and\ \bibinfo {author} {\bibfnamefont {N.}~\bibnamefont
  {Warburton}},\ }\href {\doibase 10.1103/PhysRevD.92.084019} {\bibfield
  {journal} {\bibinfo  {journal} {Phys. Rev.}\ }\textbf {\bibinfo {volume}
  {D92}},\ \bibinfo {pages} {084019} (\bibinfo {year} {2015})},\ \Eprint
  {http://arxiv.org/abs/1505.07841} {arXiv:1505.07841} \BibitemShut {NoStop}%
\bibitem [{\citenamefont {Miller}\ \emph {et~al.}(2016)\citenamefont {Miller},
  \citenamefont {Wardell},\ and\ \citenamefont {Pound}}]{Miller:2016hjv}%
  \BibitemOpen
  \bibfield  {author} {\bibinfo {author} {\bibfnamefont {J.}~\bibnamefont
  {Miller}}, \bibinfo {author} {\bibfnamefont {B.}~\bibnamefont {Wardell}}, \
  and\ \bibinfo {author} {\bibfnamefont {A.}~\bibnamefont {Pound}},\ }\href
  {\doibase 10.1103/PhysRevD.94.104018} {\bibfield  {journal} {\bibinfo
  {journal} {Phys. Rev.}\ }\textbf {\bibinfo {volume} {D94}},\ \bibinfo {pages}
  {104018} (\bibinfo {year} {2016})},\ \Eprint
  {http://arxiv.org/abs/1608.06783} {arXiv:1608.06783} \BibitemShut {NoStop}%
\bibitem [{\citenamefont {Thornburg}\ and\ \citenamefont
  {Wardell}(2017)}]{Thornburg:2016msc}%
  \BibitemOpen
  \bibfield  {author} {\bibinfo {author} {\bibfnamefont {J.}~\bibnamefont
  {Thornburg}}\ and\ \bibinfo {author} {\bibfnamefont {B.}~\bibnamefont
  {Wardell}},\ }\href {\doibase 10.1103/PhysRevD.95.084043} {\bibfield
  {journal} {\bibinfo  {journal} {Phys. Rev.}\ }\textbf {\bibinfo {volume}
  {D95}},\ \bibinfo {pages} {084043} (\bibinfo {year} {2017})},\ \Eprint
  {http://arxiv.org/abs/1610.09319} {arXiv:1610.09319} \BibitemShut {NoStop}%
\bibitem [{\citenamefont {Burko}(2000{\natexlab{a}})}]{Burko:2000a}%
  \BibitemOpen
  \bibfield  {author} {\bibinfo {author} {\bibfnamefont {L.~M.}\ \bibnamefont
  {Burko}},\ }\href@noop {} {\bibfield  {journal} {\bibinfo  {journal} {Class.
  Quantum Grav.}\ }\textbf {\bibinfo {volume} {17}},\ \bibinfo {pages} {227}
  (\bibinfo {year} {2000}{\natexlab{a}})}\BibitemShut {NoStop}%
\bibitem [{\citenamefont {Wiseman}(2000)}]{Wiseman:2000}%
  \BibitemOpen
  \bibfield  {author} {\bibinfo {author} {\bibfnamefont {A.~G.}\ \bibnamefont
  {Wiseman}},\ }\href {\doibase 10.1103/PhysRevD.61.084014} {\bibfield
  {journal} {\bibinfo  {journal} {Phys. Rev.}\ }\textbf {\bibinfo {volume}
  {D61}},\ \bibinfo {pages} {084014} (\bibinfo {year} {2000})},\ \Eprint
  {http://arxiv.org/abs/gr-qc/0001025} {arXiv:gr-qc/0001025} \BibitemShut
  {NoStop}%
\bibitem [{\citenamefont {Burko}(2000{\natexlab{b}})}]{Burko:2000b}%
  \BibitemOpen
  \bibfield  {author} {\bibinfo {author} {\bibfnamefont {L.~M.}\ \bibnamefont
  {Burko}},\ }\href@noop {} {\bibfield  {journal} {\bibinfo  {journal} {Phys.
  Rev. Lett.}\ }\textbf {\bibinfo {volume} {84}},\ \bibinfo {pages} {4529}
  (\bibinfo {year} {2000}{\natexlab{b}})}\BibitemShut {NoStop}%
\bibitem [{\citenamefont {{Price}}(1972)}]{1972PhRvD...5.2419P}%
  \BibitemOpen
  \bibfield  {author} {\bibinfo {author} {\bibfnamefont {R.~H.}\ \bibnamefont
  {{Price}}},\ }\href {\doibase 10.1103/PhysRevD.5.2419} {\bibfield  {journal}
  {\bibinfo  {journal} {\prd}\ }\textbf {\bibinfo {volume} {5}},\ \bibinfo
  {pages} {2419} (\bibinfo {year} {1972})}\BibitemShut {NoStop}%
\bibitem [{\citenamefont {Casals}\ \emph {et~al.}(2013)\citenamefont {Casals},
  \citenamefont {Dolan}, \citenamefont {Ottewill},\ and\ \citenamefont
  {Wardell}}]{Casals:2013mpa}%
  \BibitemOpen
  \bibfield  {author} {\bibinfo {author} {\bibfnamefont {M.}~\bibnamefont
  {Casals}}, \bibinfo {author} {\bibfnamefont {S.}~\bibnamefont {Dolan}},
  \bibinfo {author} {\bibfnamefont {A.~C.}\ \bibnamefont {Ottewill}}, \ and\
  \bibinfo {author} {\bibfnamefont {B.}~\bibnamefont {Wardell}},\ }\href
  {\doibase 10.1103/PhysRevD.88.044022} {\bibfield  {journal} {\bibinfo
  {journal} {Phys. Rev.}\ }\textbf {\bibinfo {volume} {D88}},\ \bibinfo {pages}
  {044022} (\bibinfo {year} {2013})},\ \Eprint {http://arxiv.org/abs/1306.0884}
  {arXiv:1306.0884} \BibitemShut {NoStop}%
\bibitem [{\citenamefont {Warburton}\ and\ \citenamefont
  {Barack}(2010)}]{Warburton:Barack:2009}%
  \BibitemOpen
  \bibfield  {author} {\bibinfo {author} {\bibfnamefont {N.}~\bibnamefont
  {Warburton}}\ and\ \bibinfo {author} {\bibfnamefont {L.}~\bibnamefont
  {Barack}},\ }\href {\doibase 10.1103/PhysRevD.81.084039} {\bibfield
  {journal} {\bibinfo  {journal} {Phys. Rev.}\ }\textbf {\bibinfo {volume}
  {D81}},\ \bibinfo {pages} {084039} (\bibinfo {year} {2010})},\ \Eprint
  {http://arxiv.org/abs/arXiv:1003.1860} {arXiv:arXiv:1003.1860} \BibitemShut
  {NoStop}%
\bibitem [{\citenamefont {Warburton}(2015)}]{Warburton:2014bya}%
  \BibitemOpen
  \bibfield  {author} {\bibinfo {author} {\bibfnamefont {N.}~\bibnamefont
  {Warburton}},\ }\href {\doibase 10.1103/PhysRevD.91.024045} {\bibfield
  {journal} {\bibinfo  {journal} {Phys. Rev.}\ }\textbf {\bibinfo {volume}
  {D91}},\ \bibinfo {pages} {024045} (\bibinfo {year} {2015})},\ \Eprint
  {http://arxiv.org/abs/1408.2885} {arXiv:1408.2885} \BibitemShut {NoStop}%
\bibitem [{\citenamefont {{Wolfram Research, Inc.}}(2016)}]{Mathematica}%
  \BibitemOpen
  \bibfield  {author} {\bibinfo {author} {\bibnamefont {{Wolfram Research,
  Inc.}}},\ }\href@noop {} {\emph {\bibinfo {title} {Mathematica}}},\ \bibinfo
  {edition} {{Version 10.4}}\ ed.\ (\bibinfo  {publisher} {Wolfram Research,
  Inc.},\ \bibinfo {address} {Champaign, Illinois},\ \bibinfo {year}
  {2016})\BibitemShut {NoStop}%
\bibitem [{\citenamefont {{Mart{\'{\i}}n-Garc{\'{\i}}a}}(2008)}]{xTensor}%
  \BibitemOpen
  \bibfield  {author} {\bibinfo {author} {\bibfnamefont {J.~M.}\ \bibnamefont
  {{Mart{\'{\i}}n-Garc{\'{\i}}a}}},\ }\href {\doibase
  10.1016/j.cpc.2008.05.009} {\bibfield  {journal} {\bibinfo  {journal}
  {Computer Physics Communications}\ }\textbf {\bibinfo {volume} {179}},\
  \bibinfo {pages} {597} (\bibinfo {year} {2008})},\ \Eprint
  {http://arxiv.org/abs/0803.0862} {arXiv:0803.0862 [cs.SC]} \BibitemShut
  {NoStop}%
\bibitem [{xTe()}]{xTensorOnline}%
  \BibitemOpen
  \href@noop {} {}\bibinfo {howpublished}
  {\url{http://metric.iem.csic.es/Martin-Garcia/xAct/}}\BibitemShut {NoStop}%
\bibitem [{Bla()}]{BlackHolePerturbationToolkit}%
  \BibitemOpen
  \href@noop {} {\enquote {\bibinfo {title} {{Black Hole Perturbation
  Toolkit}},}\ }\bibinfo {howpublished} {\url{bhptoolkit.org}}\BibitemShut
  {NoStop}%
\end{thebibliography}%

\end{document}